\documentclass[12pt, a4paper]{article} 

\usepackage[margin=1in]{geometry} %page size
\usepackage{amsfonts, amscd, amssymb, mathtools, mathrsfs, dsfont, bbm, bbding} %math and other symbols
\usepackage[amsmath, amsthm, thmmarks]{ntheorem} %theorem style, auto qedhere 
\usepackage{graphicx, xypic, color, float} %graph
\usepackage{indentfirst}%noindent
\usepackage{lmodern}%better output
\usepackage[T1]{fontenc} %better output
\usepackage{enumerate, listings, verbatim, paralist}%list%enumitem
\usepackage{setspace, xspace}%set space, shortward command
\usepackage[colorlinks=true, citecolor=blue]{hyperref}
\usepackage{extarrows}%used to define def
\usepackage{pdfpages}%insert pdf files 

\usepackage{multirow, makecell}
\usepackage{makecell}
\usepackage{url}
\usepackage[table]{xcolor}
\usepackage{booktabs}
\definecolor{gray}{rgb}{0.8, 0.8, 0.8}
\usepackage{natbib}
\bibpunct[, ]{(}{)}{,}{a}{}{,}%

\usepackage{setspace}
\AtBeginDocument{%
\addtolength\abovedisplayskip{-0.15\baselineskip}%
\addtolength\belowdisplayskip{-0.15\baselineskip}%
\addtolength\abovedisplayshortskip{-0.15\baselineskip}%
\addtolength\belowdisplayshortskip{-0.15\baselineskip}%
}

\newtheorem{theorem}{Theorem}

\newtheorem{definition}[theorem]{Definition}

\newtheorem{remark}[theorem]{Remark}

\newtheorem{proposition}[theorem]{Proposition}
\newtheorem{assumption}[theorem]{Assumption}
\newtheorem{eg}{Example}
\newcommand{\argmax}{\mathrm{arg max}}

\usepackage{amsfonts, amscd, amssymb, mathtools, mathrsfs, dsfont, bbm, bbding} %math and other symbols
\usepackage{graphicx, xypic, color, float} %graph

\newcommand{\citethm}[1]{Theorem \ref{#1}}
\newcommand{\citeprop}[1]{Proposition \ref{#1}}

\newcommand{\citeassmp}[1]{Assumption \ref{#1}}
\newcommand{\citeremark}[1]{Remark \ref{#1}}

\newcommand{\citeeg}[1]{Example \ref{#1}}
\newcommand{\citefig}[1]{Figure \ref{#1}}

\newcommand{\opfont}{\mathbb}

\newcommand{\BE}[2][]{\ensuremath{\operatorname{\opfont{E}}^{#1}\!\left[#2\right]}}
\newcommand{\bp}{\ensuremath{\opfont{P}}}

\newcommand{\BF}{\ensuremath{\mathcal{F}}}

\newcommand{\R}{\ensuremath{\operatorname{\mathbb{R}}}}

\newcommand{\BV}{\ensuremath{{V}}}%\newcommand{\BV}{\ensuremath{\mathcal{V}}}
\newcommand{\BVtwo}{\varphi}%\newcommand{\BV}{\ensuremath{\mathcal{V}}}

\newcommand{\hBV}{\ensuremath{\widehat{\BV}}}
\newcommand{\BW}{\ensuremath{{W}}}%\newcommand{\BW}{\ensuremath{\mathcal{W}}}
\newcommand{\BWtwo}{\zeta}%\newcommand{\BW}{\ensuremath{\mathcal{W}}}
\newcommand{\hBW}{\ensuremath{\widehat{\BW}}}
\newcommand{\BBW}{\ensuremath{\mathbb{W}}}
\newcommand{\BL}{\ensuremath{\mathcal{L}}}
\newcommand{\BBL}{\ensuremath{\mathbb{L}}}

\newcommand{\dd}{\ensuremath{\operatorname{d}\! }}
\newcommand{\dt}{\ensuremath{\operatorname{d}\! t}}
\newcommand{\ds}{\ensuremath{\operatorname{d}\! s}}

\newcommand{\du}{\ensuremath{\operatorname{d}\! u}}

\newcommand{\p}{\ensuremath{\partial}}

\newcommand{\setr}{\mathcal{R}}
\newcommand{\setc}{\mathcal{C}}
\newcommand{\setk}{\mathcal{K}}
\newcommand{\sets}{\mathcal{S}}

\newcommand{\id}[1]{\ensuremath{\operatorname{\mathds{1}}_{\{#1\}}}}

\usepackage[normalem]{ulem} %delete

\definecolor{mygreen}{RGB}{0, 185, 0}

\newcommand{\nn}{\nonumber}

\newcommand{\atextsc}{}
\newcommand{\atextbf}{}
\renewcommand{\geq}{\geqslant}
\renewcommand{\leq}{\leqslant}
\newcommand{\ot}{\overline{T}}

\begin{document}
%%%%%%%%%%%%%%%%
\title{Optimal Investment, Heterogeneous Consumption and Best Time for Retirement}

\author{
Hyun Jin Jang\footnote{School of Business Administration, Ulsan National Institute of Science and Technology (UNIST), Ulsan 44919, Republic of Korea, E-mail: janghj@unist.ac.kr},
\,\,
Zuo Quan Xu\footnote{Department of Applied Mathematics, The Hong Kong Polytechnic University, Kowloon, Hong Kong, E-mail: maxu@polyu.edu.hk},
\,\,
Harry Zheng\footnote{Department of Mathematics, Imperial College, London SW7 2BZ, UK, E-mail: h.zheng@imperial.ac.uk}
}

\maketitle

\begin{abstract}
This paper studies an optimal investment and consumption problem with heterogeneous consumption of basic and luxury goods, together with the choice of time for retirement. The utility for luxury goods is not necessarily a concave function. The optimal heterogeneous consumption strategies for a class of non-homothetic utility maximizer are shown to consume only basic goods when the wealth is small, to consume basic goods and make savings when the wealth is intermediate, and to consume almost all in luxury goods when the wealth is large. The optimal retirement policy is shown to be both universal, in the sense that all individuals should retire at the same level of marginal utility that is determined only by income, labor cost, discount factor as well as market parameters, and not universal, in the sense that all individuals can achieve the same marginal utility with different utility and wealth. It is also shown that individuals prefer to retire as time goes by if the marginal labor cost increases faster than that of income. The main tools used in analyzing the problem are from PDE and stochastic control theory including variational inequality and dual transformation. We finally conduct the simulation analysis for the featured model parameters to investigate practical and economic implications by providing their figures.
\end{abstract}


%%%%%%%%%%%%%%%%%%%%%%%%%%%%%%%%%
\section{Introduction}% (\emr{need update})}
%%%%%%%%%%%%%%%%%%%%%%%%%%%%%%%%%

%%%%%%%%%%%%%%%%%%%%%%%%%%%%%%%%%
\subsection{Literature Review}
%%%%%%%%%%%%%%%%%%%%%%%%%%%%%%%%%
%\noindent
There has been extensive research in optimal investment and consumption in the literature of \cite{KS99}, \cite{HP09}, and references therein for excellent expositions of the topic. In a complete market model with a fixed investment horizon, the problem can be solved with the martingale method and the optimal consumption has a representation in terms of the inverse of the marginal utility of consumption evaluated at the level of the pricing kernel. In an incomplete market model or with optimal stopping time involved, it is no longer possible to use the martingale method to solve the problem as the martingale representation theorem cannot be used or the optimal stopping time needs to be determined. For a Markov model, the optimal investment stopping problem may be studied with the stochastic control method. Using the dynamic programming principle, one can derive the Hamilton-Jacobi-Bellman (HJB) variational inequality which is in general difficult to solve as it involves the determination of the optimal stopping region and the solution of a nonlinear partial differential equation (PDE) in the continuation region. 
%\par

One effective way to solve the variational inequality is to employ the dual transformation method. For an optimal investment stopping decision problem with a mixed power and options type non-smooth non-concave utility function, \cite{GLXY17} apply the concavification and dual transformation method to convert the primal variational inequality to a dual counterpart, then analyze the properties of its solution and the multiple free boundaries, and finally characterize the optimal value and strategy for the original problem. A key advantage of the dual transformation is that one only needs to solve a linear PDE in the continuation region, which is equivalent to an optimal stopping problem and is relatively easier to solve.

Optimal investment and consumption with retirement considering with the income and labor cost factors has been actively studied in a wide range of literature. 
\cite{CS06} construct the solution of the HJB variational inequality with sophisticated technique for the optimal policies with power and log utilities. \cite{CSS08} consider a leisure component for the existing problem by using the martingale method to find the optimal solution under the constant elasticity of substitution utility. 
\cite{DL10} compare three different models (i.e., benchmark, voluntary retirement, and the one with no-borrowing constraint) in terms of optimal investment and consumption strategies. 
There is one common feature of the aforementioned papers: All models concern the lifetime consumption and investment, which lead the dual problem to a perpetual optimal stopping problem with a scalar state variable (discounted pricing kernel) and a single point free boundary determined by the continuity and the smooth pasting condition. 
The corresponding HJB variational inequality reduces to a linear second order ordinary differential equation in the continuation region. 
Unlike them, \cite{YK18} (hereafter, YK) study the optimal problem concerning with early retirement option embedded in mandatory retirement. 

%\par
The literature of optimal investment and consumption usually assumes homothetic consumption utility with single homogeneous consumption goods and no distinction on types of consumption. In the real world, however, a variety kind of consumption can exist from basic goods everyone needs to luxury goods for the rich, which supports 
heterogeneous consumption models are required to reflect non-identical types of consumption behaviors. 
Some literature of financial economics concerns such different types of consumption 
in terms of examining the explanatory power based on the empirical approach. 
\cite{APY04} evaluate the risk of holding equity by specifying a non-homothetic form of utility for both luxury and basic consumption goods and find that basic consumption overstate risk aversion, while the equity premium is not much a puzzle for the very rich. \cite{CC18} considers a model for basic and luxury goods and finds a substantial reduction in precautionary savings in an equilibrium heterogeneous agent model compared to a standard homogeneous model. \cite{WY10} use a non-homothetic consumption model to show that the expenditure share for basic goods declines in total consumption and the variance of consumption growth rises in the level of consumption, consistent with empirical findings.

%%%%%%%%%%% %%%%%%%%%%%%%%%%%%%%%%
\subsection{Motivation}% (\emr{need update} )}
%%%%%%%%%%%%%%%%%%%%%%%%%%%%%%%%%
%\noindent
The only literature on optimal heterogeneous consumption with retirement the authors are aware of is a conference presentation \citep{KRS17}, which discusses quadratic and HARA utilities for basic and luxury consumption, respectively, including voluntary retirement with the income and labor cost factor. They first solve a post-retirement problem and then the pre-retirement one, giving closed form solutions without presenting proofs. Assuming all results are correct in \cite{KRS17}, their model still contains some significant deficiencies: 
(i) no mandatory retirement age is specified, which is far from realistic; 
(ii) income and labor cost are constants over time, yet both should increase and the marginal labor cost should be greater than the marginal income in real life;
(iii) a quadratic function is not suitable for the utility as increasing and decreasing, but utility function should be non-decreasing. 
Mathematically, the model by \cite{KRS17} enables to convert the optimal stopping to a perpetual American option pricing problem. 
This ensures to solve the problem much easier than a similar finite maturity counterpart since one only needs to determine a single number to separate the continuation and exercise regions in the perpetual case.

The study of YK can be the closest work to ours in terms of methodologies -- first to solve a post-retirement problem, then move on a pre-retirement problem with dual transformation and variational inequality, finally to find the free boundary of the continuation and stopping regions. 
The analogous approach is employed by \cite{GLXY17, MXZ19}. 
Main differences of YK with our study are as follows: YK consider single homogeneous consumption with strictly concave and infinitely differential utility, whereas we have heterogeneous consumption with non-concave and non-differential utility, leading our problem to be more complicated and resulting in considerably different consumption behavior.
YK assume all constant model coefficient, whereas ours take time-dependent, which makes our model more realistic and results in a non-monotonic free boundary, in sharp contrast to the globally increasing free boundary of YK. Furthermore, 
YK has a complete market model, whereas we can handle closed convex cone control constraints.

%%%%%%%%%%%%%%%%%%%%%%%%%%%%%%%%%
\subsection{Our Contributions }
%%%%%%%%%%%%%%%%%%%%%%%%%%%%%%%%%
\noindent
This study considers two differentiated features for the dynamic portfolio and consumption problem. 
First, we build heterogeneous consumption, namely, the basic and luxury goods, where the utility over the luxury consumption can be non-concave.
This setup can reflect a phenomenon that one gets satisfaction from consuming the luxury only upon one can afford to spend sufficiently large amount of wealth on it, and the luxury utility is zero otherwise. 
Yet, it ensures to maintain the utility for total consumption with an increasing but not necessarily concave form. 
Second, our model allows early retirement option before a finite mandatory retirement age. The retirement time relies on relative tradeoff between the income and labor cost while working. 
\par
Under this framework we solve the heterogeneous consumption problem by firstly finding the optimal total consumption and then having the optimal basic and luxury consumptions by applying local optimization (see \citeeg{nonconcave}). 
Under the bisectioned consumption utility, we show the optimal consumption strategies are to consume only basic goods for the case of small wealth; to consume the basic with more investment and savings for the case of intermediate wealth; and to consume almost the luxury for the case of high wealth (see \citeprop{largeconsumption}). 
\par
On the optimal retirement, we divide the original problem into the post and the pre-retirement problems. 
{For both, the value function and the optimal strategies for wealth, trading strategy, and consumption are determined by the dual value function and the dual state process (see Theorems \ref{postsolution1} and \ref{presolution}). 
The two problems have the following key difference: 
in the post-retirement problem, the dual function satisfies a linear PDE and can be computed like European options and the dual state process can be determined by the fixed proportions of investment in the risk-free and risky assets when all coefficients are constants (see Remark \ref{onefund}).
In the pre-retirement, the dual function satisfies a variational inequality with a linear PDE in the continuation region, which can be treated as American options.} 
Since closed-form solutions are usually not available for finite horizon optimal stopping problems, we provide a numerical method to solve the variational inequality. 

Our results show that 
the optimal retirement policy is universal in the sense that all individuals should retire at the same level of the marginal utility determined by market factors and income/labor cost, whereas it is not universal in the viewpoint that individuals retire at the different wealth levels owing to their own utility preferences (see Proposition \ref{Universal_stopping_region}). 
With more realistic assumptions, we prove that individuals reaching at mandatory retirement age are more likely to retire as time goes by.

The main mathematical tools used in analyzing the problems are the stochastic control theory including variational inequality and dual transformation. 
In particular, we introduce dual transformation for a nonlinear variational inequality, turning it into a linear one. Although the main results of the study are constructed in a complete market setup, the similar ones hold for the portfolio trading strategies constrained in a closed convex cone. 
This extension covers interesting cases such as not-allowance of short selling or unavailability of certain stocks for trading \citep{KS99}. 
We emphasize that adding bounded portfolio constraints may lead to the concavification principle invalid, which could give significant economic insights \citep{DKQW19}. 

As the theoretical side, this study contributes to solve the complicated decision problem on investment and consumption considering with early retirement as well as heterogeneous consumption features by decomposing into a number of simpler decision problems:
(i) the optimal allocation of basic and luxury consumption can be decided once the total consumption is known;
(ii) the optimal stopping time and the working/retirement regions can be determined with the marginal utility (or, the dual state variable); 
(iii) the optimal wealth, investment, and total consumption can be computed with the dual variational inequality method.

Such separation principles facilitate to work with numerical analysis as a sequential procedure. With this benefit, this study provides intensive simulation results to further investigate practical economic implications. 
Testing the impact of the key factors including trade-off of working income/cost, wealth endowment, and non-concavity for utility to the consumption and early retirement decisions, our major findings are as follows: 
The net income from salary and cost required for lifetime play an important role to determine the optimal retirement boundary. 
Also, the initial wealth granted makes a significant difference of consumers' behavior for the basic and the luxury as well as the early retirement decision.
Our test demonstrates that it is an optimal decision that high-net worth individuals have no reason to be working and the poor should be working longer and harder.
The factor causing non-concavity affects heterogeneous consumers' behavior and their retirement time.
More precisely, the higher degree of non-concavity tends to decide earlier retirement, and also to allow consuming the luxury more than the basic.

\par
The rest of the paper is organized as follows. Section~\ref{Sec:Model} formulates the optimal investment consumption model with non-homothetic consumption utility.
Section~\ref{Sec:Post} solves the post-retirement optimal problem, while 
Section~\ref{Sec:Pre} discusses the pre-retirement optimal problem concerning the optimal stopping problem. 
Section\ref{Sec:Retire} finally characterizes the optimal retirement region with the free boundary independent of consumption utility. 
Section~\ref{Sec:Simulation} presents the simulation results regarding the featured variables of the proposed model, and Section~\ref{Sec:Conclusion} concludes the paper.

%%%%%%%%%%%%%%%%%%%%%%%%%%%%%%%%%%%%%%%%
%\subsection*{Notation}
%\noindent
We make use of the following notation throughout this paper:
\begin{itemize}
\item $\mathbb{R}^n$, the $n$-dimensional real Euclidean space;
\item $M^{\top}$, the transpose of a matrix or vector $M$;
\item $\Vert M\Vert=\sqrt{\sum_{i, j}m_{ij}^2}$, the $L^{2}$-norm for a matrix or vector $M=(m_{ij})$. 
\end{itemize}
The uncertainty of the financial market is generated by a standard $\{{\BF}_t \}_{t \geqslant 0}$-adapted $n$-dimensional Brownian motion $B(\cdot) \equiv (B_{1}(\cdot), \ldots, B_{n}(\cdot))^{\top}$ defined on a fixed filtered complete probability space $(\Omega, \mathbf{F}, \bp, \{{\BF}_t \}_{t \geqslant 0})$.
For two functions $f$ and $g$, we write $f(x)\ll g(x)$ if there exists a constant $C>0$ such that $|f(x)|\leqslant C g(x)$ for all $x$ sufficiently large. 
We use $C$ as a generic positive constant, which can take different values at different places.

{
\begin{definition}
For $0<q<1$, (i) a function $f: (0, +\infty)\to (0, +\infty)$ is called power-like decreasing {indexed $q$ (PLD($q$))} if it is continuous, strictly decreasing, and satisfies $\lim_{x\to 0}f(x)=+\infty$ and $f(x)\ll x^{\frac{q}{q-1}}$ as $x\to+\infty$; and
(ii) $f$ is called {power-like increasing {indexed $q$ (PLI($q$))}} if it is continuous, strictly increasing, and satisfies $\lim_{x\to 0} f(x)=0$, $\lim_{x\to+\infty} f(x)=+\infty$ and $f(x)\ll x^{q}$ as $x\to+\infty$. 
\end{definition}}

%%%%%%%%%%%%%%%%%%%%%%%%%%%%%%%%%
\smallskip
\section{Model Formulation}\label{Sec:Model}
%%%%%%%%%%%%%%%%%%%%%%%%%%%%%%%%%
\noindent 
We consider an arbitrage-free Black-Scholes financial market with one risk-free savings account satisfying $dS_0(t)=r(t)S_0(t)dt$ and $r(t)$ the risk-free interest rate at time $t$, and $n$ risky stocks satisfying the following stochastic differential equations (SDEs):
\begin{align*}
\dd S_{i}(t)=S_{i}(t)\left(b_i(t)\dt+\sum\limits_{j=1}^n\sigma_{ij}(t)\dd B_{j}(t)\right), \quad t \geqslant 0, \quad i=1, 2, \ldots, n, 
\end{align*}
where $b_i(t)$ is the growth rate of the stock $i$ and $\sigma_{ij}(t)$ the volatility coefficient between the stock $i$ and risk source $B_{j}$ at time $t$. Define the volatility matrix $\sigma(t)=(\sigma_{ij}(t))$ and the excess return vector $\mu(t)=(b_{1}(t)-r(t), \ldots, b_n(t)-r(t))^{\top}$. 
\par
Consider a representative individual (``He''). Before the retirement time $\tau$, the individual has an income 
$I(t)$ at time $t$, and has no income after retirement. At any time $t$, before or after retirement, he consumes $c(t)$ on the basic goods and $g(t)$ on the luxury goods, also invests $\pi_i(t)$ dollars in the stock $i$. His wealth process, $X(t)$, evolves according to \cite{KS99}
\begin{align}\label{wealth1}
\dd X(t)=(r(t)X(t)+\pi(t)^{\top}\mu(t)+I(t)\id{t\leq\tau}-c(t)-g(t))\dt+\pi(t)^{\top}\sigma(t)\dd B(t), 
\end{align} 
where $\pi(t):=(\pi_1(t), \ldots, \pi_n(t))^{\top}$ is called a portfolio, which is equivalent to holding $\pi_i(t)/S_{i}(t)$ units of stock $i$ at time $t$ for $i=1, \ldots n$. We assume that $\pi$ is progressively measurable and square integrable and $c, g$ are non-negative, adapted and integrable. 
There is also a finite mandatory retirement age $T$, so the individual must choose a retirement time $\tau$ no later than $T$.

The individual aims to find a strategy, which consists of a portfolio $\pi^{*}$ for stock trading, a non-negative consumption rate $c^{*}$ on the basic goods, a non-negative consumption rate $g^{*}$ on the luxury goods, and a retirement time $\tau^{*}$ no late than $T$, to maximize the expected utility, namely, 
%\blue
\begin{align}\label{p1}
\sup_{\pi, c, g, \tau}\;\BE{\int_{0}^{\ot} e^{-\int_0^t \rho(s)\ds}u(c(t), g(t))\dt-\int_{0}^{\tau} e^{-\int_0^t \rho(s)\ds}L(t)\dt }, 
\end{align}
where %\greenone
{$\ot$ is the total investment/consumption period and bigger than $T$, $\rho(s)$ is the discount rate at time $s$, } $u$ is his utility function on two types of consumption, and $L(\cdot)$ denotes the labor cost process before retirement to represent his effort to work. 
%\blue

Throughout this paper we assume that $r(\cdot)$, $\mu(\cdot)$, $\rho(\cdot)$, $I(\cdot)$, $L(\cdot)$ are (deterministic) positive continuous bounded functions on $[0,\ot]$. Moreover $\sigma(\cdot)$ is continuous and there exists a positive constant $C$ such that $ C^{-1} z^{\top}z\leq z^{\top}\sigma(t)\sigma(t)^{\top}z\leq C z^{\top}z$ for any $z\in\R^n$ and $t\in [0,\ot]$.

\begin{remark} 
%\blue
{This paper assumes $\ot$ is finite though, it can be easily extended to $\ot=\infty$ with additional requirement that $r, \mu, \sigma, \rho$ are all constants to make the model time-invariant after retirement. We can also introduce an exogenous random variable $\eta$ to represent the death time into the model. If $\eta$ takes values in $[0, \ot]$ and has the cumulative distribution function $F$ and is independent of the market, then problem \eqref{p1} becomes
\begin{align}\label{p1a}
\sup_{\pi, c, g, \tau}\;\BE{\int_{0}^{\eta} e^{-\int_0^t \rho(s)\ds}u(c(t), g(t))\dt-\int_{0}^{\tau\wedge \eta} e^{-\int_0^t \rho(s)\ds}L(t)\dt }.
\end{align}
Taking the expectation over $\eta$, we can write \eqref{p1a} equivalently as
\begin{align}\label{p1aa}
\sup_{\pi, c, g, \tau}\;\BE{\int_{0}^{\ot} e^{-\int_0^t \rho(s)\ds} (1-F(t))u(c(t), g(t))\dt-\int_{0}^{\tau} e^{-\int_0^t \rho(s)\ds} (1-F(t))L(t)\dt}.
\end{align} 
%\greenone
{If we further assume $F$ is differentiable, then \eqref{p1aa} can be written as
\begin{align}\label{p1aa2}
\sup_{\pi, c, g, \tau}\;\BE{\int_{0}^{\ot} e^{-\int_0^t (\rho(s)+\lambda(s))ds}u(c(t), g(t))\dt-\int_{0}^{\tau} e^{-\int_0^t (\rho(s)+\lambda(s))ds}L(t)\dt}
\end{align} 
where $\lambda(s)=\frac{F'(s)}{1-F(s)}$. In particular, if } $\eta$ is an exponential variable with parameter $\lambda$ (mortality rate) and $\ot=\infty$, then \eqref{p1aa} reduces to the infinite horizon problem with $\rho$ replaced by $\rho+\lambda$ (assuming $\rho$ is a constant). }
\end{remark}

\begin{remark}
There is no control constraint in \eqref{wealth1}, that is, $\pi(t)\in \R^n$ for all $t$. We can show the results of the paper also hold with control constraints $\pi(t)\in\setk$ for all $t$, where $\setk$ is a closed convex cone in $\R^n$. There is a unique pricing kernel in the complete market model, but there are infinitely many of them in the presence of control constraints. With a closed convex cone constraint, one can {choose the} so-called \emph{minimum pricing kernel} with which one can find the replicating portfolio trading strategy and the model is just like a standard complete market model, see \cite[Section 4]{HZ11}, Remarks \ref{K1} and \ref{K2} below. 
There is no utility of terminal wealth for problem (\ref{p1}), which would make the optimal terminal wealth zero as one has no incentive to leave any wealth and would rather consume it all by terminal time $\ot$. We can easily include utility of terminal wealth as the post-retirement problem can be solved with standard utility maximization, see \cite{KS99}.
\end{remark}

Constant elasticity of substitution function is usually used to combine multiple consumption goods or production inputs into an aggregate quantity in economics. This aggregator function exhibits constant elasticity of substitution. It also arises as a utility function in consumer theory. Due to different nature of consumptions on basic and luxury goods, in this paper, we restrict ourselves to the following type of {non-homothetic utility function}
\[u(c, g)=U\bigg(\frac{c^{\alpha}}{\alpha}+\frac{((g-a)^{+})^{\beta}}{\beta}\bigg), \]
where $0<\alpha<\beta<1$, $U$ is a differentiable, strictly increasing, {unbounded} concave function, and $a>0$ represents the starting level of luxury consumption. The function $u(c, g)$ is increasing in $c$ and $g$, respectively. It is {strictly} concave in $c$ globally on $(0, \infty)$. It is also {strictly} concave in $g$ on $(a, \infty)$, {but not concave} on $(0, \infty)$.

We reformulate problem \eqref{p1} to make it more attractable and study the behavior of optimal consumption. 
Define the total utility of consumption at level $k\geq 0$ by
\begin{align*}
\bar{u}(k)=\max_{\substack{c, \;g\geq 0, \\c+g=k}}u(c, g). 
\end{align*}
Once the best total consumption $k$ at a time is determined, then the best consumption $c^{*}(k)$ on the basic goods and $g^{*}(k)$ on the luxury goods at that time are determined by the above optimization problem. Therefore, there is no need to consider separately the best consumption on basic and that on luxury goods; we only need to consider the best total consumption. 
\par 
Clearly, $\bar{u}(\cdot)$ is a strictly increasing, unbounded function. Because one consumes no luxury goods when $k<a$, the total utility function $\bar{u}(k)$ is equal to $u(k, 0)$; thus, 
\begin{align}\label{unboundedu'}
\lim_{k\to 0^{+}}\bar{u}'(k)=+\infty.
\end{align} 
Economically speaking, it motives the individual to consume basic goods even if his wealth is very small. Furthermore, it is easily seen that 
\begin{align}\label{barugrowth}
\bar{u}(k)\ll k^{\beta}
\end{align} 
for sufficiently large $k$, so $\bar{u}(\cdot)$ satisfies the power growth condition, {in particular, we have the Inada condition $\lim_{k\to\infty} \bar{u}'(k)=0$}. Therefore, the function $\bar{u}(\cdot)$ is of PLI type. 
%\par 
The following shows our first important conclusion: 
\begin{proposition}\label{largeconsumption}
A non-homothetic utility maximizer only consumes basic goods when the total consumption is small.
His consumption on both basic and luxury goods tends to infinity as the total consumption tends to infinity, but the latter accounts for almost all. Specifically, we have $\lim_{k\to\infty}{c^*(k)}/{k^{\frac{1-\beta}{1-\alpha}}}=1$ and $\lim_{k\to\infty}{g^*(k)}/{c^*(k)}=\infty$. 
\end{proposition}

We now rewrite problem \eqref{p1} as follows 
\begin{align}\label{p2}
\sup_{\tau, k, \pi}\; &\BE{\int_{0}^{\ot} e^{-\int_0^t \rho(s)\ds}\bar{u}(k(t))\dt-\int_{0}^{\tau}e^{-\int_0^t \rho(s)\ds}L(t)\dt}, 
\end{align}
subject to the process 
\begin{align}\label{wealth2}
\dd X(t)=(r(t)X(t)+\pi(t)^\top\mu(t)+I(t)\id{t\leq\tau}-k(t))\dt+\pi(t)^\top\sigma(t)\dd B(t), 
\end{align} 
with a positive initial endowment $X(0)=x$\footnote{The wealth process is normally assumed to be nonnegative for investment consumption problem when there is no income, but can be negative if there is a deterministic income which may offset the negative wealth, see \cite{DMZ} for discussions on that point. 
In our model the wealth process can go negative range within a certain boundary determined by considering the current value of net-income amount being able to earn for a whole lifetime. We assume that the initial wealth $x>0$ without loss of generality as it cannot be generally allowed for all negative values.}.
\par
As the total utility function $\bar{u}(\cdot)$ is not necessarily global concave on $[0, \infty)$, the problem \eqref{p2} is a non-concave utility maximization problem. Non-concave utility maximization problems in general are hard to deal with (e.g., \cite{BCX18} for a similar non-concave utility maximization problem, and \cite{DKQW19} for the case with trading constraint).

\begin{eg}\label{nonconcave} 
The following shows an example of non-homothetic utility function for the total utility function with the optimal basic/luxury consumptions as an explicit form. 
Let $u(c, g)=2c^{1/2}+\frac{4}{3} ((g-a)^{+})^{3/4}$. Then 
{\begin{align*}
\bar{u}(k)=\max_{\substack{c, g\geq 0, \\c+g=k}}u(c, g)=
\begin{cases}
2k^{1/2}, &\quad 0\leq k\leq k_{0};\\
{4\over 3}\sqrt{\sqrt{k-a+\frac{1}{4}}-\frac{1}{2}}\bigg(\sqrt{k-a+\frac{1}{4}}+1\bigg), &\quad k>k_{0}, 
\end{cases} 
\end{align*}
}
where $k_{0}>a$ is the unique solution for 
{
\[{4\over 3}\sqrt{\sqrt{k-a+\frac{1}{4}}-\frac{1}{2}}\bigg(\sqrt{k-a+\frac{1}{4}}+1\bigg)=2k^{1/2}.\]
}
Given a total consumption $k$, the optimal consumption pair on basic and luxury goods is 
\[(c^{*}(k), g^{*}(k))
=\begin{cases}
(k, 0), &\quad 0\leq k\leq k_{0};\\
\left(\sqrt{k-a+\frac{1}{4}}-\frac{1}{2}, \;k+\frac{1}{2}-\sqrt{k-a+\frac{1}{4}}\right), &\quad k> k_{0}.
\end{cases}\]
\citefig{fig:nonconcave} demonstrates the shapes of $\bar{u}(\cdot)$ and its the concave envelope $\widetilde{u}(\cdot)$ (namely, the smallest concave function dominating $\bar{u}(\cdot)$). We can see that $\bar{u}(\cdot)$ is not globally concave. 
%\blue
{Simple computation shows that the concave envelope $\widetilde{u}(\cdot)$ is given by
$$
\widetilde{u}(k)=\left\{\begin{array}{ll}
2k^{1/2}, &\quad 0\leq k\leq k_-;\\
(3a)^{-1/4}k+(3a)^{1/4}, &\quad k_-\leq k\leq k_+;\\
{4\over 3}\sqrt{\sqrt{k-a+\frac{1}{4}}-\frac{1}{2}}\bigg(\sqrt{k-a+\frac{1}{4}}+1\bigg), &\quad k\geq k_+
\end{array}\right.
$$
with $k_-=(3a)^{1/2}$ and $k_+=(3a)^{1/2}+4a$.
}
Later, we will show that the optimal total consumption should never {be} in the range where $\bar{u}$ is apart from its concave envelope, i.e. $(k_{-}, k_{+})$ in \citefig{fig:nonconcave}.

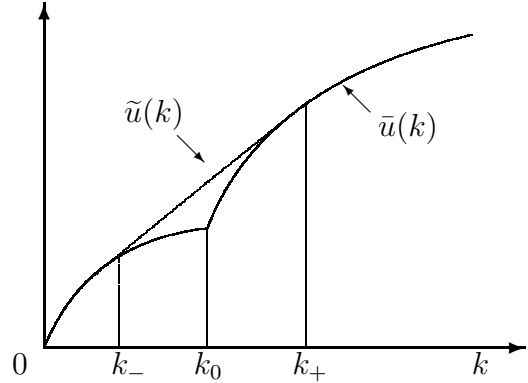
\begin{figure}[H]
\begin{center}
\vspace{-30pt}
\begin{picture}(345, 150)
\thicklines\put(80, -30){\vector(1, 0){180}}
\put(80, -30){\vector(0, 1){130}}\thinlines
\put(68, -40){$0$}
\qbezier(80, -30)(96, 10)(141, 15) 
\qbezier[150](96, -5)(140, 32)(179, 63)
\put(130, 50){\vector(1, -1){10}}
\put(110, 55){$\widetilde{u}(k)$}
\qbezier(141, 15)(162, 70)(240, 88) 
\put(202, 59){\vector(-1, 1){10}}
\put(205, 50){$\bar{u}(k)$}
\put(105, -40){$k_{-}$}
\qbezier[40](108, -30)(108, -14)(108, 4)
\put(136, -40){$k_{0}$}
\qbezier[60](141, -30)(141, -10)(141, 15)
\put(173, -40){$k_{+}$}
\qbezier[120](178, -30)(178, 10)(178, 62)
\put(240, -40){$k$}
\end{picture}\vspace{45pt}
\caption{An example of the total utility function $\bar{u}(\cdot)$ and its concave envelope $\widetilde{u}(\cdot)$.}
\label{fig:nonconcave}\vspace{-20pt}
\end{center}
\end{figure}
\end{eg}

%%%%%%%%%%%%%%%%%%%%%%%%%%%%%%%%%
\smallskip
\section{Post-Retirement Problem}\label{Sec:Post}
%%%%%%%%%%%%%%%%%%%%%%%%%%%%%%%%%
\noindent 
%\blue

In this section we discuss the problem \eqref{p2}. 
After retirement, since there is no income, the wealth process \eqref{wealth2} reduces to 
\begin{align}\label{wealth2-1}
\dd X(t)=(r(t)X(t)+\pi(t)^\top\mu(t)-k(t))\dt+\pi(t)^\top\sigma(t)\dd B(t), \quad \text{for $\tau\leq t\leq \ot$}. 
\end{align} 
Define the value function for the post-retirement problem as 
\begin{align}\label{retirep1}
V(t, x)=\sup_{k, \pi}\; \BE{\int_{t}^{\ot} e^{-\int_t^s \rho(u)du}\bar{u}(k(s))\ds\;\Big|\;X(t)=x}, \quad \tau\leq t\leq \ot,\quad x>0.
\end{align}
The value function $V$ satisfies the following HJB equation:
\begin{multline}\label{hjb1}
\p_t V(t, x)+\sup\limits_{\pi}\left\{\tfrac{1}{2}\Vert\pi^\top\sigma(t)\Vert^{2}\p_{xx} V(t, x)+(r(t)x+\pi^\top\mu(t)) \p_x V(t, x)\right\}\\
-\rho(t) V(t, x)+h(\p_x V(t, x))=0,
\end{multline} 
where $\p_t V$ is the partial derivative of $V$ with respect to $t$, $\p_x V, \p_{xx}V$ are similarly defined, and \[h(y):=\sup_{k\geq0} \big(\bar{u}(k)-ky\big), \quad y>0, \] 
is a convex {PLD($\beta$) function. }
\begin{remark}
One can prove that 
\[h(y)=\sup_{k\geq0} \big(\widetilde{u}(k)-ky\big), \]
where $\widetilde{u}$ is the concave envelope of $\bar{u}$ (see \citefig{fig:nonconcave}). If we replace $\bar{u}$ in problem \eqref{retirep1} by $\widetilde{u}$, the value function $V$ does not change, which implies that one should not consume at the level $k$ if $\bar{u}(k)<\widetilde{u}(k)$; in other words, the optimal total consumption $k^{*}(t)$ should satisfy $\bar{u}(k^{*}(t))\equiv\widetilde{u}(k^{*}(t))$ for all $t$. In \citefig{fig:nonconcave}, this means one consumes either no more than $k_{-}$ or no less than $k_{+}$. If the individual can afford to consume ({say $k_{0}$}) in the middle range $(k_{-}, k_+)$, he would prefer to {consume only to the level $k_-$ and use the unconsumed amount $k_0-k_-$ for investment or saving until he can afford to consume at least to the level $k_+$} to buy an expensive luxury goods later, rather than to buy any luxury goods and satisfy his basic needs at present. 
Non-homothetic consumers would not consider of consuming luxury goods unless their wealth reaches the upper bound of $k_+$.
\end{remark} 

\begin{remark}
By convexity, $h$ is always continuous, but may not be differentiable. In fact, $h$ is differentiable if and only if $\bar{u}$ is strictly concave; see, e.g., \cite[Lemma 7.2]{XY16}. In particular, $h$ is not differentiable at $y_0$ that is the slope of the tangent line of the concave envelope $\widetilde{u}$, connecting the two pieces of strictly concave curves of $\bar{u}$. 
\end{remark}

Using the dual control theory, see, e.g., \cite[Remark 2.7, Theorem 3.8]{BMZ11}, we characterize 
the explicit relations between
the value function $V$, the optimal wealth $X^*$, the optimal portfolio $\pi^*$ and the optimal total consumption $k^*$ with the dual value function $\hBV$ and the dual state process $Y$ in the next theorem. 
In this paper the process $Y$ evolves according to a geometric Brownian motion 
\begin{align}\label{Y}
\dd Y(t)=Y(t)( (\rho(t)-r(t))\dt-\vartheta(t)^\top\dd B(t)),
\end{align} 
where $\vartheta(t)$ is the market prices of risk and unique solution to $\sigma(t)\vartheta(t)=\mu(t)$. 
To ensure the dual functions to be considered below have finite values, we assume 
\begin{align}\label{growthcondition}
\int_{0}^{\ot} e^{\int_0^s\left(\frac{1}{\beta-1}\rho(u)-\frac{\beta}{\beta-1} r(u)+ \frac{1}{2}\frac{\beta}{(\beta-1)^2}|\vartheta(u)|^2\right)\du}\ds<\infty, 
\end{align} 
which is true when $\ot<\infty$ or the discount rate is sufficiently large compared to the interest rate and the market price of risk.

%\blue
{
\begin{theorem}[Solution for the post-retirement problem]\label{postsolution1}

Define 
\begin{align}\label{hbv}
\hBV(t, y)&=\BE{\int_{t}^{\ot}e^{-\int_t^s \rho(u)du}h( Y(s))\ds \;\bigg|\; Y(t)=y}. 
\end{align} 
Assume \eqref{growthcondition} holds. Then {$\hBV\in C^{1, 2}([0, \ot)\times (0, \infty))$ is convex and {PLD($\beta$)} in $y$, and} the value function $V$ for the post-retirement problem \eqref{retirep1} is given by 
\begin{align} 
\BV(t, x)&=\inf_{y>0}(\hBV(t, y)+xy), \quad x>0.\label{def:bvdual}
\end{align} 
Moreover, the optimal wealth, the optimal portfolio and the optimal total consumption at time $t$ after retirement time $\tau$ are given respectively by, for $t\geq \tau,$
\begin{equation} \label{XinY}
X^*(t)=-\p_y\hat V(t, Y(t)), \quad \pi^{*}(t)=(\sigma(t)^T)^{-1}\vartheta(t) Y(t)\p_{yy} \hat V(t, Y(t)), \quad k^{*}(t)=-\p_y h(Y(t)), 
\end{equation}
where $Y(t)$ is the solution of (\ref{Y}) with the initial condition $Y(\tau)=y$ and $y$ is the solution of $\p_y \hat V(\tau, y)+x=0$ with $x$ the initial wealth at retirement time $\tau$. 
\end{theorem}
}

%\blue
{\begin{remark}\label{remarkmono}
Both $-\partial_y\hat V(t, \cdot)$ and $-\partial_y h$ are nonnegative decreasing functions, which implies the optimal wealth $X^*(t)$ and the optimal total consumption $k^*(t)$ are positively correlated (comonotonic) and both are negatively correlated to $Y(t)$ (counter-monotonic), that is, the higher the wealth $X^*(t)$, the lower the dual state value $Y(t)$, and the higher the total consumption $k^*(t)$. 
Furthermore, since $\bar{u}$ satisfies the Inada condition, we have $X^*(t)\to\infty$ is equivalent to $Y(t)\to0$ and $X^*(t)\to0$ is equivalent to $Y(t)\to\infty$. It may be difficult to find a closed-form formula for the dual value function $\hBV(t, y)$ in (\ref{hbv}) and its derivatives due to integration being involved, but it is straightforward to compute their values with a numerical integration scheme.
\end{remark}}

%\blue
{
\begin{remark}\label{onefund}
When all the parameters are constants, \eqref{XinY} shows that the amount invested in stocks is a multiple of $(\sigma^T)^{-1}\vartheta$. This means the proportion invested in each stock is fixed. Therefore it can be regarded as investing in a fund, we only need to determine the total amount invested in this fund. The optimal portfolio and total consumption after $\tau$ depend on this fund only. The size of investment for this fund depends on the dual process $Y$, calculated by
%\blue
{\begin{align}\label{Eq:Y_solution}
\log Y(t)&=\log Y(\tau)+{c\over r} \log {S_0(t)\over S_0(\tau)}-\vartheta^{\top}\sigma^{-1}\bigg(\log \frac{S_1(t)}{S_1(\tau)}, \cdots, \log \frac{S_n(t)}{S_n(\tau)}\bigg)^{\top}, 
\end{align}
for $t\geq \tau$, where $c=\rho-r-\frac{1}{2}\Vert\vartheta\Vert^{2}+\vartheta^{\top}\sigma^{-1}\Big(b_1-\frac{1}{2}\sum\limits_{j=1}^n\sigma_{1j}^2, \cdots, b_n-\frac{1}{2}\sum\limits_{j=1}^n\sigma_{nj}^2\Big)^{\top}
$}.
In practice, we have the data of each term on the right hand, so one can determine the dynamics of $Y$ by the above equation and thus apply the strategy given by \eqref{XinY}.
\end{remark}
}

%\blue
{

\begin{remark}\label{nonconcave1} Note that $k^{*}(t)$ in (\ref{XinY}) is well-defined as $h$ is continuously differentiable everywhere except at one point $\overline{Y}$ (the slope of the straight line of the concave envelope $\widetilde u$), but $Y(t)$ is a lognormal variable and the probability of $Y(t)$ taking the value $\overline{Y}$ is zero. One can find a threshold 
%\blue
{$\overline{X}(t)=-\partial_y \hat V(t, \overline{Y})$ at time $t$} from \eqref{XinY}. If the optimal wealth $X^*$ is less than $\overline{X}$, then the optimal total consumption $k^*$ is kept at a lower level (basic consumption only), whereas if $X^*$ is greater than $\overline{X}$, then the optimal total consumption $k^*$ jumps to a higher level (mainly luxury consumption). This jump consumption phenomenon only happens when there exists a non-concave consumption utility. The gap or the unconsumed part is used for additional investment or savings in the hope of achieving large wealth. 
As an example, assume $u(c, g)$ is given as in Example \ref{nonconcave}. Simple computation shows that for $y>0$, 
$$h(y)=y^{-1}+\Big(\tfrac{1}{3}y^{-3}-ay\Big)1_{y< (3a)^{-1/4}}, $$
where $1_S$ is an indicator that equals 1 if $S$ is true and 0 otherwise. Clearly $h$ is continuous, strictly decreasing, strictly convex, and continuously differentiable everywhere except at point $y=(3a)^{-1/4}$. This shows the optimal total consumption is given by
$$k^*(t)=-\p_y h(Y(t))=Y(t)^{-2}+\big(Y(t)^{-4}+a\big)1_{Y(t)< (3a)^{-1/4}}, $$
which consists of consumption $Y(t)^{-2}$ and additional consumption $Y(t)^{-4}+a$ that happens only when $Y(t)$ is below the threshold $\overline{Y}=(3a)^{-1/4}$ or, equivalently, the wealth $X(t)$ is above the threshold $\overline{X}(t)=-\p_y\hat V\big(t, (3a)^{-1/4}\big)$. There is a jump of the total consumption level from $(3a)^{1/2}$ to $(3a)^{1/2}+4a$ at the threshold. The higher the value $a$, the higher the jump magnitude; for instance, if $a=3$, then the total consumption would jump from 3 to 15 at the threshold $\overline{Y}=3^{-1/2}$, but if $a=300$, then from 30 to 1230 at the threshold $\overline{Y}=(30)^{-1/2}$. 
\end{remark}

}

{
\begin{remark}\label{pension} 
No income (such as pension) after retirement is considered in our setup. 
If deterministic income $I(s)$ exists after retirement, one can adjust the initial wealth at $\tau$ to $x+\int_\tau^{\ot} e^{-\int_\tau^s r(u)du}I(s)\ds$, where $x$ is the wealth immediately before $\tau$, as if there is no income after retirement. This is because one can borrow at retirement time against future income when there is a certainty of amount of future income \citep{VD09}. Theorem \ref{postsolution1} still holds with everything the same except that the initial wealth $x$ is replaced by $x+\int_\tau^{\ot} e^{-\int_\tau^s r(u)\du}I(s)\ds$ which is used to compute the initial value $Y(\tau)$ for the dual process $Y$. Introducing the income after retirement clearly affects the retirement decision. The higher the income, the more likely to retire early. We may absorb the income after retirement as an additional labor cost or reduced labor income while working and still treat the problem as if there is no income after retirement. 
\end{remark}
}

\begin{theorem}[Solution for the post-retirement problem with infinite horizon]\label{postsolution}
Suppose $\ot=\infty$, $r, \mu, \sigma, \rho$ are all constants with
\begin{align}\label{bigrho}
{\rho>r\beta+\frac{\beta}{2(1-\beta)}\Vert\vartheta\Vert^{2}.}
\end{align} 
Then the dual value function $\hBV$ is independent of time $t$ and given by 
$$\hBV(y)=\BE{\int_{0}^{\infty}e^{-\rho t}h( Y(t))\dt \;\bigg|\; Y(0)=y}. $$
We have {$\hBV\in C^2(0, \infty)$ is convex, PLD, and} the value function $V$ for the post-retirement problem \eqref{retirep1} is independent of time $t$ and is given by (\ref{def:bvdual}), the optimal wealth $X^*$, the optimal portfolio $\pi^*$, and the optimal total consumption $k^*$ are given by (\ref{XinY}), 
where $Y(t)$ is the solution of the equation (\ref{Y}) with the initial condition $Y(0)=y$ which is the solution of the equation 
$\p_y \hat V(y)+x=0$ with $x$ the initial wealth at time of retirement. 
\end{theorem}

%\blue
Condition (\ref{bigrho}) ensures the value function for the post-retirement problem is finite when $\ot=\infty$. 

\begin{remark} \label{K1}
In the presence of closed convex cone control constraint $\pi(t)\in\setk$, one may define the positive polar cone of $\setk$ by
$\widetilde{\setk}:=\{v\in\R^{n}\mid v^\top \pi \geq 0, \;\forall\; \pi \in \setk\}.$ 
Let $\hat v(t)$ be the minimum solution of the quadratic function $\Vert\sigma(t)^{-1}(v+\mu(t))\Vert^{2}$ over $v\in \widetilde{\setk}$. Denote by $\hat \vartheta(t)=\sigma(t)^{-1}(\hat v(t)+\mu(t))$. Assume $\hat \vartheta(t)\ne0$ for all $t$. Replacing $\vartheta(t)$ in \eqref{Y} by $\hat\vartheta(t)$, $\hBV(t, y)$ in \eqref{hbv} is the corresponding dual value function for control constrained problem \cite[Section~4]{HZ11}. If the set $K$ is a general closed convex set, not necessarily a cone, especially if $K$ is a bounded set, then the dual control problem is as hard to solve as the primal one due to the presence of the support function of the dual control in the drift term of the dual process $Y$, which results in the dual HJB equation a fully nonlinear PDE and there is no representation of the dual value function $\hat V$ as in (\ref{hbv}). 

\end{remark}

\begin{eg} \label{eg:3}
This example shows the dual value function for the post-retirement problem can be explicitly given when $\ot=\infty$.
\cite{APY04} consider the following utility function: 
\[u(c, g)=\frac{(c-c_{0})^{1-\phi}}{1-\phi}+\frac{(g+b_{0})^{1-\psi}}{1-\psi}, \]
where the basic consumption $c$ is at least $c_{0}$. This utility is slightly different from the non-homothetic utility. In this case 
%\blue
{$\bar{u}$ is a continuously differentiable, strictly increasing and strictly concave function. In fact, we do not need to compute $\bar{u}$ to find $h$ as $u$ is additively separable.} We have 
\[\hBV(y)=C_{1}y^{1-\frac{1}{\phi}}+C_{2}y+C_{3}+(C_{4}y^{1-\frac{1}{\psi}}+C_{5}y+C_{6})\id{y< b^{-\psi}}, \]
where the parameters $C_{i}$, $i=1, \cdots 5$, are explicitly given constants. Although an explicit expression for $\BV$ is not available, we can determine its value via \eqref{def:bvdual} by effective numerical scheme. 
%\greenone
{It is unnecessary to compute $\BV$ in practice, since our optimal strategies given by \eqref{XinY} only rely on $\hBV$. In fact, we can compute the strategies explicitly.}
\end{eg}

%%%%%%%%%%%%%%%%%%%%%%%%%%%%%%%%%
\smallskip
\section{Pre-Retirement Problem}\label{Sec:Pre}
%%%%%%%%%%%%%%%%%%%%%%%%%%%%%%%%%
\noindent
Now we go back to problem \eqref{p2}, which may be rewritten as a pre-retirement problem
%\blue
\begin{align}\label{p3}
\sup_{\tau, k, \pi}\; &\BE{\int_{0}^{\tau} e^{-\int_0^t \rho(s)\ds}(\bar{u}(k(t))-L(t))\dt+e^{-\int_0^\tau \rho(s)\ds} \BV(\tau, X(\tau))}, 
\end{align}
where $\BV$ is given by \eqref{def:bvdual}. %The last term represents the total utility after retirement. 
Then, the wealth process $X$ before retirement follows 
\begin{align}\label{wealth3}
\dd X(t)=(r(t)X(t)+\pi(t)^\top\mu(t)+I(t)-k(t))\dt+\pi(t)^\top\sigma(t)\dd B(t), \quad \text{for $t\leq\tau$}.
\end{align} 
We use the dynamic programming to solve problem \eqref{p3} by denoting its value function by $W$. We consider the following dual variational inequality 
\begin{align} \label{hjb5}
\begin{cases}
\min \left\{-(\p_{t}+\BBL)\hBW(t, y)+\rho(t)\hBW(t, y)-yI(t)-h(y)+L(t), \;\hBW(t, y)-\hBV(t, y)\right\}=0, \\
\hBW(T, y)=\hBV(T, y), \hfill (t, y)\in\sets:=[0, T)\times(0, \infty), 
\end{cases}
\end{align}
where
\(\BBL :=\tfrac{1}{2}\Vert\vartheta(t)\Vert^{2}y^{2}\p_{yy}+(\rho(t)-r(t))y\p_{y}.\)
Then, the dual optimal stopping problem corresponding to (\ref{hjb5}) is given by
\begin{align}\label{p3d}
\sup_{0\leq \tau\leq T}\; &\BE{\int_{0}^{\tau} e^{-\int_0^t \rho(s)\ds}(Y(t) I(t)+h(Y(t))-L(t)) \dt 
+e^{-\int_0^\tau \rho(s)\ds} \hBV(\tau, Y(\tau))}, 
\end{align}
where $Y$ is the dual process satisfying the SDE (\ref{Y}).

In general, there does not exist a classical solution to \eqref{hjb5}, but we can find a solution that allows us to give the optimal strategy for the pre-retirement problem \eqref{p3}. 

\begin{theorem}[Solution for the pre-retirement problem]\label{presolution} 
The PDE \eqref{hjb5} has a unique solution $\hBW$ such that $\hBW$ and $\p_{y}\hBW $ are continuous in $\sets$; the free boundary, defined by the boundary of $\{\hBW=\hBV\}$, is Lipschitz in both $t$ and $y$; 
$\p_{t}\hBW$ and $\p_{yy}\hBW $ are continuous in $\sets$ excluding the free boundary.
Moreover, the function
\begin{align*}
\BW(t, x)=\inf_{y>0}(\hBW(t, y)+xy), \quad (t, x)\in\sets, 
\end{align*}
is the value function of the pre-retirement problem \eqref{p3}.
The optimal wealth, portfolio and total consumption at time $t$ before retirement time $\tau$ are respectively given by, for $t<\tau$, 
%\blue
\[X^*(t)=-\p_y\hBW (t, Y(t)), \quad \pi^{*}(t)=(\sigma(t)^T)^{-1}
\vartheta(t) Y(t)\p_{yy} \hBW (t, Y(t)), \quad k^{*}(t)=-\p_y h(Y(t)), \]
where $Y(t)$ is the solution of the equation (\ref{Y}) with the initial condition $Y(0)=y^*$ and $y^*$ is the solution of the equation $\p_y \hBW (0, y)+x=0$ with $x$ the initial wealth at time 0.

\end{theorem}

\begin{remark} \label{K2} 
In the presence of a closed convex cone control constraint $\pi(t)\in \setk$, similar to the case of post-retirement problem, we may replace $\vartheta(t)$ in \eqref{hjb5} by $\hat \vartheta(t)$ and get the corresponding dual value function for control constrained problem, see \citeremark{K1} and \cite[Sec.4]{HZ11}. 
\end{remark}
% 

%%%%%%%%%%%%%%%%%%%%%%%%%%%%%%%%%
\smallskip
\section{Optimal Retirement Region}\label{Sec:Retire}
%%%%%%%%%%%%%%%%%%%%%%%%%%%%%%%%%
\noindent
In this section we study the optimal retirement time. Even though the dual problem (\ref{p3d}) is much simpler than the primal problem (\ref{p3}), one still needs to determine the optimal stopping time $\tau$, which is a free-boundary problem. We can nevertheless simplify this problem further. 
Define 
%\blue
{\begin{equation} \label{WnW}
\BBW(t, y):=e^{-\int_0^t \rho(s)\ds}(\hBW(t, y)-\hBV(t, y)), \quad (t, y)\in\sets.
\end{equation}
Then the variational inequality \eqref{hjb5} in terms of $\BBW$ becomes 
\begin{align} \label{hjb6}
\begin{cases}
\min\left\{-(\p_{t}+\BBL)\BBW(t, y)-e^{-\int_0^t \rho(s)\ds}(yI(t)-L(t)), \;\BBW(t, y)\right\}=0, \\
\BBW(T, y)=0, \hfill (t, y)\in\sets.
\end{cases}
\end{align} The optimal stopping problem corresponding to the dual variational inequality (\ref{hjb6}) is given by
\begin{align}\label{p3dd}
\sup_{0\leq \tau\leq T}\; &\BE{\int_{0}^{\tau} e^{-\int_0^t \rho(s)\ds}(Y(t) I(t)-L(t)) \dt }, 
\end{align}
where $Y$ is the dual process satisfying the SDE (\ref{Y}). }

We define the retirement region and the working region, respectively, as 
\begin{align*} 
\setr=\{(t, y)\in\sets\mid \BBW(t, y)=0\}, \quad 
\setc=\{(t, y)\in\sets\mid \BBW(t, y)>0\}.
\end{align*}
{Let \[b(t)=\inf\{y>0\mid \BBW(t, y)>0\}, \quad 0\leq t\leq T, \] with the convention that $\inf\emptyset=+\infty$. Then we have the following characterization of the optimal stopping region:
\begin{proposition}[Universal stopping region] \label{Universal_stopping_region}
The free boundary $b(\cdot)$ is irrelevant to the individual's utility function, and
\begin{align*} 
\setr=\{(t, y)\in\sets\mid y\leq b(t)\}, \quad \setc=\{(t, y)\in\sets\mid y> b(t)\}.
\end{align*}
Furthermore, bigger $I(\cdot)$ or smaller $L(\cdot)$ leads to smaller $b(\cdot)$, which implies one defers retirement as the income increases or labor cost drops.
\end{proposition}
}

Proposition \ref{Universal_stopping_region} tells us one should retire if his marginal utility is small enough (i.e. his wealth is large enough). 
This result is universal in the sense that all individuals choose to retire at the same level of marginal utility. The free boundary $b(t)$ does not depend on $h(y)$ (or the personal preference $u$) and only depends on $\rho$, $r$, $I$, $L$, and market parameters. Meanwhile, the result is also not universal in the sense that different individuals may choose to retire at different levels of wealth. This is a crucial consequence of our model. For example, two individuals A and B will retire at the same marginal utility, say, $1/8$, but A and B have different total utility functions with $\bar{u}_A(x)=4x^{1/4}$ and $\bar{u}_B(x)=2x^{1/2}$. Then A will retire when the wealth reaches 16 whereas B will not retire unless the wealth is at least 64. 

%\greenone
{
As technology development, labor cost for the same job will reduce gradually, Proposition \ref{Universal_stopping_region} suggest people tend to retire later. Governors face with two facts: (i) less labors are needed for the same job; (ii) more people prefer to postpone retirement. As a consequence, more labors may be available than they would be needed, which is a challenge for governors as they have to create more jobs. Demographic dividend may change to demographic burden due to surplus labors. 
It also tells us that governors should not set a universal retirement age; it is better to let each individual decide his own retirement time based on his health situation, family burden, and so on. For instance, in Japan, many people aged 80 or above are still working because technology development reduces work intensity and young labors are not enough as Japan is an aged society. 
}

Intuitively speaking, everyone should retire earlier if he has arrived at a good economic situation. The following confirms this conjecture {and provides an explicit upper bound for the retirement. }
\begin{proposition}[Everybody may retire earlier] \label{upperboundfreeboundary}
We have $b(t)\leq {L(t)}/{I(t)}$ for all $t\in[0, T]$.
\end{proposition}
Note that this upper bound only depends on the labor cost and income and is independent of individual's utility. We may conclude that it is likely a good time to take early retirement when labor cost is high or income is low, which provides sensible economic appeal. 

We propose the following hypothesis on the income and labor cost processes, which is not only realistic but also economically important. It will play an important role in determining the monotonicity of the optimal retirement time. 
\begin{assumption}[Growth condition]\label{h1}
We have 
%\blue
{$\frac{I'(t)}{I(t)}\leq \rho(t)\leq \frac{L'(t)}{L(t)}$}
for $t\in[T-\ell, T]$ with $0<\ell\leq T$. 
\end{assumption}
%\greenone
This condition means the discounted labor cost is increasing and discounted income is decreasing. 
The economic interpretation of this hypothesis is given as follows. For a young person, his marginal labor cost is decreasing as he is getting more skilled. By contrast, for an older one, his marginal labor cost is increasing as he is becoming aged with less energy and more burdens such as illness, family issue, child care. Therefore, as an individual becomes aging, his marginal labor cost increases faster than his marginal income. It will be shown that the growth condition guarantees the monotonicity of the retirement boundary for $t\in[T-\ell, T]$, but this conclusion is not necessarily true before $T-\ell$ since we cannot compare the income and labor cost in that range. %A numerical example will be given to demonstrate this importance phenomenon.
%\par
The left of \citefig{Fig:bt_example} shows an example of the income process $I(\cdot)$ and the labor cost process $L(\cdot)$ satisfing \citeassmp{h1}. 
%\red {Their shapes before $T-\ell$ can be arbitrary. }

\begin{figure}[h]
\begin{center}
\includegraphics[width=0.48\textwidth]{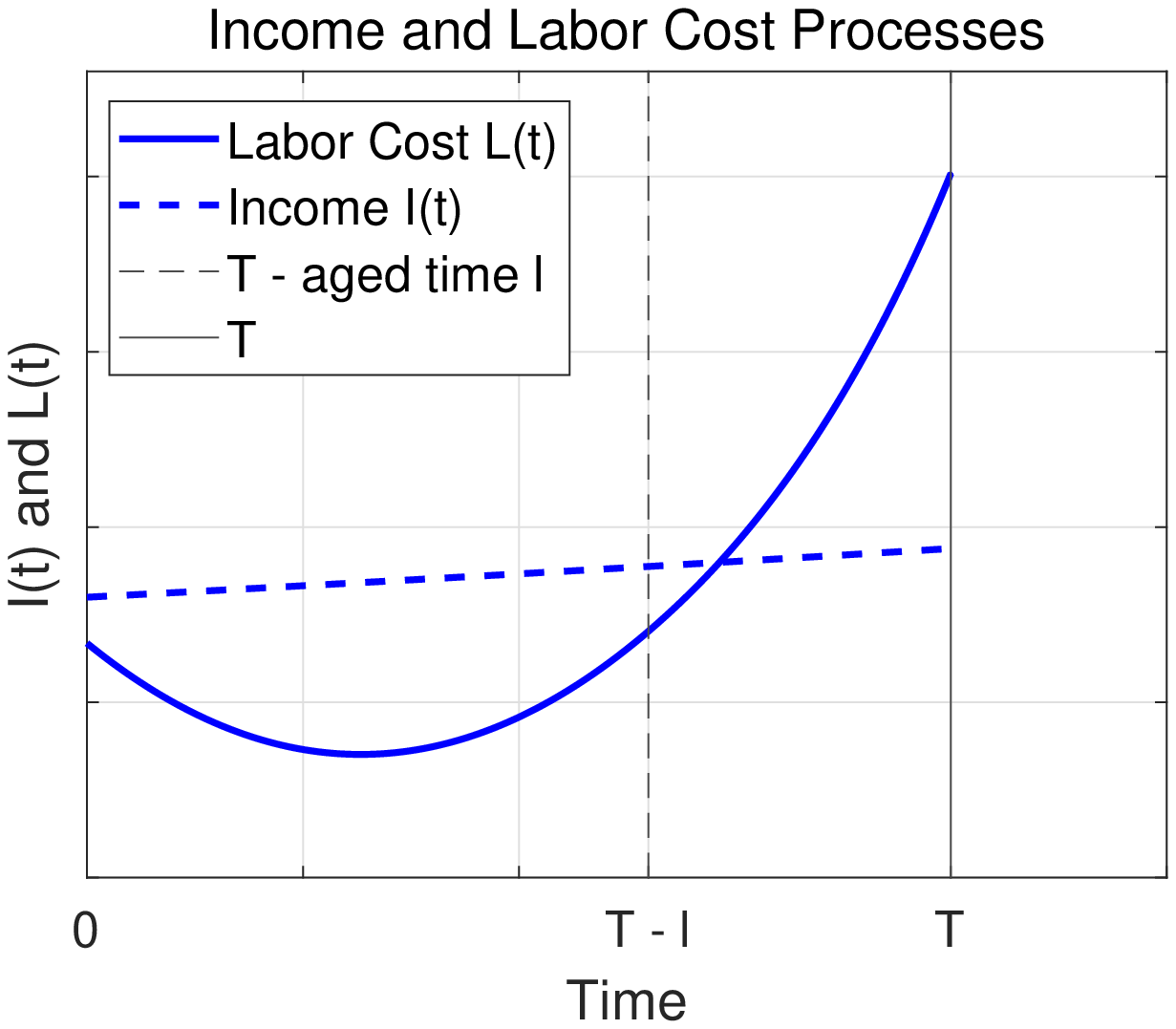}
\includegraphics[width=0.48\textwidth]{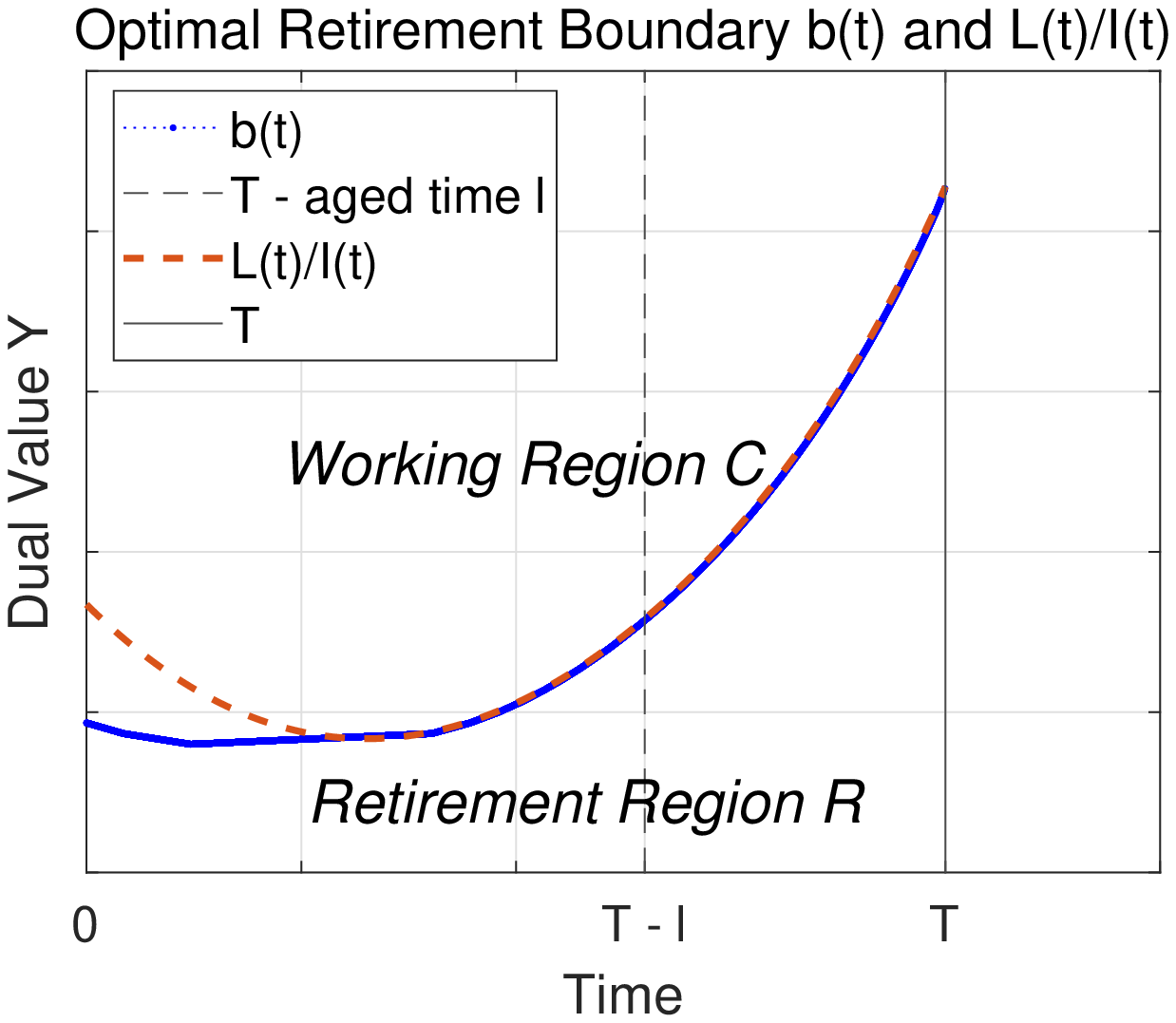}
\end{center}
\caption{An example of the income process $I(\cdot)$ and the labor cost process $L(\cdot)$ (left) and an example of the retirement region $\setr$ and the working region $\setc$ under \citeassmp{h1} (right).}
\label{Fig:bt_example}
\end{figure}

\begin{theorem}[Monotonic free boundary]\label{monotone}
Assume \citeassmp{h1} holds. Then $b(t)$ is increasing for $t\in[T-\ell, T]$, with terminal value $b(T-):=\lim_{t\to T}b(t)=\frac{L(T)}{I(T)}.$
\end{theorem}

Economically speaking, older individuals (less than $\ell$ years to the mandatory retirement age) are more like to retire as time goes by, but for younger individuals, the marginal income may growth faster than marginal labor cost, so they may not prefer to retire. In particular, if $T-\ell=0$, then everyone is more like to retire as time flies. 
%\greenone
{We call the interval $[T-\ell, T]$ the aged region. The above result tells us if an institute or company wants to postpone the retirement time for a skilled individual in the aged region, they can either reduce his/her labor cost or increase his/her income significantly (compared to the inflation). This makes perfect economic meaning. } 
%\par 
The right of \citefig{Fig:bt_example} illustrates the results of \citeprop{upperboundfreeboundary} and \citethm{monotone}.
The free boundary $b(t)$ is shown in \citethm{monotone} under \citeassmp{h1} to be monotonically increasing as time approaches to $T$. 
The next section gives an example to show $b(t)$ is not necessarily monotonically increasing globally.

\smallskip

{
\section{Numerical Simulation}\label{Sec:Simulation} 

In this section, we implement the estimates for the theoretical results derived with introducing the numerical procedure for estimating the value functions for the pre- and post-retirement problems. 
Regarding the numerical scheme for solving variational inequality, 
\cite{J91} and \cite{P05} show that the free boundary is the unique solution to a nonlinear integral equation by using the smooth pasting condition and the It\^{o}-Doob-Meyer decomposition of the value function. 
\cite{CV05} reformulate the dual variational inequality as a linear complementarity problem and solve it numerically with the implicit-explicit method.

Since the free boundary is unknown a priori, the pre-retirement problem \eqref{p3} is known to be difficult as one not only needs to find the free boundary but also solves a nonlinear PDE in the continuation region. In our approach, however, the dual formulation \eqref{p3d} significantly simplifies the problem as it is an optimal stopping problem of one state variable.
The numerical procedure we adopt is described in \ref{EC:VFE} in detail.

To implement this procedure, we construct the functions $I(\cdot)$ and $L(\cdot)$ satisfying \citeassmp{h1}. 
Define $I(t)=Ce^{K^\prime t}$; and $L(t)=a_0+a_1t+\frac{1}{2}a_2t^2$, where 
$$a_0 =e^{K(T-\ell)}\left(1-K(T-\ell)+\tfrac{1}{2}K^2(T-\ell)^2\right), \,\,
a_1 =Ke^{K(T-\ell)}\left(1-K(T-\ell)\right), \,\,
a_2 =K^2e^{K(T-\ell)},$$
if $t\leq T-\ell$; $L(t)=e^{Kt}$, otherwise. 
This setup ensures continuity up to the second order derivative at $T-\ell$. 
The condition $0<1/K < T-\ell$ implying $a_1<0$ and $a_2(T-\ell)+a_{1}>0$ ensures the labor cost function is convex, and the condition $K^\prime<\rho<K$ is required to satisfy \citeassmp{h1}.

\subsection{Optimal Strategies}\label{Sec:value_function}

We estimate the value function surface on the pre- and post-retirement regions with finding the optimal retirement boundary.
The detailed estimate results are displayed in \ref{EC:Figures}.
Considering all the requirements, we set the market and model parameters as follows:
\begin{equation}\label{Eq:baseline_gen}
\sigma = 0.4,\, \mu = 0.1,\, r = 0.03,\, \rho = 0.1,\, T=2,\, \bar{T}=2.5,\, a = 10,\,
K = 1.3,\, K' = 0.08,\, C = 5,\, \ell = 1.
\end{equation}
Figure~\ref{Fig:bt} shows the income/labor cost processes and the corresponding optimal retirement boundary $b(t)$ with a sample path of the dual process $Y(t)$. 
The tests in this section assume (\ref{Eq:baseline_gen}) with the initial wealth of $x = 10$.
The unit of all the computed optimal values are considered to dollar value.

Moreover, we demonstrate sample paths of the optimal strategies of wealth $X^*(t)$, the stock investment dollar value $\pi^*(t)$, and consumption $k^*(t)$ numerically obtained (see \ref{EC:Figures}). 
Figure~\ref{Fig:Sample_Path} illustrates the relevant paths with the optimal retirement.
As the initial wealth is \$10, the initial revenue is about \$3, that is, $I(0)-L(0)$. 
The amount of $x$ can be interpreted as legacy or seed fund (not a consequence of own labor); thus indicating the wealth granted is given about three times larger than the initial net income. 
This setup employs the luxury consumption trigger of $a=\$10$, which 
carries the non-consumption range from $k_{-}\approx \$5.5$ to $k_{+}\approx \$45.5$, indicating that no one is willing to consume the amount in this range to be maximizing his satisfaction from consumption. 

The optimal strategy steers the direction to expanding the amount of stock investment rather than consuming to grow wealth unless current wealth is high enough to be able to consume luxury.
This sample path spends only basic goods since wealth grows not sufficiently high from stock investment (at least up to $k_+\approx 45.5$) as much as to consume the luxury. 
As the consuming pattern stays at basic goods prior to $X^*$ reaches $k_{+}$, 
this scenario has no chance to consume the luxury.

\begin{figure}[H]
\begin{center}
\includegraphics[width=0.48\textwidth]{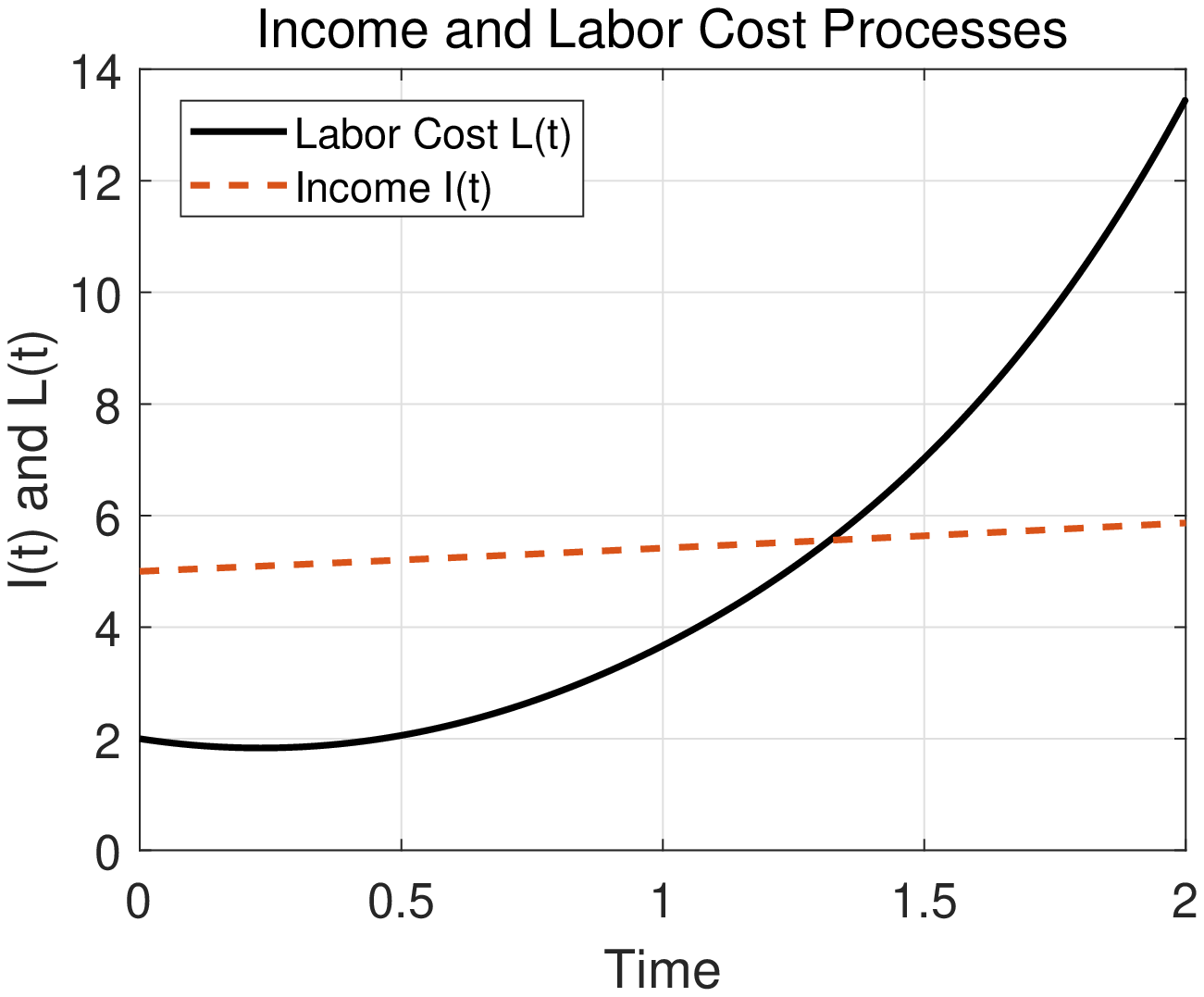}
\includegraphics[width=0.48\textwidth]{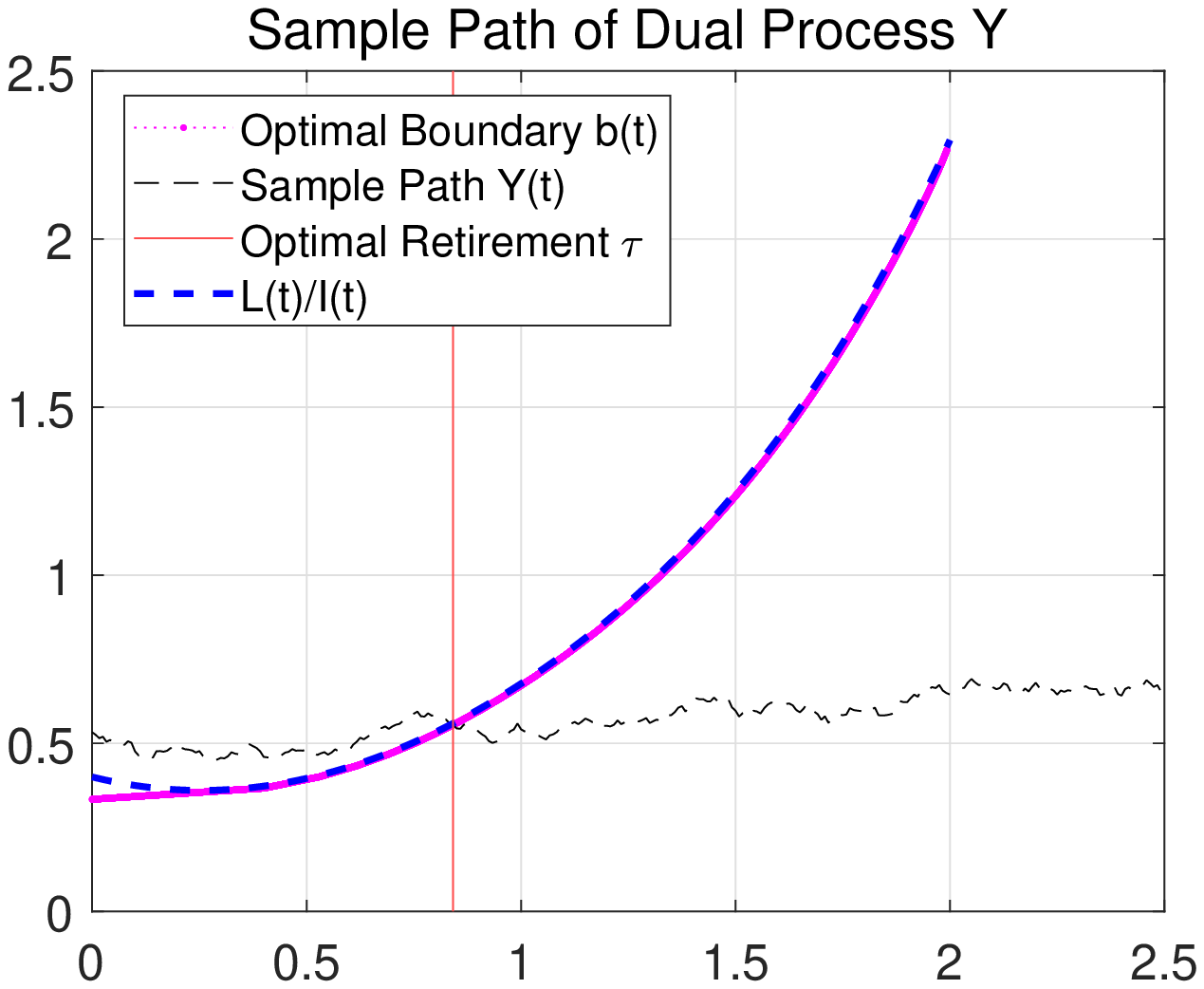}
\end{center}
\caption{The income $I(t)$ and labor cost $L(t)$ processes under the basic setting (left) and the corresponding optimal retirement boundary $b(t)$ with a sample path of the dual process $Y(t)$ and the ratio of $L(t)$ over $I(t)$(right). The vertical red line indicates the optimal retirement time $\tau$. }
\label{Fig:bt}
\end{figure}

\subsection{Sensitivity Test}

In this section, we conduct sensitivity analysis for the optimal strategies discussed in Section~\ref{Sec:value_function} under diverse parameter assumptions.
This section considers the following model parameters required to implement as a baseline:
\begin{equation}\label{Parameters}
a = 10, \quad K = 1.5, \quad C = 8, \quad \ell = 0.7, \quad x = 10.
\end{equation}
The market variables and the lifetime scale remain unchanged with the previous subsection. 
\smallskip\noindent
{\it The optimal strategies with respect to the passage of time}.

We consider various initial amount $C$ in the income processes $I(t)=Ce^{K't}$ while keeping $L(t)$ the same with (\ref{Parameters}). 
This test is to investigate how the relation between income and labor cost affects the decision making for retirement and individuals' wealth as an individual is aging.
We choose the initial income levels of $C=2, 4, 6, 8, 10, 15$ with the aged variable of $\ell = 0.7$ in the working cost, which situates more curvy shape for $L(t)$ than $\ell = 1$ used in Figure~\ref{Fig:bt}. 

These parameter sets enable to construct diverse relative scale of the income against the labor cost, which is represented as its difference, that is, $I(t) - L(t)$.
In Figure~\ref{Fig:Low-High}, the left presents the difference curves depending on the income values chosen, 
and the right shows the corresponding optimal retirement boundary $b(t)$ for each case. 
We can see that high salary-paying jobs (or, relatively high income to labor cost required jobs) encourage to be working longer, 
whereas lower salary-paying jobs (or, relatively low income to labor cost) impels to retire earlier. 

\begin{figure}[H]
\begin{center}
\includegraphics[width=0.48\textwidth]{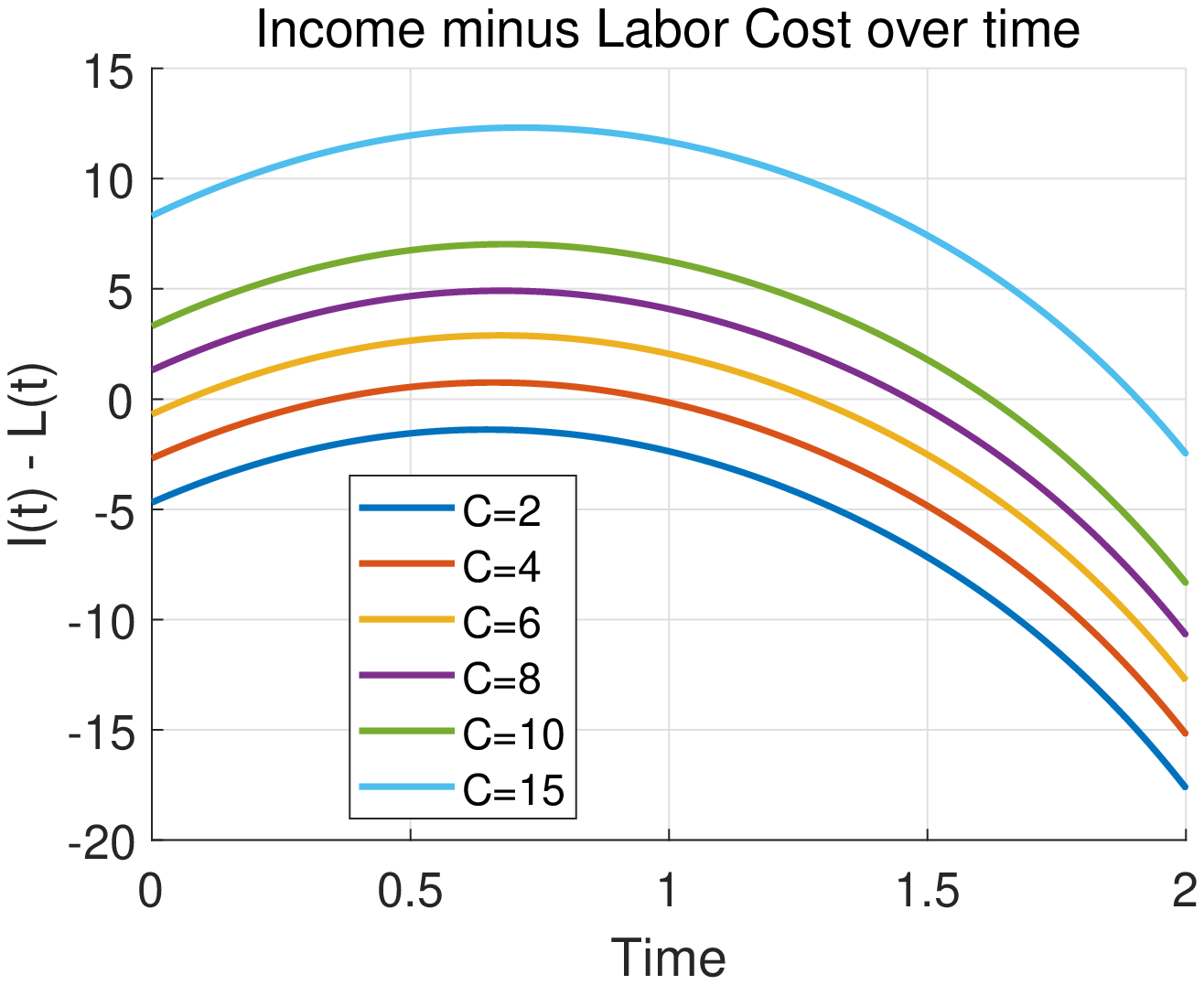}
\includegraphics[width=0.48\textwidth]{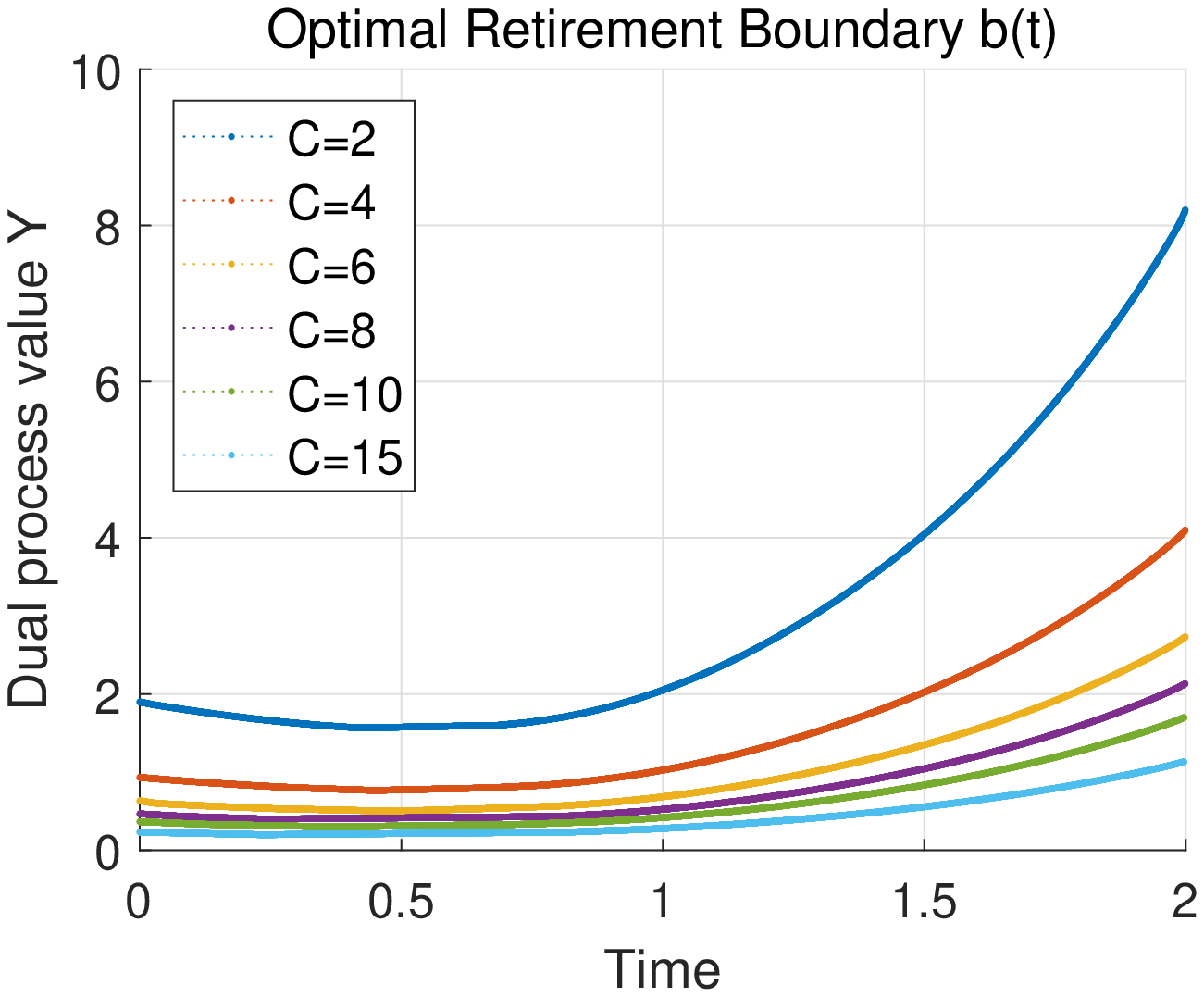}
\end{center}
\caption{The optimal retirement boundaries with respect to the difference between the income and labor cost processes, $I(t)-L(t)$ with $C=2,4,\cdots,10$ and 15.}
\label{Fig:Low-High}
\end{figure}

In addition, we investigate statistical figures for the optimal strategies to lifetime depending on the relative scale of income and labor cost.
Table~\ref{Table:Income} displays the sample mean and sample standard deviation of $X^*(T)$ for each time $T$.
Since the time $T=2$ represents the whole working period up to mandatory retirement (commonly age of 60), the scales of $T=0.5, 1.0$, and 1.5 can be interpreted as 20's, 30's, and 40's, respectively. More distributional evolutions appear in \ref{EC:Figures}.

As income increases, the optimal retirement time tends to be deferred. Especially, not working may be an optimal decision for someone who has the initial income lower than \$6 in terms of maximizing satisfaction from consumption. 
Moreover, the expected optimal wealth tends to increase as the income rises, and its standard deviation follows to rise. 
Until approaching the optimal retirement time, both return and risk of individuals' wealth tend to be increasing; however, when turning to post-retirement age, the wealth and its deviation are likely to decline over the rest of whole life.

\begin{table}
\renewcommand{\arraystretch}{0.7}
\centering
\caption{The sample mean (Mean) and standard deviation (Std) for the optimal wealth $X^*(T)$ at the life time $T=0.5,1.0,1.5,2.0$ (that can be interpreted as age from 20's to 50's) with respect to the initial income condition $C$ and the estimate of the optimal retirement time $\tau$ with its standard error (s.e.). }
\bigskip
\begin{tabular}{cccccccc}
\hline \\[1pt]
$T$&Initial income $C$&$15$ & $10$ & \cellcolor{gray}$8$& $6$ & $4$ & $2$\\[6pt]
\hline\\[1pt]
{0.5} &Mean&15.192& 10.630& \cellcolor{gray}11.916&9.4284& 8.157 & 9.120\\ [6pt]
{(20's)}&Std &29.129& 18.827& \cellcolor{gray}17.239&8.6606&6.900&8.164\\[6pt]
{1.0} &Mean&17.695& 13.215& \cellcolor{gray}13.874&8.3123& 7.083 & 7.966\\[6pt]
{(30's)}&Std &34.066& 22.692& \cellcolor{gray}21.428&12.069&10.623&11.508\\[6pt]
{1.5} &Mean&17.994& 11.724& \cellcolor{gray}10.988&6.2088& 5.405 & 6.015\\[6pt]
{(40's)}&Std &29.651& 20.143& \cellcolor{gray}19.472&11.708&10.562&11.299\\[6pt]
{2.0} &Mean&10.122& 6.471 & \cellcolor{gray}6.052 &3.4278& 3.002 & 3.344\\[6pt]
{(50's)}&Std &18.025& 13.228& \cellcolor{gray}12.642&7.7416&6.863&7.522\\[6pt]
\hline\\[1pt]
&{$\tau$} &1.3992& 1.1776 & \cellcolor{gray}0.9788 &0& 0 & 0 \\[6pt]
&{(s.e.)} &(0.0153)& (0.0131)& \cellcolor{gray}(0.0225) &-& - & - \\[6pt]
\hline
\end{tabular}\label{Table:Income}
\end{table}

Next simulation examines the sensitivity of the optimal wealth and retirement time with respect to the initial wealth as time goes by. 
We consider the initial wealth $x=5,10,15,20$ dollars, while the other conditions stay the same at the baseline setup.
The result is displayed in Table~\ref{Table:InitialWealth},
which shows as the initial wealth increases, the optimal retirement time tends to be earlier. The initial wealth greater than \$30 results in $\tau=0$ indicating no one would not desire to be working in the case of large initial legacy granted, which can be optimal in terms of maximizing satisfaction from consumption. 
In opposite, if the initial wealth granted is smaller, working lifetime gets longer. 

Another implication is that a large initial fund puts high deviation of the optimal wealth owing to be able to invest risky assets utilizing abundant financial resource in order to grow an individual's wealth. 
However, debt at the early age of working does not allow to put the current fund to explore any risky investment opportunities as consuming basic goods is prioritized. 
For this reason, uncertainty for the optimal wealth amount for the passage of time is relatively low for a low-net worth group; and it is highly predictable to build a low amount of wealth at the old age.

\begin{table}
\renewcommand{\arraystretch}{0.7}
\centering
\caption{The sample mean (Mean) and standard deviation (Std) for the optimal wealth $X^*(T)$ at the lifetime $T=0.5,1.0,1.5,2.0$ to the initial wealth and the estimate of the optimal retirement $\tau$ with its standard error (s.e.).}
\bigskip
\begin{tabular}{cccccc}
\hline\\[1pt]
$T$&Initial wealth $x$& $5$ &\cellcolor{gray} $10$ & $15$& $20$ \\[6pt]
\hline\\[1pt]
{0.5} &Mean& 9.3007 &\cellcolor{gray} 11.916 &16.851&22.823\\ [6pt]
{(20's)} &Std & 14.868 &\cellcolor{gray} 17.239 &21.567&26.148\\[6pt]
{1.0} &Mean& 11.539 &\cellcolor{gray} 13.874 &17.748&21.865\\[6pt]
{(30's)} &Std & 18.818 &\cellcolor{gray} 21.428 &25.192&29.062\\[6pt]
{1.5} &Mean& 9.5184 &\cellcolor{gray} 10.988 &13.479&16.277\\[6pt]
{(40's)} &Std & 18.162 &\cellcolor{gray} 19.472 &22.069&25.695\\[6pt]
{2.0} &Mean& 5.3197 &\cellcolor{gray} 6.0524 &7.3482&8.8727\\[6pt]
{(50's)} &Std & 11.883 &\cellcolor{gray} 12.642 &14.181&16.901\\[6pt]
\hline\\[1pt]
&{$\tau$}&1.0141&\cellcolor{gray} 0.9788 & 0.8537&0.4833\\[6pt]
&{(s.e.)}&(0.0172) &\cellcolor{gray} (0.0225) & (0.0323)&(0.0482)\\[6pt]
\hline
\end{tabular}\label{Table:InitialWealth}
\end{table}

%\smallskip
\noindent
{\it Optimal retirement time and the optimal strategies}.

This section conducts the analysis for the expected optimal retirement time with respect to the change of the featured parameters in our model, the luxury trigger level $a$ and the aged time $\ell$.
We estimate the optimal retirement time within a significant confidence level, and compute sample mean and standard deviation of the optimal wealth $X^*(\tau)$ based on the estimate of $\tau$ for different selection of the luxury consumption level of $a=2, 5, 10, 15, 20$. 
This is to demonstrate the impact of $\tau$ and the expected optimal wealth at $\tau$ when the non-consumption range are getting higher and widened owing to reluctant and bisectional behavior at consuming luxury goods. 
In Table~\ref{Table:lux_trigger}, Panel A displays
the results of the test while the other conditions stay the same, and it shows the corresponding non-consumption range $(k_{-}, k_{+})$ in the last row.
We also demonstrate the mean estimate of the optimal total consumption $k^*(\tau)$ composed of the basic $c^*(\tau)$ and the luxury $g^*(\tau)$ amounts 
under the change of $a$, which is illustrated in Panel B of Table~\ref{Table:lux_trigger}.

We can see that as the luxury trigger higher and widened, the optimal retirement time tends to be earlier, and the optimal wealth at that moment of retirement tends to increase even with not very considerable difference depending on the level of trigger $a$.
The total consumption amount tends to decrease on average as the luxury consumption trigger level increases. 
A notable investigation on the consumption pattern is the change of the consumption proportion to the basic and the luxury with respect to $a$:
The lower luxury trigger level (the thinner non-consumption range), the higher the proportion to the luxury consumption while the lower to the basic. 
In an opposite way, the higher luxury trigger level (the thicker non-consumption range), the lower the proportion to the luxury consumption while the higher to the basic. 

If an individual has an appropriate consumption proportion to the basic and the luxury, the luxury trigger level can be set conversely.
For example, if the proportion between the basic against the luxury is preferred to be around 40\% to 60\%, the luxury consumption needs to be consumed from \$10 in its wealth, under the condition given by the initial income of \$8 and the legacy fund (initial wealth) of \$10. 

\begin{table}
\renewcommand{\arraystretch}{0.7}
\centering
\caption{\emph{Panel A}: The optimal retirement time $\tau$ and the optimal wealth at the estimate of $\tau$ to the luxury trigger $a$, and the corresponding non-consumption range $(k_{-}, k_{+})$.
\emph{Panel B}: The optimal total consumption $k^*(\tau)$ at $\tau$ and the corresponding basic $c^*(\tau)$ and luxury $g^*(\tau)$ consumption with its proportion (Prop) of the total amount.}
\bigskip
\begin{tabular}{cccccccc}
\hline\\[1pt]
&Luxury trigger&$a$&$2$ &$5$& \cellcolor{gray}$10$& $15$ & $20$\\[6pt]
\hline\\[1pt]
{\it Panel A}&{Optimal retire}&{$\tau$} & 1.1656 &1.0536&\cellcolor{gray} 0.9788 & 0.8769 & 0.6748 \\[6pt]
&&{(s.e.)} & (0.0174)&(0.0201)&\cellcolor{gray} (0.0225) & (0.0304) & (0.0433) \\[6pt]
%\hline
&{$X^*(\tau)$}&{Mean} & 10.7865&12.8480& \cellcolor{gray}14.0976& 14.3883 &16.1415\\[6pt]
&&{Std} & 12.0302&16.9628& \cellcolor{gray}21.3372& 23.2820 &20.5534\\[6pt]
%\hline
&{No-consume}&{$k_{-}$} & 2.4495&3.8730& \cellcolor{gray}5.4772& 6.7082& 7.746\\ [6pt]
&&{$k_{+}$} & 10.449&23.873& \cellcolor{gray}45.477& 66.708& 87.746\\[6pt]
\hline\\[1pt]
\hline\\[1pt]
{\it Panel B}&Total consume &{$k^*(\tau)$} & 9.3978 &9.7077 &\cellcolor{gray}9.0440 &7.9753 &6.1934\\[6pt] 
%\hline
&Basic &{$c^*(\tau)$} & 2.5821 &3.1481 &\cellcolor{gray}3.7137 &4.1205 &4.4502\\[6pt]
&&(Prop) & 27.5\% &32.5\% &\cellcolor{gray}41.1\% &51.7\% &71.8\% \\[6pt]
%\hline
&Luxury &{$g^*(\tau)$} & 6.8157 &6.5596 &\cellcolor{gray}5.3303 &3.8548 &1.7432\\[6pt]
&&(Prop)&72.5\% &67.5\% &\cellcolor{gray}58.9\% &48.3\% &28.2\%\\[6pt]
\hline\\[1pt]
\end{tabular}\label{Table:lux_trigger}
\end{table}

Next simulation examines the impact on the aged time $\ell$ of the working cost $L(t)$ to the optimal retirement and the optimal wealth at $\tau$.
Figure~\ref{Fig:Curves_aged} illustrates the difference between $I(t)$ and $L(t)$ (i.e., $I(t) - L(t)$) chosen from the aged time $\ell=1.9, 1.0, 0.7, 0.5$ 
and the relevant optimal retirement boundary curves.
A small $\ell$ indicates requiring high working cost in early lifetime; and the high $\ell$ means relatively low working cost in early lifetime.
As $\ell$ is smaller, the difference curve tends to be more bending at its concavity.
Requiring higher working cost, as $\ell$ is smaller, drives forcing earlier retirement. 
Similarly, a lower working cost taking from a larger $\ell$ leads to encourage working lifetime to be longer. 
This tendency appears in the right of Figure~\ref{Fig:Curves_aged}, where the optimal retirement boundary goes down as $\ell$ is smaller.
Meanwhile, the optimal wealth at $\tau$ varies not significantly to the change of $\ell$, as shown in Panel A of Table~\ref{Table:aged}.
One notable thing is that \emph{not-working} can be optimal if a job with $\ell$ less than 0.65, which implies a working (especially initial) cost exceeding a certain threshold, may not motivate to work for a whole life, in terms of satisfaction from consumption.

In Table~\ref{Table:aged}, Panel B exhibits the total optimal consumption $k^*(\tau)$ composed of $c^*(\tau)$ and $g^*(\tau)$ at $\tau$ based on its estimate.
We can see the total consumption and its basic and luxury good amounts have no significant difference depending on the change of $\ell$. 
Regardless of high or low working cost for the young ages, ones expect to do consuming basic and luxury goods with about four to six ratio.
It implies the extent to a working cost required at the young age affects little influence to individuals' consumption behavior on average.

\begin{figure}[H]
\begin{center}
\includegraphics[width=0.48\textwidth]{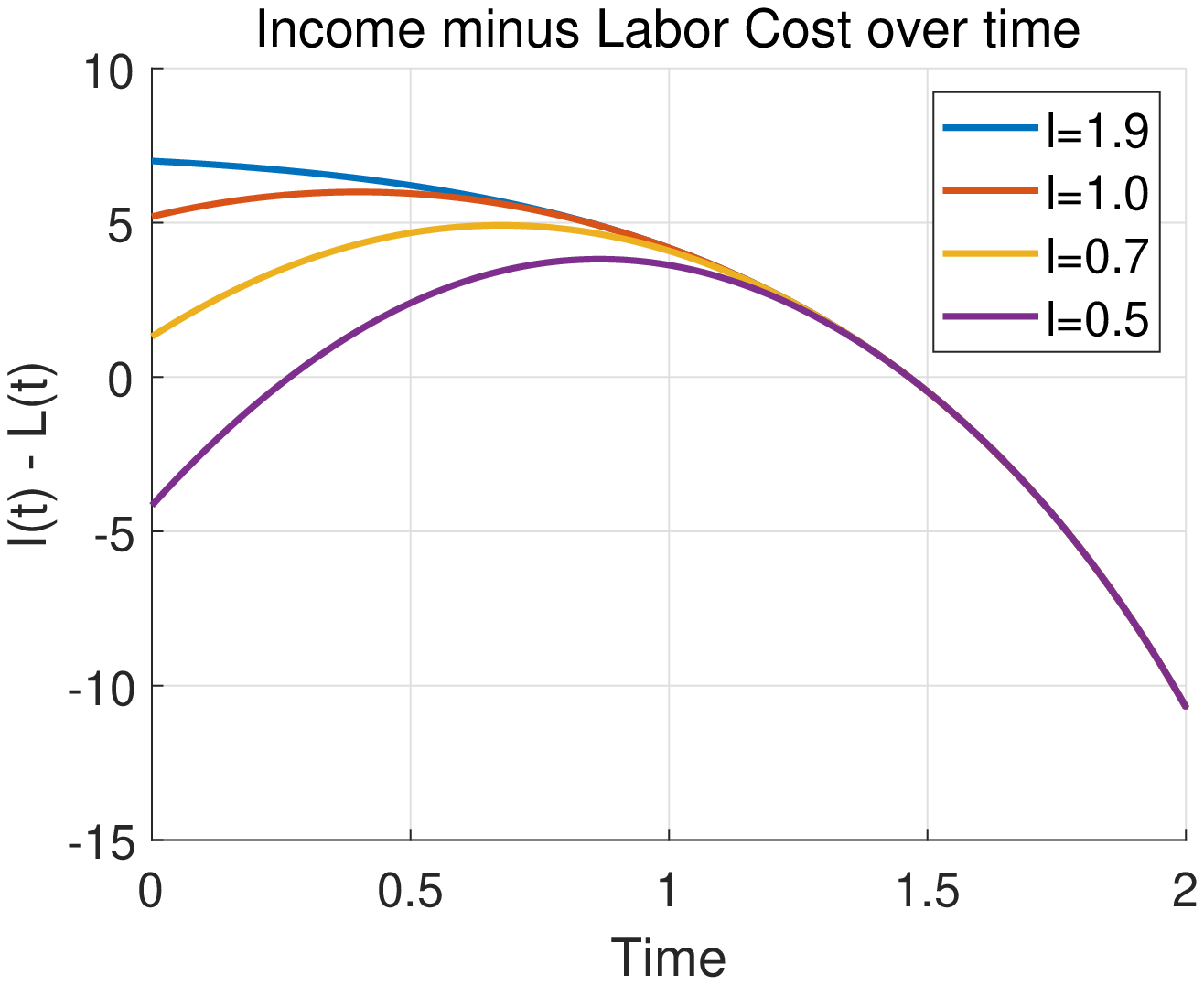}
\includegraphics[width=0.48\textwidth]{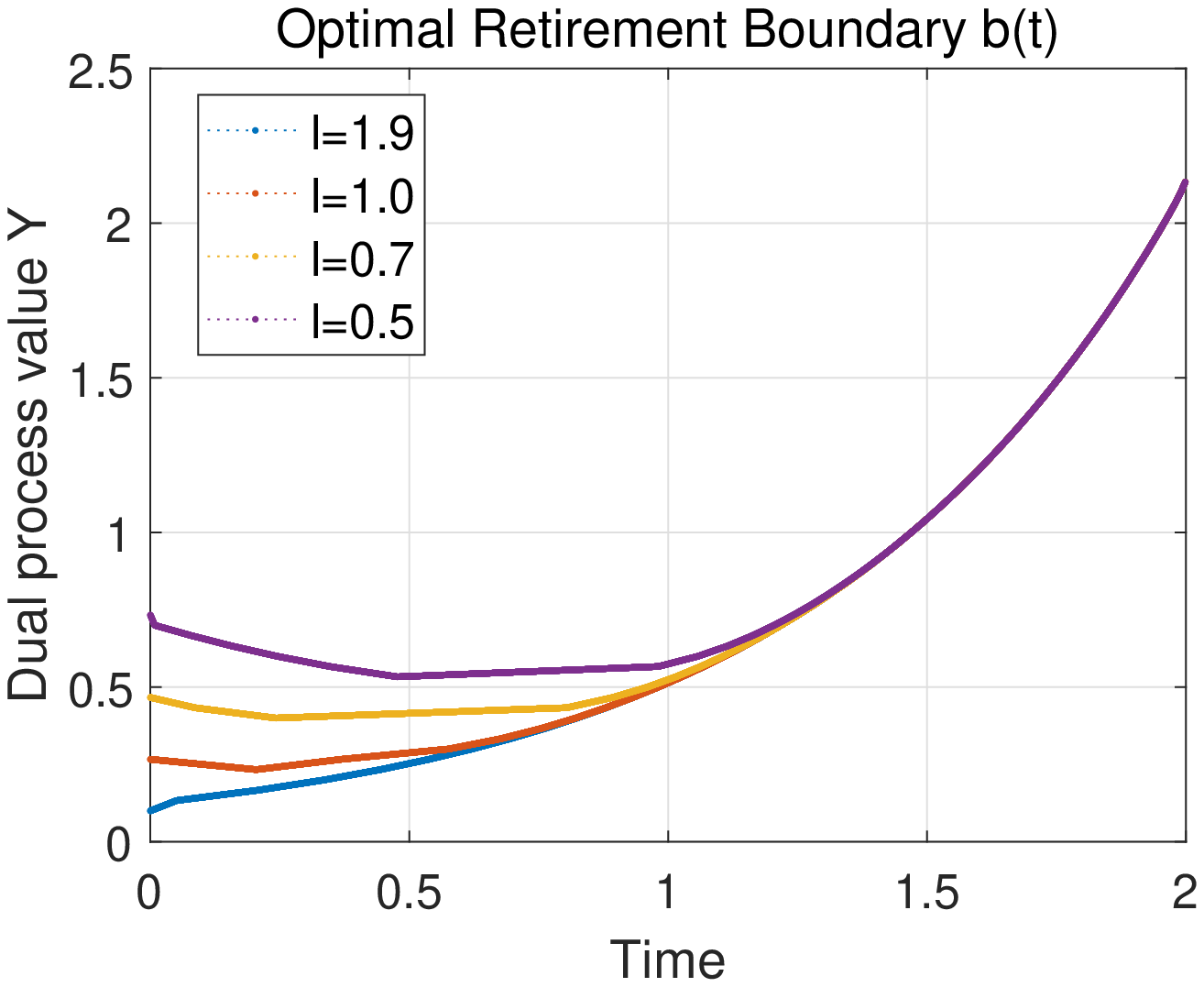}
\end{center}
\caption{Difference between the income $I(t)$ and labor cost $L(t)$ with respect to the aged time $\ell=1.9, 1.0, 0.7, 0.5$ and the corresponding optimal retirement boundary curves.
}
\label{Fig:Curves_aged}
\end{figure}

\begin{table}[h]
\renewcommand{\arraystretch}{0.7}
\centering
\caption{\emph{Panel A}: The optimal retirement time $\tau$ with respect to the aged time $\ell$ and the corresponding optimal wealth at $\tau$. 
\emph{Panel B}: The optimal total consumption $k^*(\tau)$ at the estimate of $\tau$ and the corresponding basic $c^*(\tau)$ and luxury $g^*(\tau)$ consumption with its proportion (Prop) of the total amount with respect to $\ell$.}
\bigskip
\begin{tabular}{ccccccccc}
\hline\\[1pt]
&Aged time&$\ell$&\cellcolor{gray}$0.7$& $1.0$ & $1.3$&$1.7$&$1.9$\\
\hline\\[1pt]
{\it Panel A}&{Optimal retire}& {$\tau$} & \cellcolor{gray} 0.9788 & 1.0376 & 0.9932 & 1.0380& 1.0277\\[6pt]

&&{(s.e.)} &\cellcolor{gray} (0.0225)& (0.0122)&(0.0119) & (0.0140)& (0.0136)\\[6pt]
%\hline
%\cmidrule(r){2-8}
&{$X^*(\tau)$}&{Mean} & \cellcolor{gray}14.097& 16.600&16.491&15.240&14.910\\[6pt]
&&{Std} & \cellcolor{gray}21.337& 23.981 &23.904&22.860&22.430\\[6pt]
\hline\\[1pt]
\hline\\[1pt]
{\it Panel B}&Total consume &{$k^*(\tau)$} &\cellcolor{gray}9.0440 &10.999 &10.675 &10.005 &9.8639\\ [6pt]
%\cmidrule(r){2-7}
%\hline
&Basic &{$c^*(\tau)$} &\cellcolor{gray} 3.7137 &3.9020 &3.8938 &3.8023 &3.7670\\[6pt]
&&(Prop) &\cellcolor{gray}41.1\% &35.5\% &36.5\% &38.0\% &38.2\% \\[6pt]
%\hline
&Luxury &{$g^*(\tau)$} &\cellcolor{gray}5.3303 &7.0971 &6.7814 &6.2028 &6.0969\\[6pt]
&&(Prop)&\cellcolor{gray}58.9\% &64.5\% &63.5\% &62.0\% &61.8\%\\[6pt]
\hline\\[1pt]
\end{tabular}\label{Table:aged}
\end{table}

}

\smallskip
\section{Concluding Remarks}\label{Sec:Conclusion}
\noindent

{

In this study, we discussed an optimal investment, heterogeneous consumption problem with the conditions of mandatory retirement age and options for early retirement. We analytically solved the problem with the dual control method and characterized the free boundary separated by the working/retirement regions. 
We derived the optimal wealth, trading strategy, and basic and luxury consumption for optimal decision making. 

By means of the sequential procedure shown by the theoretical results, we conducted the sensitivity tests for the key model factors including trade-off income/working cost, wealth endowment, and non-concavity of utility. 
At this end, the theoretical results were clearly visualized to further intuitively examine the impact to the consumption and early retirement decisions. 
Our main findings were as follows:
The higher net-income the retirement time defers later. 
The higher wealth granted, the earlier retirement is optimal, and it tends to consume the luxury more.
The luxury trigger level causing non-concavitiy for utility 
really matters to determine overall consumption patterns for the basic and the luxury -- the lower luxury trigger level allows to consume the luxury relatively more than the basic. 

Our simulation showed that high-net worth individuals have no reason to be working and the poor should be working longer depending on their wealth level. 
Such even too obvious phenomena can be understood as one of the optimal decision making results to individuals.
The optimal retirement time is determined to late 30's or early 40's under a plausible parameter set appliable average individuals, which implies that it is the optimal decision that average persons may retire earlier before the mandatory age. 
Our findings could have important policy implications for government in making pension and retirement age decisions. 
}

{There remain many open questions, for example, models with random 
parameters, general utility functions for heterogeneous consumption and terminal wealth, 
testable empirical hypotheses and implications, etc. We leave these questions and others to future research.
}

\smallskip
%\theendnotes

\section*{Acknowledgements}
The authors are grateful to the two anonymous reviewers, Associate Editor and Area Editor, whose constructive comments and suggestions have helped to improve the paper of the previous four versions. 
The first author was partially supported by the National Research Foundation of Korea grant funded by the Korea government (MSIT) (No. 2021R1C1C1004647).
The second author was partially supported by the National Natural Science Foundation of China (No. 11971409) and Hong Kong GRF (Nos. 15204216, 15202817, 15202421).
The third author was partially supported by the Engineering and Physical Sciences Research Council (UK) grant (EP/V008331/1).

\newpage
\appendix

\section{Proof of \citeprop{largeconsumption}} 
For any non-homothetic utility function
\[u(c, g)=U\bigg(\frac{c^{\alpha}}{\alpha}+\frac{((g-a)^{+})^{\beta}}{\beta}\bigg), \]
we {first} show \(\lim_{k\to\infty}\frac{g^{*}(k)}{k}=1,\) 
%\blue
{which means the consumption on luxury goods dominates as the total consumption increases to infinity.}
Suppose this was not true. Then, because $0\leq g^{*}(k)\leq k$, there would exist a constant $0<\epsilon<1$ and an unbounded sequence of $k_{n}$ such that $g^{*}(k_{n})<(1-\epsilon)k_{n}$ for all $n$. Thus, 
\[\bar{u}(k_{n})=u(c^{*}(k_{n}), g^{*}(k_{n}))\leq u(k_{n}, g^{*}(k_{n}))\leq U\bigg(\frac{k_{n}^{\alpha}}{\alpha}+\frac{(((1-\epsilon) k_{n}-a)^{+})^{\beta}}{\beta}\bigg).\]
On there other hand, since $\beta>\alpha$, we have 
\[\frac{(((1-\epsilon/2) k_{n}-a)^{+})^{\beta}}{\beta}>\frac{k_{n}^{\alpha}}{\alpha}+\frac{(((1-\epsilon) k_{n}-a)^{+})^{\beta}}{\beta}\]
for sufficiently large $k_{n}$, which leads to 
\[u\big(\epsilon k_{n}/2, (1-\epsilon/2)k_{n}\big)\geq U\bigg(\frac{(((1-\epsilon/2) k_{n}-a)^{+})^{\beta}}{\beta}\bigg)>U\bigg(\frac{k_{n}^{\alpha}}{\alpha}+\frac{(((1-\epsilon) k_{n}-a)^{+})^{\beta}}{\beta}\bigg)
\geq \bar{u}(k_{n}).\]
This clearly contradicts the definition of $\bar{u}$. 

%\blue
{
We next show that $c^*(k)=k$ when the total consumption $k$ is small. When $k\leq a$, 
\[\bar{u}(k)=\max_{0\leq c\leq k}u(c, k-c)=\max_{0\leq c\leq k}U\bigg(\frac{c^{\alpha}}{\alpha}\bigg)=u(k, 0)\] with $c^*(k)=k$. When $k>a$, 
$$\bar{u}(k)=\max\Big\{\max_{0\leq c\leq k-a}u(c, k-c), \max_{k-a\leq c\leq k}u(c, k-c)\Big\}=\max\big\{u(c(k), k-c(k)), u(k, 0)\big\}, $$
where $c(k)\in (0, k-a)$ is the unique root of the equation $c^{\alpha-1}-(k-c-a)^{\beta-1}=0$. 
Evidently $k-a=c(k)+(c(k))^{\frac{1-\alpha}{1-\beta}}$, so $\lim_{k\to\infty}c(k)=\infty$. 
And consequently $\lim_{k\to\infty}\frac{k-a}{(c(k))^{\frac{1-\alpha}{1-\beta}}}=\lim_{k\to\infty}\frac{c(k)+(c(k))^{\frac{1-\alpha}{1-\beta}}}{(c(k))^{\frac{1-\alpha}{1-\beta}}}=1$, which yields $\lim_{k\to\infty}\frac{c(k)}{k^{\frac{1-\beta}{1-\alpha}}}=1$. 
Also when $c(k)\geq 1$, $k-a=c(k)+(c(k))^{\frac{1-\alpha}{1-\beta}}\geq 2c(k)$, so 
$c(k)\leq (k-a)/2$. As $\beta>\alpha$, it follows 
\begin{align*}
\liminf_{k\to\infty} \left({1\over \alpha} c(k)^{\alpha}+{1\over\beta} (k-c(k)-a)^{\beta}-{1\over \alpha}k^\alpha\right) 
&\geq \liminf_{k\to\infty} \left({1\over\beta} ((k-a)/2)^{\beta}-{1\over \alpha}k^\alpha\right)=\infty, 
\end{align*}
which implies $u(c(k), k-c(k))>u(k, 0)$ for sufficiently large $k$. 
As $\lim_{k\to a+}c(k)=0$, we have
\begin{align*}
&\lim_{k\to a+}\left({1\over \alpha} c(k)^{\alpha}+{1\over\beta} (k-c(k)-a)^{\beta}-{1\over \alpha}k^\alpha\right)=-{1\over \alpha}a^\alpha<0, 
\end{align*}
which implies $u(c(k), k-c(k))<u(k, 0)$ when $k$ is sufficiently close to $a$. 
Because 
$$\left({1\over \alpha} c(k)^{\alpha}+{1\over\beta} (k-c(k)-a)^{\beta}-{1\over \alpha}k^\alpha\right)'=-k^{\alpha-1}<0, $$ we conclude that $u(c(k), k-c(k))-u(k, 0)=0$ admits a unique root $k_0>a$. 
Furthermore, $c^*(k)=k $ for $k\leq k_0$ and $c^*(k)=c(k)$ for $k\geq k_0$. This implies one should only have basic consumption when the total consumption $k$ is small, and the consumption on basic goods tends to infinity as the total consumption increases to infinity due to the relation $\lim_{k\to\infty}c(k)=\infty$. The concave envelope $\widetilde u$ of $\bar{u}$ has one straight line segment that connects two curves at tangent points.
}

\section{Example \ref{nonconcave}}
Let 
\begin{align*}
f(c)=2c^{1/2}+\frac{4}{3} ((k-c-a)^{+})^{3/4}=\begin{cases}
2c^{1/2}+\frac{4}{3} (k-c-a)^{3/4}, &\quad 0\leq c\leq k-a;\\
2c^{1/2}, & c\geq k-a.
\end{cases}
\end{align*}
\begin{itemize}
\item When $0\leq k\leq a$, we have
\[\bar{u}(k)=\max_{0\leq c\leq k}f(c)=\max_{0\leq c\leq k}2c^{1/2}=2k^{1/2}, \] 
and the optimal pair is
\[c^{*}(k)=k, \quad g^{*}(k)=0.\]
\item When $k>a$, let 
{\[c^{*}=\sqrt{k-a+\frac{1}{4}}-\frac{1}{2}.\]
Then $0\leq c^{*}<k-a$ and $f'(c^{*})=0$. Because $f''\leq 0$, $f$ is concave on $[0, k-a]$. 
Hence 
\begin{align*}
\max_{0\leq c\leq k-a}f(c)&=f(c^{*})=2(c^{*})^{1/2}+\frac{4}{3}(c^{*})^{3/2}={4\over 3}\sqrt{\sqrt{k-a+\frac{1}{4}}-\frac{1}{2}}\bigg(\sqrt{k-a+\frac{1}{4}}+1\bigg).
\end{align*}
Consequently, 
\begin{align*}
\bar{u}(k)&=\max_{0\leq c\leq k}f(c)=\max\Big\{\sup_{0\leq c< k-a}f(c), \;\sup_{k-a\leq c\leq k}f(c)\Big\}\\
&=\max\left\{\tfrac{4}{3}\sqrt{\sqrt{k-a+\tfrac{1}{4}}-\tfrac{1}{2}}\bigg(\sqrt{k-a+\tfrac{1}{4}}+1\bigg), \; 2k^{1/2}\right\}\\
&=\begin{cases}
2k^{1/2}, &\quad a<k\leq k_{0};\\
{4\over 3}\sqrt{\sqrt{k-a+\frac{1}{4}}-\frac{1}{2}}\bigg(\sqrt{k-a+\frac{1}{4}}+1\bigg), &\quad k> k_{0}, 
\end{cases}
\end{align*}
where $k_{0}$ is the unique solution for 
\[{4\over 3}\sqrt{\sqrt{k-a+\frac{1}{4}}-\frac{1}{2}}\bigg(\sqrt{k-a+\frac{1}{4}}+1\bigg)=2k^{1/2}, \]
such that the continuity condition $\bar{u}(k_{0}+)=\bar{u}(k_{0}-)$ holds. {Combining the discussion on $0\leq k\leq a$, we have} the optimal pair is given by
\[c^{*}(k)
=\begin{cases}
k, &\quad a<k\leq k_{0};\\
\sqrt{k-a+\frac{1}{4}}-\frac{1}{2}, &\quad k> k_{0}, 
\end{cases}\]
and
\[g^{*}(k)=\begin{cases}
0, &\quad a<k\leq k_{0};\\
k+\frac{1}{2}-\sqrt{k-a+\frac{1}{4}}, &\quad k> k_{0}.
\end{cases}\]
}
\end{itemize}

\section{Post-Retirement Problem}

\textbf{On the function $V$: } 
%\blue
{Choose the policy $\pi(t)=0$ and $k(t)=r(t)X(t)/2$ for $t\geq \tau$, then the wealth process follows 
\begin{align*} 
\dd X(t)=r(t)X(t)/2\dt, \quad \text{for $t>\tau$}. 
\end{align*} 
It follows that $X(s)=X(t)e^{\int_t^s r(u)\du/2}$ and $k(s)=r(s)X(t)e^{\int_t^s r(u)\du/2}/2$ for $s\geq t\geq \tau$. Hence, 
\begin{align*} 
V(t, x)&\geq \BE{\int_{t}^{\ot} e^{-\int_t^s \rho(u) \du }\bar{u}\Big(r(s)X(t)e^{\int_t^s r(u)\du/2}/2\Big)\ds\;\bigg|\;X(t)=x}\\
&=\int_{t}^{\ot} e^{-\int_t^s \rho(u) \du }\bar{u}\Big(r(s)xe^{\int_t^s r(u)\du/2}/2\Big)\ds, 
\end{align*}
which shows $V$ is unbounded as $\bar{u}$ is unbounded. }
\par
\textbf{On the function $h$: }
The convexity of $h$ follows from its definition and this implies its continuity on $(0, \infty)$. The estimate $h(y)\ll y^{\frac{\beta}{\beta-1}}$ as $y\to+\infty$ follows from \eqref{barugrowth}; and $h(y)\to+\infty$ as $y\to 0^{+}$ follows from \eqref{unboundedu'}. Obviously $h$ is decreasing. If it were not strictly decreasing, then $h$ would attain a local (positive) minimum at some point, which would also be the global minimum by its convexity, but this contradicts the estimate $h(y)\ll y^{\frac{\beta}{\beta-1}}\to 0$ as $y\to+\infty$.

%\par 
{
\par
\textbf{On Remark \ref{onefund}: }
By It\^{o}'s lemma, 
\begin{align*}
\dd \;(\log Y(t))=(\rho-r-\frac{1}{2}\Vert\vartheta\Vert^{2})\dt-\vartheta^\top\dd B(t), 
\end{align*}
and 
\begin{align*}
\dd \;(\log S_{i}(t))=\left(b_i-\frac{1}{2}\sum\limits_{j=1}^n\sigma_{ij}^2\right)\dt+\sum\limits_{j=1}^n\sigma_{ij}\dd B_{j}(t), \quad i=1, 2, \ldots, n.
\end{align*}
This gives the equation. 
}

%\blue
{
\textbf{Proof of \citethm{postsolution1}: }
%\blue
{We first notice that the condition \eqref{growthcondition} implies 
\begin{align*}%\label{growthcondition}
&\quad\;\BE{\int_{0}^{\ot}e^{-\int_0^s \rho(u)du} Y(s)^{\frac{\beta}{\beta-1}}\ds \;\bigg|\; Y(0)=1}\\
&=\BE{\int_{0}^{\ot}e^{-\int_0^s \rho(u)du} e^{\frac{\beta}{\beta-1}\big(\int_0^s(\rho(u)-r(u)-\frac{1}{2}\Vert\vartheta(u)\Vert^2)\du-\int_0^s\vartheta(u)^\top\dd B(u)\big)}\ds}\\
&=\int_{0}^{\ot} e^{-\int_0^s \rho(u)du+\frac{\beta}{\beta-1}\int_0^s(\rho(u)-r(u)-\frac{1}{2}\Vert\vartheta(u)\Vert^2)\du+\frac{1}{2}\frac{\beta^2}{(\beta-1)^2}\int_0^s\Vert\vartheta(u)\Vert^2\du}\ds\\
&=\int_{0}^{\ot} e^{\int_0^s(\frac{1}{\beta-1}\rho(u)-\frac{\beta}{\beta-1} r(u)+ \frac{1}{2}\frac{\beta}{(\beta-1)^2}\Vert\vartheta(u)\Vert^2\du}\ds<\infty.
\end{align*} 
Together with the upper bound $h(y)\ll y^{\frac{\beta}{\beta-1}}$, we obtain 
$\hBV(t,y)\ll y^{\frac{\beta}{\beta-1}}$.}
It is not hard to verify that $\hBV(t,y)$ is a convex solution to 
\begin{align}\label{hjb2}
\partial_{t} \hBV(t,y)+\frac{1}{2}\Vert\vartheta(t)\Vert^{2}y^{2} \partial_{yy} \hBV(t,y)+(\rho(t)-r(t))y \partial_{y} \hBV(t,y)-\rho(t)\hBV(t,y)+h(y)=0, 
\end{align} 
on $[0,\ot)\times(0, \infty)$. Using the fact that $h$ is a convex PLD function, we see that $\hBV$ is strictly convex w.r.t. $y$. Furthermore, when $\ot=+\infty$, 
\begin{align}\label{growthy1}
\lim_{N\to+\infty}\BE{e^{-\int_{\tau}^{\tau+N} \rho(t)\dt}\hBV(\tau+N, Y(\tau+N)) \;\big|\; Y(\tau)=y}=0
\end{align} 
for any stoping time $\tau$ and $y>0$. %Here we define $\hBV(t,y)=\hBV(\ot,y)$ for $t>\ot$. 
%By means of viscosity solution, we can show $\hBV\in C^{0,1}([0,\ot)\times (0, \infty))$. 
Since $\vartheta(\cdot)$, $\rho(\cdot)$, $r(\cdot)$ and $h(\cdot)$ are all continuous and $\Vert\vartheta(\cdot)\Vert$ is uniformly positive, 
we have $\hBV\in C^{1,2}([0,\ot)\times (0, \infty))$ by standard PDE theory (see, e.g., \cite{K80,K87}).
Define 
\begin{align*} 
\BVtwo(t,x)=\inf_{y>0}(\hBV(t,y)+xy), \quad (t,x)\in [0,\ot]\times (0, \infty), 
\end{align*} 
then we have the dual relationship $\hBV(t,y)=\sup_{x>0}(\BVtwo(t,x)-xy)$. 
Moreover, $\BVtwo(t,x)=\hBV(y^*)+xy^*$ with $y^*=y^*(t,x)$ being the solution of the equation $\hBV'(t,y)+x=0$, which leads to $\BVtwo\in C^{1}([0,\ot)\times (0, \infty))$ by the implicit function theorem and $\BVtwo'(t,x)=y^*$ and $\hBV''(t,y)\partial_x y^*+1=0$. Thus $\BVtwo''(t,x)=\partial_x y^*<0$. So $\BVtwo$ is a concave PLI solution for the HJB equation \eqref{hjb1}.
Since the coefficients of \eqref{hjb1} are all continuous and $\vartheta\neq 0$, we see $\BVtwo\in C^{2}([0,\ot)\times (0, \infty))$. 
Furthermore, when $\ot=+\infty$, since $0\leq \BVtwo(t,x)\leq \hBV(t,y)+xy$, it follows from \eqref{growthy1} that 
\begin{align}\label{growthx1}
\lim_{N\to+\infty}\BE{e^{-\int_{\tau}^{\tau+N}}\BVtwo(\tau+N, X(\tau+N)) \;\big|\; X(\tau)=x}=0 
\end{align} 
for any stoping time $\tau$ and $x>0$. 
\par
For any admissible stopping time $\tau$ and strategy $(k(\cdot), \pi(\cdot))$, let 
\[\theta_{n, N}=\inf\{t\geq \tau: |X(t)-X(\tau)|+|\pi(t)|\geq n\}\wedge (\tau+N)\wedge \ot.\]
Then by It\^{o}'s lemma and \eqref{hjb1}, 
\begin{align}\label{ver1}
\nn &\quad\dd\; (e^{-\int_0^t\rho(s)\ds}\BVtwo(t,X(t)))\\
\nn &=e^{-\int_0^t\rho(s)\ds}(-\rho(t) \BVtwo(t,X(t))+\tfrac{1}{2}\Vert\pi^\top(t)\sigma(t)\Vert^{2} \partial_{xx} \BVtwo(t,X(t))\\
\nn &\quad+(r(t)X(t)+\pi(t)^\top\mu(t)-k(t)) \partial_{x} \BVtwo(t,X(t)))\dt+e^{-\int_0^t\rho(s)\ds}\BVtwo_x(t,X(t))\pi(t)^\top\sigma\dd B(t)\\
&\leq-e^{-\int_0^t\rho(s)\ds}\bar{u}(k(t))\dt+e^{-\int_0^t\rho(s)\ds} \partial_{x}\BVtwo(t,X(t))\pi(t)^\top\sigma(t)\dd B(t).
\end{align}
Hence, 
\begin{align} \label{ver2}
\BVtwo(\tau,X(\tau)))& \geq e^{-\int_{\tau}^{\theta_{n, N}}\rho(s)\ds}\BVtwo(\theta_{n, N}, X(\theta_{n, N}))+\int_{\tau}^{\theta_{n, N}}e^{-\int_{\tau}^t\rho(s)\ds}\bar{u}(k(t))\dt\nn \\
&\quad-\int_{\tau}^{\theta_{n, N}}e^{-\int_{\tau}^t\rho(s)\ds}\BVtwo' (X(t))\pi(t)^\top\sigma(t)\dd B(t)\nn\\
& \geq \int_{\tau}^{\theta_{n, N}}e^{-\int_{\tau}^t\rho(s)\ds}\bar{u}(k(t))\dt -\int_{\tau}^{\theta_{n, N}}e^{-\int_{\tau}^t\rho(s)\ds}\BVtwo' (X(t))\pi(t)^\top\sigma(t)\dd B(t).
\end{align}
Taking the conditional expectation on both sides, 
\[\BVtwo(\tau,x)\geq \BE{\int_{\tau}^{\theta_{n, N}}e^{-\int_{\tau}^t\rho(s)\ds}\bar{u}(k(t))\dt\;\bigg|\; X(\tau)=x}.\]

Letting $n\to\infty$ and $N\to\infty$ in above, by Fatou's lemma we obtain 
\begin{align*} 
\BVtwo(\tau,x)& \geq \BE{\int_{\tau}^{\ot}e^{-\int_{\tau}^t\rho(s)\ds}\bar{u}(k(t))\dt\;\bigg|\; X(\tau)=x}.
\end{align*}
Maximizing the right hand side over $(k(\cdot), \pi(\cdot))$, we get $\BVtwo(\tau,x)\geq V(\tau,x)$.
\par
On the other hand, we use feedback controls %\emr{needs dual argument}%
\[\pi^{*}(t)=-\tfrac{\partial_{x}\BVtwo(t,X(t))}{\partial_{xx}\BVtwo(t,X(t))}\vartheta(t), 
\quad k^{*}(t)=\argmax_{k\geq 0}(\tilde u(k)-k\partial_{x}\BVtwo(t,X(t)))\]
in the wealth process \eqref{wealth2-1}. The corresponding inequalities in \eqref{ver1} and \eqref{ver2} become identities, so 
\begin{align*} 
\BVtwo(\tau,x)&=\BE{e^{-\int_{\tau}^{\theta_{n, N}}\rho(s)\ds}\BVtwo(\theta_{n, N}, X(\theta_{n, N}))\;\bigg|\; X(\tau)=x}+\BE{\int_{\tau}^{\theta_{n, N}}e^{-\int_{\tau}^t\rho(s)\ds}\bar{u}(k(t))\dt\;\bigg|\; X(\tau)=x}.
\end{align*}
Letting $n\to\infty$ in above, by the dominated convergent theorem, we obtain 
\begin{align*} 
\BVtwo(\tau,x)&=\BE{e^{-\int_{\tau}^{(\tau+N)\wedge \ot}\rho(s)\ds}\BVtwo((\tau+N)\wedge \ot,X((\tau+N)\wedge \ot))\;\bigg|\; X(\tau)=x}\\
&\quad\;+\BE{\int_{\tau}^{(\tau+N)\wedge \ot}e^{-\int_{\tau}^t\rho(s)\ds}\bar{u}(k(t))\dt\;\bigg|\; X(\tau)=x}. 
\end{align*}
Letting $N\to\infty$, it follows from \eqref{growthx1} that
\begin{align*} 
\BVtwo(\tau,x)&=\BE{\int_{\tau}^{\ot}e^{-\rho (t-\tau)}\bar{u}(k(t))\dt\;\bigg|\; X(\tau)=x}\leq V(\tau,x).
\end{align*}
This shows $\BVtwo=V$. By virtue of this, we see
\[\pi^{*}(t)=-\tfrac{\partial_{x}V(t,X(t))}{\partial_{xx}V(t,X(t))}\vartheta(t)=(\sigma(t)^T)^{-1}\vartheta(t) Y(t)\p_{yy} \hat V(t,Y(t)).\]
Similarly 
\[ k^{*}(t)=\argmax_{k\geq 0}(\tilde u(k)-k\partial_{x}V(t,X(t)))=-\p_y h(Y(t)).\]
This completes the proof. 

\par
\textbf{Proof of \citethm{postsolution}: }
This is similar to the proof of \citethm{postsolution1}, so we omit the details.
}
\par

\textbf{On \citeeg{eg:3}: }
Let 
\[u_{1}(c)=\frac{(c-c_{0})^{1-\phi}}{1-\phi}, \quad u_{2}(g)=\frac{(g+b_{0})^{1-\psi}}{1-\psi}.\]
Then $u(c, g)=u_{1}(c)+u_{2}(g)$. 
%\blue
{Simple computation shows that $\bar{u}(k)=u_1(k)+u_2(0)$ if $c_0\leq k\leq c_0+b_0^{\psi/\phi}$ and $\bar{u}(k)=u_1(c^*)+u_2(k-c^*)$ if $k> c_0+b_0^{\psi/\phi}$, where $c^*\in (c_0, k)$ is the unique root of the equation $(c-c_0)^{-\phi}-(k-c+b_0)^{-\psi}=0$, and $\bar{u}$ is a continuously differentiable, strictly increasing and strictly concave function. In fact, we do not need to compute $\bar{u}$ to find $h$ as discussed next. 
}
The dual functions of $u_{1}$ and $u_{2}$ are respectively given by 
\[\widehat{u}_{1}(y)=\sup_{\substack{c\geq c_{0}}}\big(u_{1}(c)-cy\big)=\tfrac{\phi}{1-\phi}y^{1-\frac{1}{\phi}}-c_{0}y, \]
and
\begin{align*}
\widehat{u}_{2}(y)=\sup_{\substack{g\geq 0}}\big(u_{2}(g)-gy\big)=\begin{cases}
\tfrac{\psi}{1-\psi}y^{1-\frac{1}{\psi}}+b_{0}y, &\quad y< b_{0}^{-\psi};\\
\tfrac{1}{1-\psi}b_{0}^{1-\psi}, &\quad y\geq b_{0}^{-\psi}.
\end{cases}
\end{align*}
Therefore, 
\begin{align*}
h(y)&=\sup_{k\geq c_{0}} \sup_{\substack{c\geq c_{0}, \;g\geq 0, \\c+g=k}}\big(u(c, g)-ky\big)=\sup_{\substack{c\geq c_{0}, g\geq 0}}\big(u_{1}(c)+u_{2}(g)-(c+g)y\big)\\
&=\sup_{\substack{c\geq c_{0}}}\big(u_{1}(c)-cy\big)+\sup_{\substack{g\geq 0}}\big(u_{2}(g)-gy\big)\\
&=\widehat{u}_{1}(y)+\widehat{u}_{2}(y)\\
&=\tfrac{\phi}{1-\phi}y^{1-\frac{1}{\phi}}-c_{0}y+\tfrac{1}{1-\psi}b_{0}^{1-\psi}+\big(\tfrac{\psi}{1-\psi}y^{1-\frac{1}{\psi}}+b_{0}y-\tfrac{1}{1-\psi}b_{0}^{1-\psi}\big)\id{y< b_{0}^{-\psi}}, 
\end{align*}
which is in $C(0, \infty)$ but not in $C^{1}(0, \infty)$. 
\par
We now turn to find a special solution of \eqref{hjb2} in the form
\[\hBV_{0}(y)=C_{1}y^{1-\frac{1}{\phi}}+C_{2}y+C_{3}+(C_{4}y^{1-\frac{1}{\psi}}+C_{5}y+C_{6})\id{y< b_{0}^{-\psi}}, \]
which satisfies 
\[\hBV_{0}(y)\ll y^{\frac{p}{p-1}}.\]
The parameters can be easily determined to be 
\begin{align*}
C_{1}&=-\tfrac{\phi}{1-\phi}\tfrac{\phi^{2}}{\frac{1}{2}\Vert\vartheta\Vert^{2}(1-\phi)-\rho\phi-r(\phi^{2}-\phi)}, 
& C_{2}&=-\tfrac{c_{0}}{r}, & C_{3}&=\tfrac{1}{\rho(1-\psi)}b_{0}^{1-\psi}, \\
C_{4}&=-\tfrac{\psi}{1-\psi}\tfrac{\psi^{2}}{\frac{1}{2}\Vert\vartheta\Vert^{2}(1-\psi)-\rho\psi-r(\psi^{2}-\psi)}, 
& C_{5}&=\tfrac{b_{0}}{r}, & C_{6}&=-\tfrac{1}{\rho(1-\psi)}b_{0}^{1-\psi}.
\end{align*}
Since $h\in C(0, \infty)$, by \eqref{hjb2}, we see that $\hBV_{0}\in C^{2}(0, \infty)$. One can show that $\hBV_{0}$ satisfies the requirements.

\section{Pre-Retirement Problem}%\label{Sec:Pre}
%\blue
{
In general, there is no classical solution to (\ref{hjb5}), so we need to work on derivatives in some weak sense. 
A natural space for that is the Sobolev space, used extensively in the PDE theory, see \cite{K80} for details. Denote by $W^{1, 2}_{p, loc}(\sets)$ the local Sobolev space where $p>3$.

We have the following comparison principle for non linear equations; see \cite{L96}.
\begin{theorem}[Comparison principle]\label{cp}
Let $u_i(t, y)$, $ i=1, 2$, be the solutions in $W^{1, 2}_{p, loc}(\sets)$ of the following variational inequalities
\begin{align*}
\begin{cases}
\min\big\{-(\p_{t}+{\cal M})u_i-f_i(t, y), \;u_i-g_i(t, y)\big\}=0, \quad\;(t, y)\in\sets, \\
u_i(T, y)=h_i(y), 
\end{cases}
\end{align*}
where ${\cal M}$ is a linear elliptic operator on $y$. If $f_1\geq f_2$, $g_1\geq g_2$, $h_1\geq h_2, $ and for some $C>0$, 
$|u_1(t, y)|+|u_2(t, y)|\leq C e^{Cy^2}$ on $\sets$, 
then $u_1 \geq u_2$ on $\sets$.
\end{theorem}

\textbf{Proof of \citethm{presolution}: }
%\greenone
Thanks to $h(y)\ll y^{\frac{\beta}{\beta-1}}$ and \eqref{growthcondition}, we see $\hBW<\infty$. 
The convexity and PLD property w.r.t. $y$ variable follows immediately from its definition. 
We can find a solution to \eqref{hjb5} in $W^{1,2}_{p, loc}$ by standard penalty approximation method. We just give the main idea here; details can be found, for instance, in \cite{DY09}. Consider the following penalty approximation problem to \eqref{hjb5}:
\[-(\p_{t}+\BBL)\hBW+\rho(t)\hBW-yI(t)-h(y)+L(t)+\beta_{\epsilon}(\hBW-\hBV)=0, \quad\;(t, y)\in\sets, \]
where $\beta_{\epsilon}$ is certain penalty function.
Using the Leray-Schauder fixed point theorem (see \cite{GT83}) and the embedding theorem (see \cite{Fr64}), we obtain a $C^{\frac{1+\alpha}{2},1+\alpha}$ (for some $0<\alpha<1$) solution $\hBW^\epsilon$ to the above penalty approximation problem. By the Schauder estimation (see \cite{SL68}), it follows that $\hBW^\epsilon \in C^{1+\frac{\alpha}{2}, 2+\alpha} $. Then applying $C^{\frac{\alpha}{2},\alpha}$ interior estimate and $W^{1,2}_p$ interior estimate (with $p>3$), we affirm that $\hBW^\epsilon$ weakly converges to some $\hBW$ in $W^{1,2}_{p, loc}$. Finally one can show $\hBW$ is the desired solution to \eqref{hjb5} in $W^{1,2}_{p, loc}$. Moreover, by embedding theorem, we have $\hBW$ and $\p_{y}\hBW $ are continuous in $\sets$; the free boundary, defined as the boundary of the region $\{\hBW=\hBV\}$, is Lipschitz in both $t$ and $y$; and $\p_{t}\hBW$ and $\p_{yy}\hBW $ are continuous in $\sets$ excluding the free boundary. For the proof of the Lipschitz continuity of the free boundary, we refer to \cite[Theorem 16 and Remark 2]{N07}. 
The uniqueness of the solution follows from the above comparison principle Theorem \ref{cp}. 
}

Define 
\begin{align*}
\BWtwo(t, x)=\inf_{y>0}(\hBW(t, y)+xy), \quad (t, x)\in\sets, 
\end{align*} which is clearly non-decreasing w.r.t. $x$.
Similar to the proof of $\varphi''(x)<0$ in \citethm{postsolution1}, we have $\partial_{xx} \zeta<0$. 
We next show that 
\begin{align} \label{hjb3}
%\greenone
{\begin{cases}
\min\left\{-\sup\limits_{k, \pi}\left\{(\p_{t}+\BL) \BWtwo-\rho(t) \BWtwo+\bar{u}(k)-L(t)\right\}, \BWtwo-V\right\}=0, \\
\BWtwo(T, x)=V(T, x), \quad \hfill (t, x)\in\sets.
\end{cases}}
\end{align}
where
{\[\BL=\frac{1}{2}\Vert\pi^\top\sigma(t)\Vert^{2}\p_{xx}+\big(r(t)x+\pi^\top\mu(t)+I(t)-k\big)\p_{x}, \] }
which can be reformulated as 
{\begin{align} \label{hjb4}
\begin{cases}
\min\left\{-\p_{t}\BWtwo+\tfrac{\p_{x}^{2}\BWtwo}{2\p_{xx}\BWtwo}\Vert\vartheta(t)\Vert^{2}-(r(t)x+I(t))\p_{x}\BWtwo-h(\p_{x}\BWtwo)+\rho(t) \BWtwo+L(t), \;\BWtwo-V\right\}=0, \\
\BWtwo(T, x)=V(T, x), \hfill (t, x)\in\sets.
\end{cases}
\end{align}}
Let $y=y(t, x)=\p_{x}\BWtwo (t, x)$ for $(t, x)\in\sets$. Then $\p_{x}y=\p_{xx}\BWtwo (t, x)$ and
\begin{align}\label{bww} 
\hBW(t, y)=\BWtwo(t, x)-xy.
\end{align} 
Differentiating the last equation w.r.t. $x$ gives 
\begin{align*} 
\p_{y}\hBW (t, y)\p_{xx}\BWtwo (t, x)=\p_{x}\BWtwo (t, x)-y-x\p_{xx}\BWtwo (t, x)=-x\p_{xx}\BWtwo (t, x), 
\end{align*}
that is, 
\begin{align}\label{bwy}
\p_{y}\hBW (t, y)=-x, 
\end{align}
Differentiating it w.r.t. $x$ again gives
\begin{gather*}
\p_{yy}\hBW (t, y)\p_{xx}\BWtwo (t, x)=-1.
\end{gather*}
Differentiating \eqref{bww} w.r.t. $t$ gives
\begin{gather*}
\p_{t}\hBW (t, y)+\p_{y}\hBW (t, y)\BWtwo_{tx}(t, x)=\p_{t}\BWtwo (t, x)-x\BWtwo_{tx}(t, x), \quad \p_{t}\hBW (t, y)=\p_{t}\BWtwo (t, x), 
\end{gather*}
in view of \eqref{bwy}. Simple calculation shows that 
%\greenone
{\begin{multline}
\Big(-\p_{t}\BWtwo+\tfrac{\p^{2}_{x}\BWtwo}{2\p_{xx}\BWtwo }\Vert\vartheta(t)\Vert^{2}-(r(t)x+I(t))\p_{x}\BWtwo-h(\p_{x}\BWtwo )+\rho(t) \BWtwo+L(t)\Big)\Big|_{(t, x)}\\
=\Big(-\p_{t}\hBW-\frac{1}{2}\Vert\vartheta(t)\Vert^{2}y^{2}\p_{yy}\hBW-(\rho(t)-r(t))y\p_{y}\hBW+\rho(t)\hBW-yI(t)-h(y)+L(t)\Big)\Big|_{(t, y)}, \label{lw} 
\end{multline}}
which is nonnegative by \eqref{hjb5}. Using \eqref{hjb5} again, $\hBW\geq \hBV$ on $\sets$, so $\BWtwo\geq \BV$ by the dual expressions. To prove \eqref{hjb4}, it is only left to show 
%\greenone
{\begin{align}\label{zero} 
\left(-\p_{t}\BWtwo+\tfrac{\p^{2}_{x}\BWtwo }{2\p_{xx}\BWtwo }\Vert\vartheta(t)\Vert^{2}-(r(t)x+I(t))\p_{x}\BWtwo-h(\p_{x}\BWtwo )+\rho(t) \BWtwo+L(t)\right)\big(\BWtwo-\BV\big)\Big|_{(t, x)}=0.
\end{align}}
There are two cases:
\begin{itemize}
\item If $ \hBW(t, y)>\hBV(t, y)$, then by \eqref{hjb5} and \eqref{lw}, we have 
%\greenone
{\begin{align*}
\left(-\p_{t}\BWtwo+\tfrac{\p^{2}_{x}\BWtwo }{2\p_{xx}\BWtwo }\Vert\vartheta(t)\Vert^{2}-(r(t)x+I(t))\p_{x}\BWtwo-h(\p_{x}\BWtwo )+\rho(t) \BWtwo+L(t)\right)\Big|_{(t, x)}=0.
\end{align*}}
So \eqref{zero} holds. 
\item If $\hBW(t, y)=\hBV(t, y)$, then since $\hBW\geq \hBV$ on $\sets$ by \eqref{hjb5}, we see that the function $\hBW-\hBV$ attaints its minimum value 0 at $(t, y)$. Hence the first order condition gives \[\p_{y}\hBV(t, y)=\p_{y}\hBW(t, y)=-x, \] by \eqref{bwy}. This means $y$ is a stationary point of the convex function $z\mapsto \hBV(t, z)+xz$, so
\begin{align*}
\hBV(t, y)+xy=\min_{z}(\hBV(t, z)+xz)=\BV(t, x), 
\end{align*}
by \eqref{def:bvdual}. Together with \eqref{bww}, we obtain \[\BWtwo(t, x)=\hBW(t, y)+xy=\hBV(t, y)+xy=\BV(t, x), \] and thus \eqref{zero} holds. 
\end{itemize}
\par
We now show that $\BWtwo=\BW$. It suffices to prove that $\BWtwo(0, x)=\BW(0, x)$ for any $x>0$.
%\blue
{By the smoothness of $\BW$, we have $\BWtwo$ and $\p_{y}\BWtwo $ are continuous in $\sets$; the free boundary, defined as the boundary of the region $\{\BWtwo=\BV\}$, is Lipschitz in both $t$ and $y$; and $\p_{t}\BWtwo$ and $\p_{yy}\BWtwo $ are continuous in $\sets$ excluding the free boundary. 
The above smoothness of $\BWtwo$ allows us to apply the It\^{o}-Krylov formula (see \cite{K80}) to it.}

For any stopping time $\tau\leq T$ and strategy $(k(\cdot), \pi(\cdot))$, let 
\[\theta_{n}=\inf\{t\geq0: X(t)\geq x+n\mbox{ or } |\pi(t)|\geq n\}\wedge \tau.\]%
%\greenone 
{By the It\^{o}-Krylov formula (see \cite{K80}) and \eqref{hjb3}, 
\begin{align}\label{ineq0}
\nn \dd\; (e^{-\int_0^t \rho(s)\ds}\BWtwo(t, X(t)))&=e^{-\int_0^t \rho(s)\ds}(\p_{t}\BWtwo+\BL \BWtwo-\rho(t)\BWtwo)\dt+e^{-\int_0^t \rho(s)\ds}\p_{x}\BWtwo\pi(t)^\top\sigma(t)\dd B(t)\\
&\leq-e^{-\int_0^t \rho(s)\ds}(\bar{u}(k(t))-L(t))\dt+e^{-\int_0^t \rho(s)\ds}\p_{x}\BWtwo\pi(t)^\top\sigma(t)\dd B(t).
\end{align}
Taking integral on both sides and rearranging terms, we have 
\begin{align} 
\BWtwo(0, x)& \geq e^{-\int_0^{\theta_{n}} \rho(s)\ds}\BWtwo(\theta_{n}, X(\theta_{n}))+\int _{0}^{\theta_{n}}e^{-\int_0^t \rho(s)\ds}(\bar{u}(k(t))-L(t))\dt\nn\\
&\quad\;-\int_{0}^{\theta_{n}}e^{-\int_0^t \rho(s)\ds}\p_{x}\BWtwo\pi(t)^\top\sigma(t)\dd B(t), \label{ineq1}
\end{align}
which, after taking expectation, leads to
\begin{align} \label{ineq2}
\BWtwo(0, x)& \geq \BE{e^{-\int_0^{\theta_{n}\rho(s)\ds}}\BWtwo(\theta_{n}, X(\theta_{n}))}+\BE{\int_{0}^{\theta_{n}}e^{-\int_0^t \rho(s)\ds}(\bar{u}(k(t))-L(t))\dt}.
\end{align}
Sending $n$ to $\infty$ in above, by Fatou's lemma and \eqref{hjb3}, we have
\begin{align*} 
\BWtwo(0, x)& \geq \BE{e^{-\int_0^{\tau}\rho(s)\ds}\BWtwo(\tau, X(\tau))}+\BE{\int_{0}^{\tau}e^{-\int_0^t \rho(s)\ds}(\bar{u}(k(t))-L(t))\dt}\\
& \geq \BE{e^{-\int_0^{\tau}\rho(s)\ds}V(\tau, X(\tau))}+\BE{\int _{0}^{\tau}e^{-\int_0^t \rho(s)\ds}(\bar{u}(k(t))-L(t))\dt}.
\end{align*}
By arbitrariness of $\tau$ and strategy $(k(\cdot), \pi(\cdot))$, we obtain $\BWtwo(0, x)\geq \BW(0, x)$.
\par
Use the feedback controls 
{
\[\pi^{*}(t)=-\tfrac{\p_{x}\BWtwo (t, X(t))}{\p_{xx}\BWtwo (t, X(t))}\vartheta, 
\quad k^{*}(t)=\argmax_{k\geq 0}(\tilde u(k)-k\p_{x}\BWtwo (t, X(t))), \]
} 
in \eqref{wealth3}. Let 
\[\tau^{*}=\inf\{t\geq 0: \BWtwo(t, X(t))=V(t, X(t))\}, \]
then $\tau^{*}\leq T$ by \eqref{hjb3}. Furthermore the inequalities in \eqref{ineq0}, \eqref{ineq1}, and \eqref{ineq2} become identities under the above controls. Let $n$ go to $\infty$ in \eqref{ineq2}, by the dominated convergence theorem, 
\begin{align*} 
\BWtwo(0, x) &=\BE{e^{-\int_0^{\tau^*}\rho(s)\ds}\BWtwo(\tau^{*}, X(\tau^{*}))}+\BE{\int _{0}^{\tau^{*}}e^{-\int_0^{t}\rho(s)\ds}(\bar{u}(k^{*}(t))-L(t))\dt}\\
&=\BE{e^{-\int_0^{\tau^*}\rho(s)\ds}V(\tau^{*}, X(\tau^{*}))}+\BE{\int_{0}^{\tau^{*}}e^{-\int_0^{t}\rho(s)\ds}(\bar{u}(k^{*}(t))-L(t))\dt}\leq W(0, x). 
\end{align*}
Therefore, we conclude that $\BWtwo$ is equal to the value function $\BW$. Moreover, $(\tau^{*}, \pi^{*}(\cdot), k^{*}(\cdot))$ defined above is an optimal control. 
}

%\greenone
{\section{Optimal Retirement Region}%\label{Sec:Retire}

{\textbf{Proof of \citeprop{Universal_stopping_region}: }}
By \eqref{hjb6}, $\BBW$ is independent of $h(\cdot)$, so the stopping region is irrelevant to the individual's utility function. 
We next show that the function $\BBW$ is increasing w.r.t. $y$ for each fixed $t\in[0, T)$.
For any fixed $\kappa>1$, and define $\BBW^{\kappa}(t, y)=\BBW(t, \kappa y)$ for $(t, y)\in\sets$. Then 
\begin{align*} 
\begin{cases}
\min\left\{-(\p_{t}+\BBL)\BBW^{\kappa}-e^{-\int_0^{t}\rho(s)\ds}(\kappa y I(t)-L(t)), \;\BBW^{\kappa}\right\}=0, \quad (t, y)\in \sets;\\
\BBW^{\kappa}(T, y)=0.
\end{cases}
\end{align*} 
Recalling $I(t)\geq 0$ and applying \citethm{cp}, we obtain $\BBW^{\kappa}(t, y)\geq \BBW(t, y)$. Because $\kappa>1$ is arbitrarily chosen, it follows that $\BBW$ is increasing w.r.t. $y$, {which ensures the existence and uniqueness of the free boundary $b(t)$. Furthermore, }
\citethm{cp} implies that a bigger $I(\cdot)$ or a smaller $L(\cdot)$ leads to a smaller $b(\cdot)$.

\textbf{Proof of \citeprop{upperboundfreeboundary}: }
In $\{(t, y)\in\sets\mid y< b(t)\}\subset\setr$, we have $\BBW=0$, so the variational inequality \eqref{hjb6} yields 
\[-(\p_{t}+\BBL)\BBW-e^{-\int_0^{t}\rho(s)\ds}(yI(t)-L(t))=e^{-\int_0^{t}\rho(s)\ds}(L(t)-yI(t))\geq 0, \]
namely, $L(t)\geq yI(t)$. Hence the claim follows.

\textbf{Proof of \citethm{monotone}: }
For any small $\varepsilon>0$, set $\BBW^{\varepsilon}(t, y)=\BBW(t-\varepsilon, y)$ for $(t, y)\in [(T-\ell)\vee\varepsilon, T]\times(0, \infty)$. Then 
\begin{align*} 
\begin{cases}
\min\left\{-(\p_{t}+\BBL)\BBW^{\varepsilon}-e^{-\int_0^{t-\varepsilon}\rho(s)\ds}(y I(t-\varepsilon)-L(t-\varepsilon)), \;\BBW^{\varepsilon}\right\}=0, \qquad \\ 
\hfill (t, y)\in [(T-\ell)\vee\varepsilon, T]\times(0, \infty);\\
\BBW^{\varepsilon}(T, y)\geq 0.
\end{cases}
\end{align*}
\citeassmp{h1} implies 
\[e^{-\int_0^{t-\varepsilon}\rho(s)\ds}(y I(t-\varepsilon)-L(t-\varepsilon))\geq e^{-\int_0^{t}\rho(s)\ds}(y I(t)-L(t)), \quad (t, y)\in [(T-\ell)\vee\varepsilon, T]\times(0, \infty).\]
By \citethm{cp}, we see that $\BBW^{\varepsilon}(t, y)\geq \BBW(t, y)$, which implies $\BBW$ is decreasing w.r.t. $t$ and thus $b(t)$ is increasing.
\par
Now suppose $b(T-)<\tfrac{L(T)}{I(T)}$ (see \citefig{fig:0}).
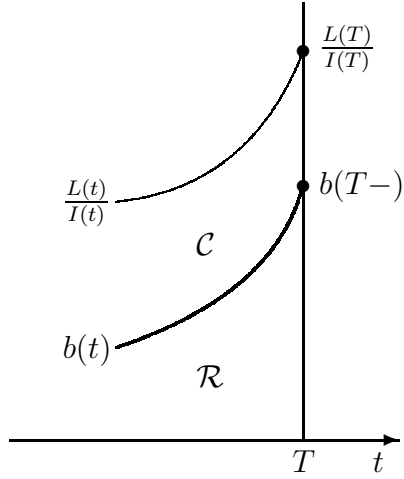
\begin{figure}[H]
\begin{center}
\begin{picture}(785, 150)
{\line(1, 0){0.5}}
{\line(1, 0){0.5}}
\thicklines\put(140, -30){\vector(1, 0){146}}
%\blue
{\qbezier(180, 5)(240, 25)(250, 67)}\thinlines
\put(256, 115){$\frac{L(T)}{I(T)}$}
\put(247, 114){$\bullet$}

\put(160, 1){$b(t)$}
\put(160, 57){$\frac{L(t)}{I(t)}$}

\put(256, 63){$b(T-)$}
\put(247, 63){$\bullet$}

\qbezier[200](180, 60)(230, 65)(250, 117)
\put(276, -42){$t$}
\qbezier(250, -30)(250, 45)(250, 135)
\put(246, -42){$T$} 
\put(210, -10){$\setr $}
\put(210, 40){$\setc $}
\end{picture}\vspace{45pt}
\caption{When $b(T-)<\tfrac{L(T)}{I(T)}$.}
\label{fig:0}\vspace{-20pt}
\end{center} 
\end{figure}

For $(t, y)\in \left\{\big[T, T-\varepsilon\big)\times \left(b(T-), \tfrac{L(T)}{I(T)}\right)\right\}\cap \setc$, we have
\[-(\p_{t}+\BBL)\BBW-e^{-\int_0^{t}\rho(s)\ds}(yI(t)-L(t))=0, \quad \BBW(t, y)>0, \quad \BBW(T, y)=0, \]
hence, 
\[-\p_{t}\BBW\big|_{(T, y)}=e^{-\int_0^{T}\rho(s)\ds}(yI(T)-L(T))<0, \]
contradicting that $\BBW$ is decreasing w.r.t. $t$.
}

%\blue

\section{Value Function Estimation in Section \ref{Sec:Simulation}} \label{EC:VFE}

This section explains the numerical scheme used to estimate the value functions on the pre-retirement and post-retirement regions with finding the optimal retirement curve $b(t)$.
In our model, solving the dual variational inequality \eqref{hjb6} is essential, which is independent of dual utility function, so solving one optimal stopping problem \eqref{p3dd} is equivalent to solving infinitely many optimal stopping problems \eqref{p3d} with a simple connection \eqref{WnW}.

Once the value function $\hBV$ on the post-retirement case is implemented using the Monte Carlo (MC) method, we estimate the solution of the variational inequality 
$\BBW$ using the finite difference method (FDM), where
$$\BBW(t, y) = \sup_{t\leq \tau \leq T}\mathbb{E}\left[\int_t^{\tau} e^{-\int_t^u\rho(s)\, ds } (Y(u) I(u) - L(u)) \,du \;\Big|\;Y(t)=y\right],$$
with the final condition $\BBW(T, y) = 0$.
Without losing generality, $\mu, \rho, \sigma$, and $r$ are assumed to be constant,
which leading $\vartheta$ is also constant.

\begin{figure}[H]
\begin{center}
\includegraphics[width=0.48\textwidth]{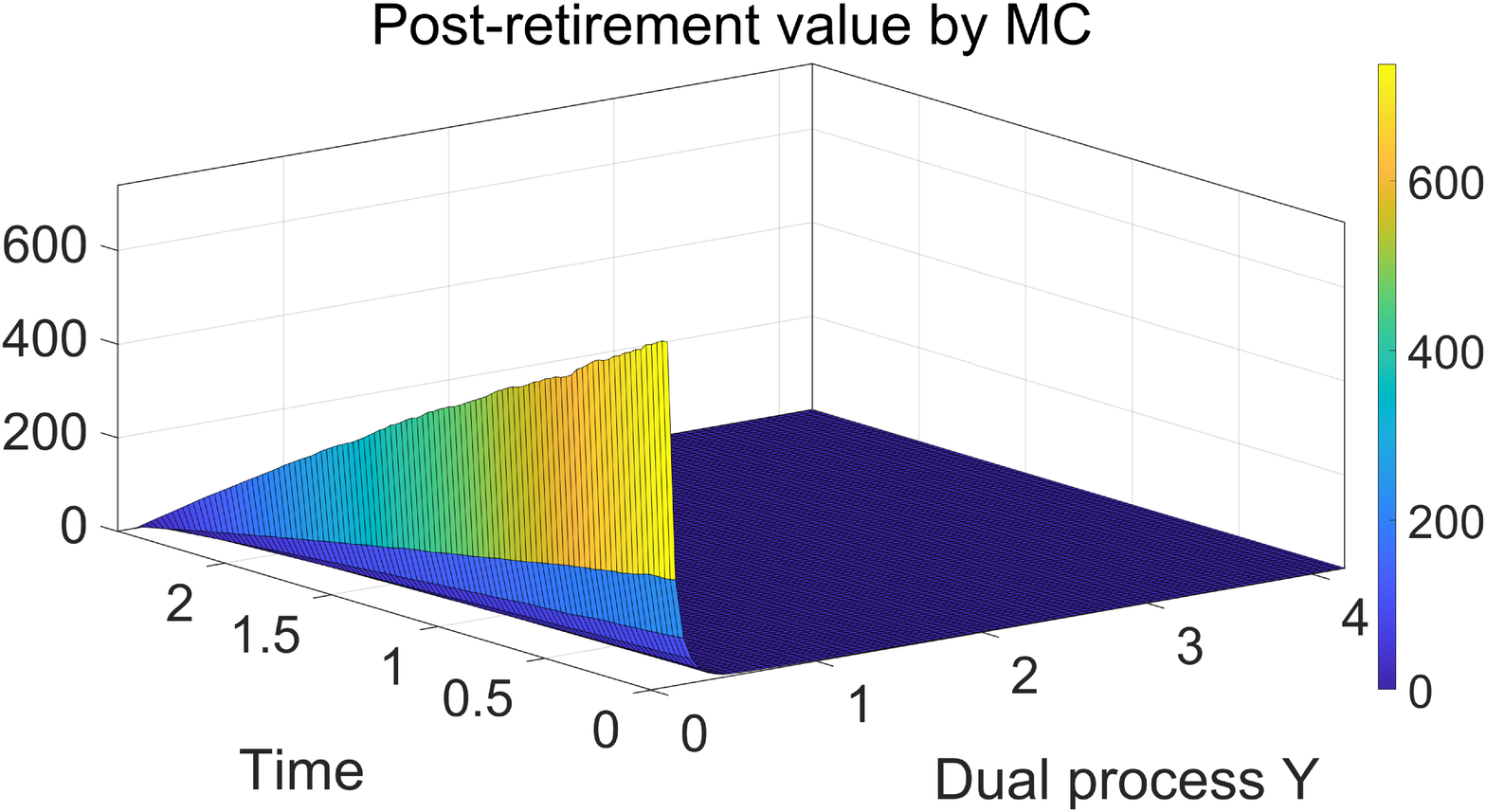}
\includegraphics[width=0.48\textwidth]{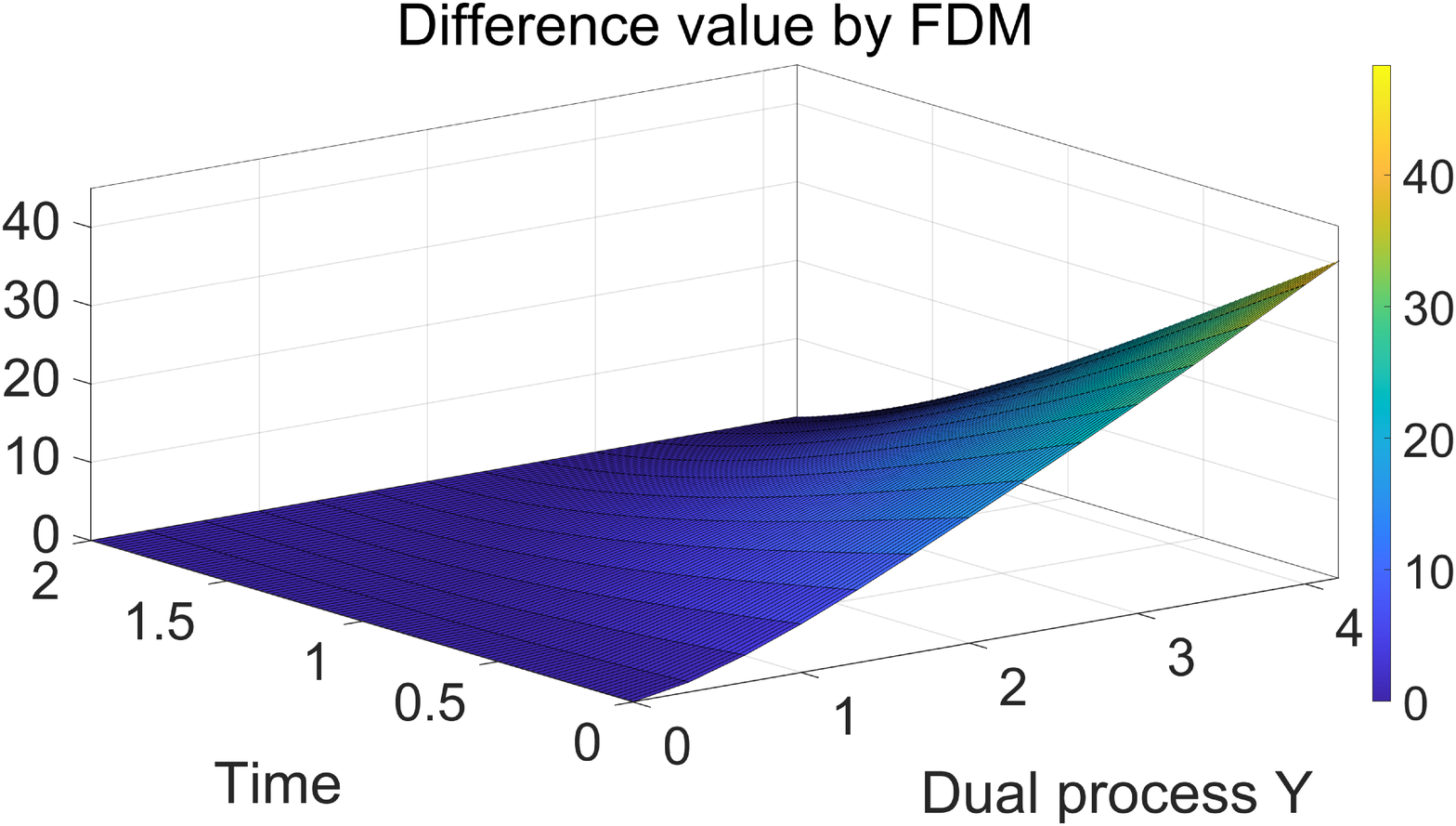}
\includegraphics[width=0.48\textwidth]{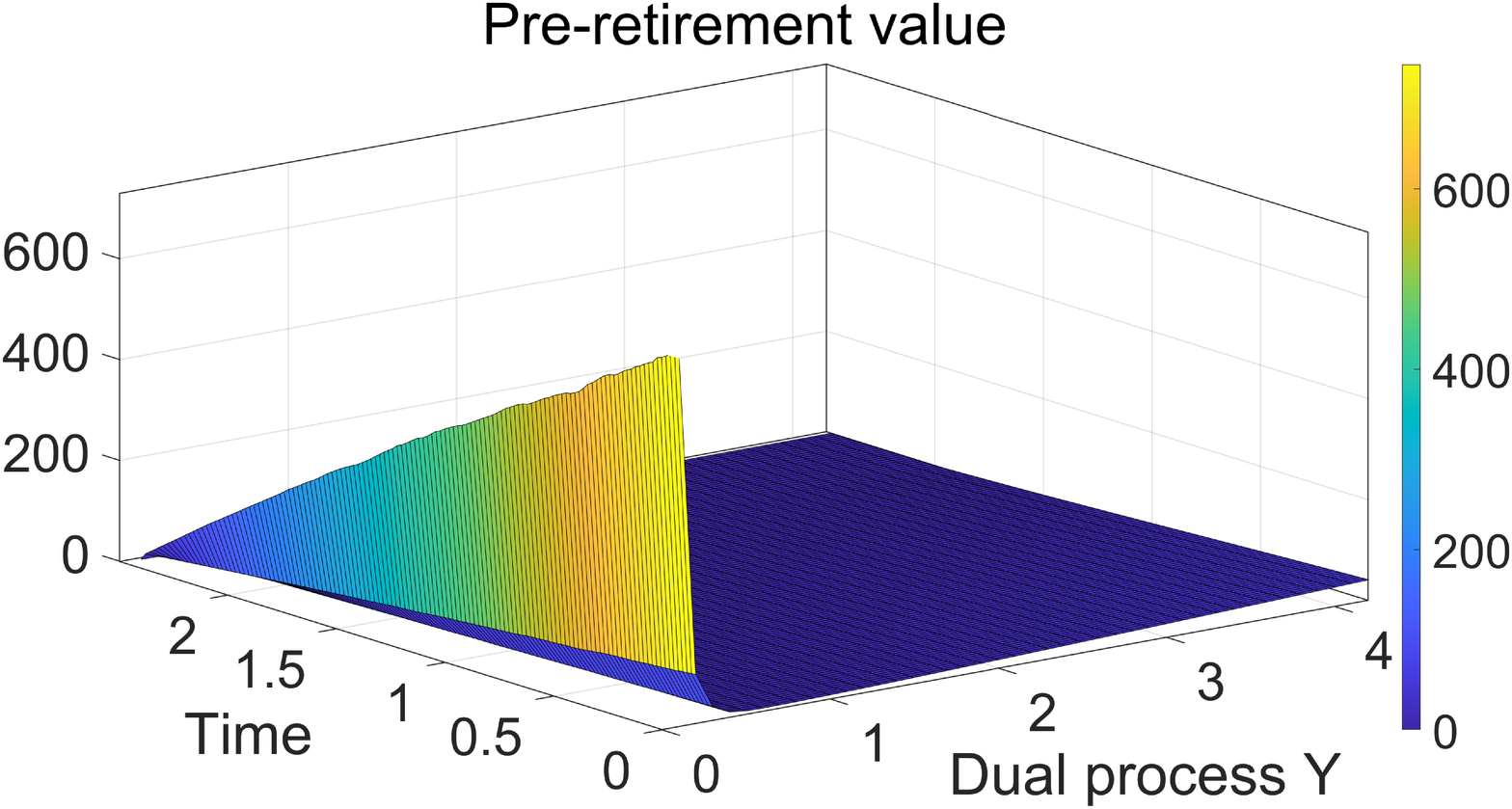}
\end{center}
\caption{The surface of the dual value function $\hBV(t, y)$ on the post retirement region by the MC (top, left) and that of $\BBW(t, y)$ by the FDM (top, right), the difference between the post- and pre-retirement value functions.
The surface of the pre-retirement value function $\hBW(t, y)$ (bottom).}
\label{Fig:FDMSolutions}
\end{figure}

Since the dual process $Y(u)$ is given as an explicit form 
$$Y(u) = Y(t)\exp\left\{ \big(\rho - r - \frac{1}{2}\Vert\vartheta\Vert^2\big) (u-t) - \vartheta(B(u) - B(t))\right\} $$
the lower boundary condition needs to be $\BBW(t, 0) = 0$ because $y=0$ always has a negative value for the optimal stopping problem of $\BBW$, which implying $\tau = 0$.
For a sufficiently large number $y_{\max}$
the upper boundary needs to take $\tau = T$ as the objective of the optimal problem is highly likely to be positive. We can compute the upper boundary condition $\BBW(t, y_{\max})$ by substituting $Y(u)$ and $\tau = T$ in expectation; thus, we finally obtain 
$$\BBW(t, y_{\max}) = y_{\max} \int_t^T e^{-r(u-t)}I(u) du - \int_t^T e^{-\rho(u-t) }L(u) du. $$
With the boundary and final conditions, 
we use the explicit FDM by defining a forward-moving time variable $\bar{t} = T-t$.
The estimates of $\hBV$ and $\BBW$ enable to compute the value function $\hBW$ on the pre-retirement case in (\ref{WnW}), in a reverse manner.
\begin{equation}\label{Eq:Pre}
\hBW(t, y) = \hBV(t, y) + e^{\rho t} \BBW(t, y).
\end{equation}
The numerical computation in Section~\ref{Sec:Simulation} employs 
the number of MC samples of 100, $y_{\max}=10$, the number of FDM grids of a time variable of 6000 and that of a state variable of 300, which satisfies the stability condition for FDM. 
The parameter set (\ref{Eq:baseline_gen}) produces the surfaces of Figure~\ref{Fig:FDMSolutions} depicting the estimates of the surface of $\hBV(t, y)$ by MC and $\BBW(t, y)$ by FDM, and the pre-retirement value function $\hBW(t, y)$ obtained by (\ref{Eq:Pre}).

The red vertical line indicates the optimal retirement time for this sample path. 

\section{Figures in Section \ref{Sec:Simulation}} \label{EC:Figures}

We implement the sample paths of $X^*(t)$, $\pi^*(t)$, and $k^*(t)$ with $c^*(t)$ and $g^*(t)$, which all result from the generated sample path of $Y(t)$ and the optimal retirement time shown in Figure~\ref{Fig:bt}.
\begin{figure}[H]
\begin{center}
\includegraphics[width=0.48\textwidth]{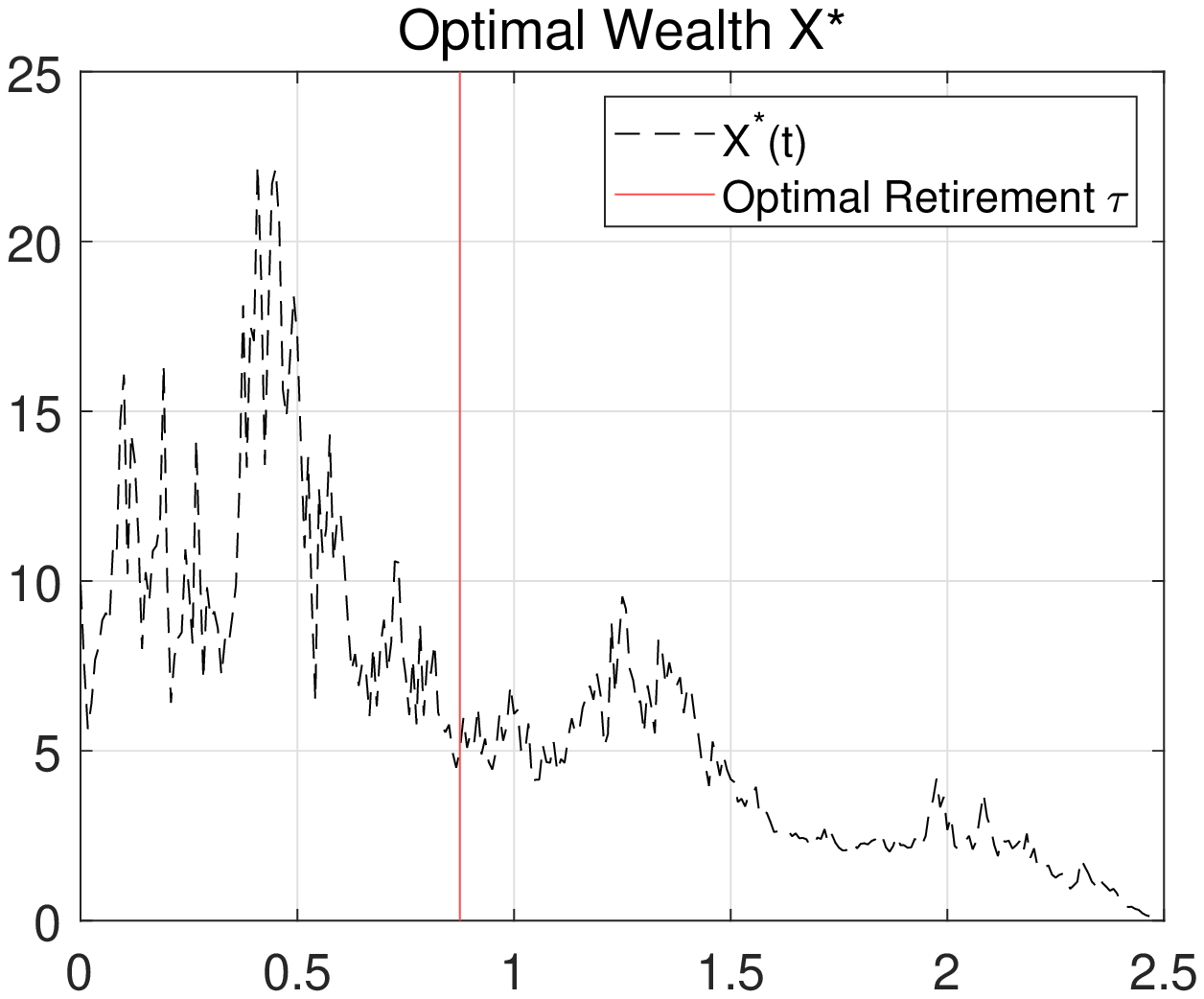}
\includegraphics[width=0.48\textwidth]{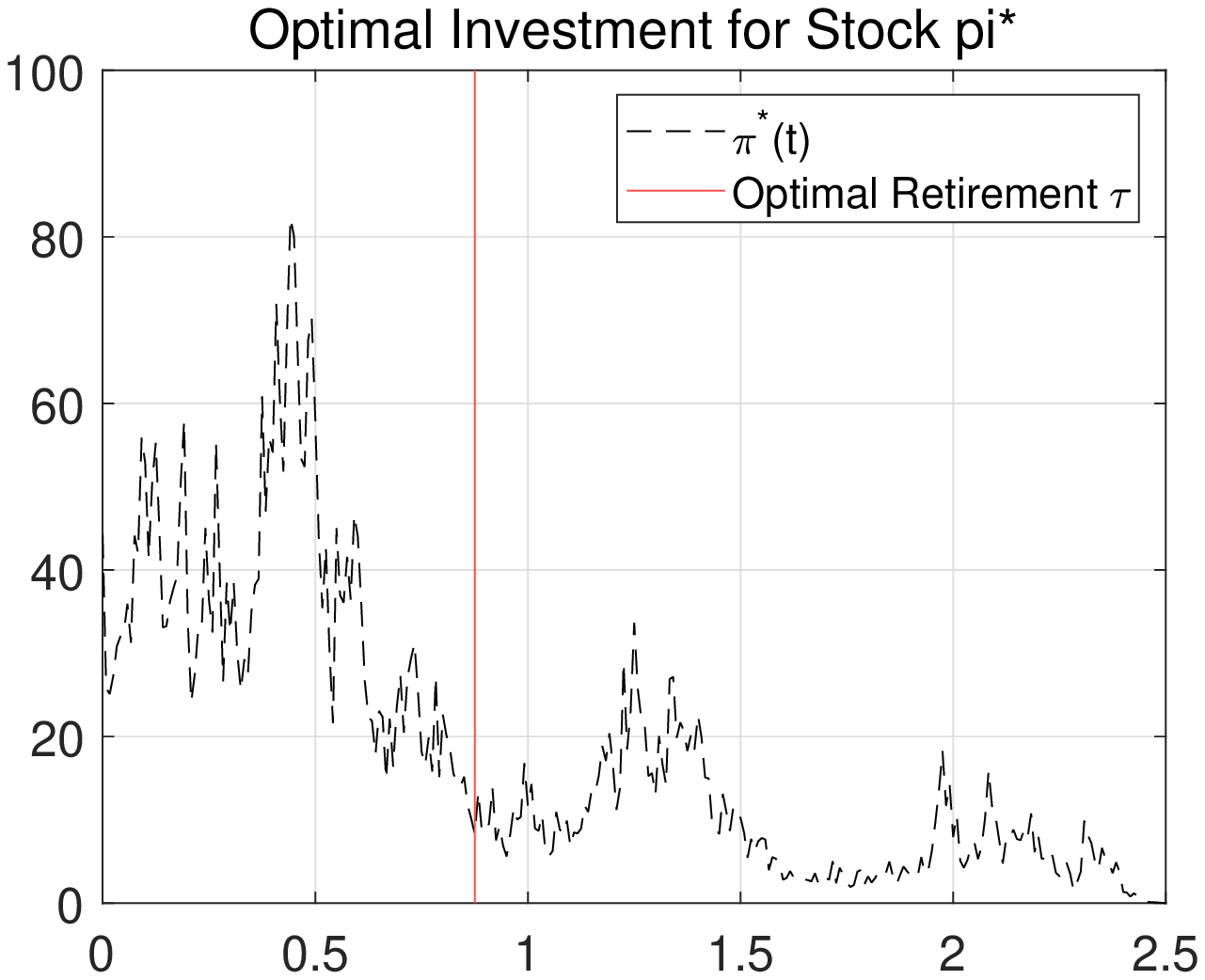}
\includegraphics[width=0.48\textwidth]{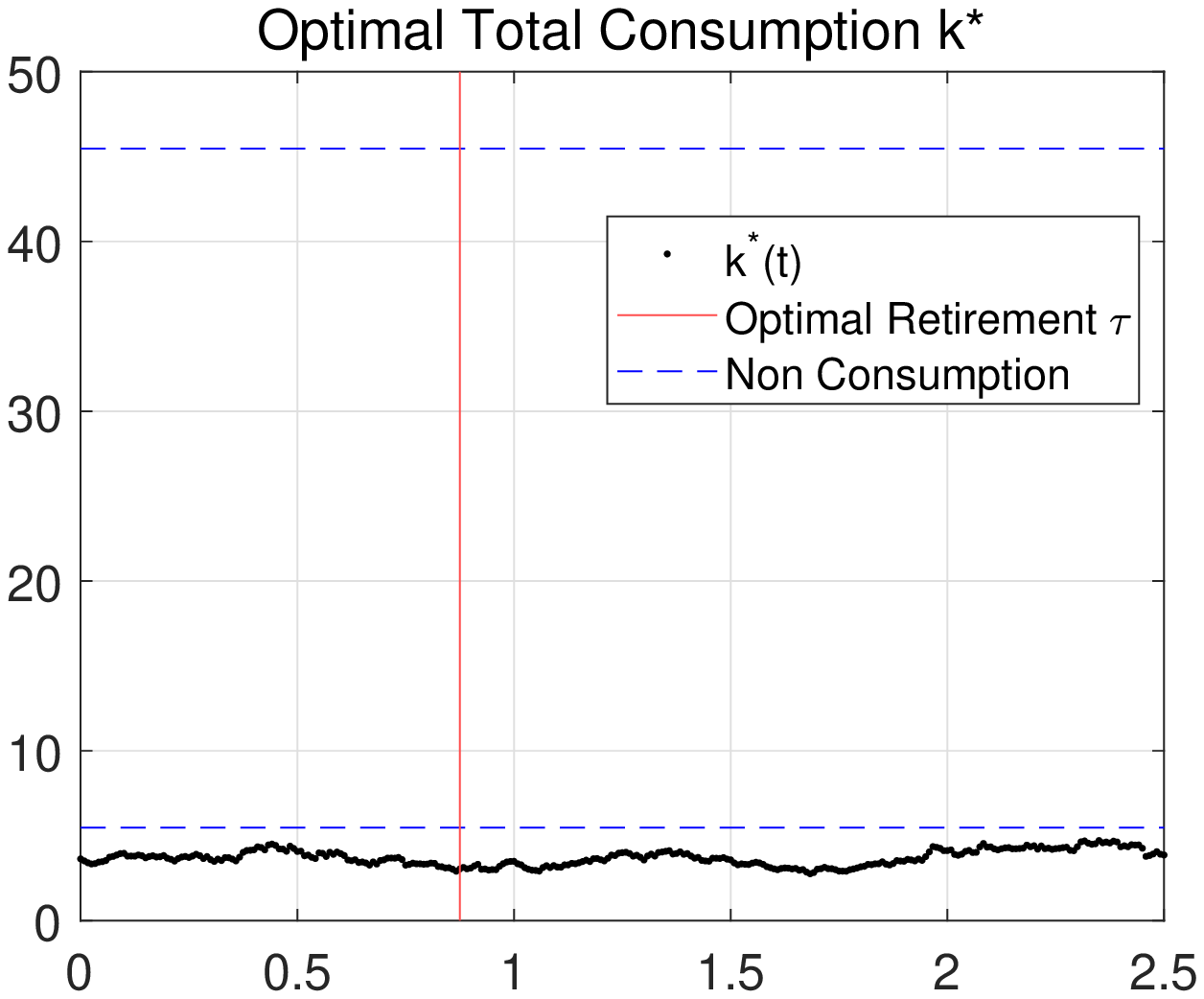}
\includegraphics[width=0.48\textwidth]{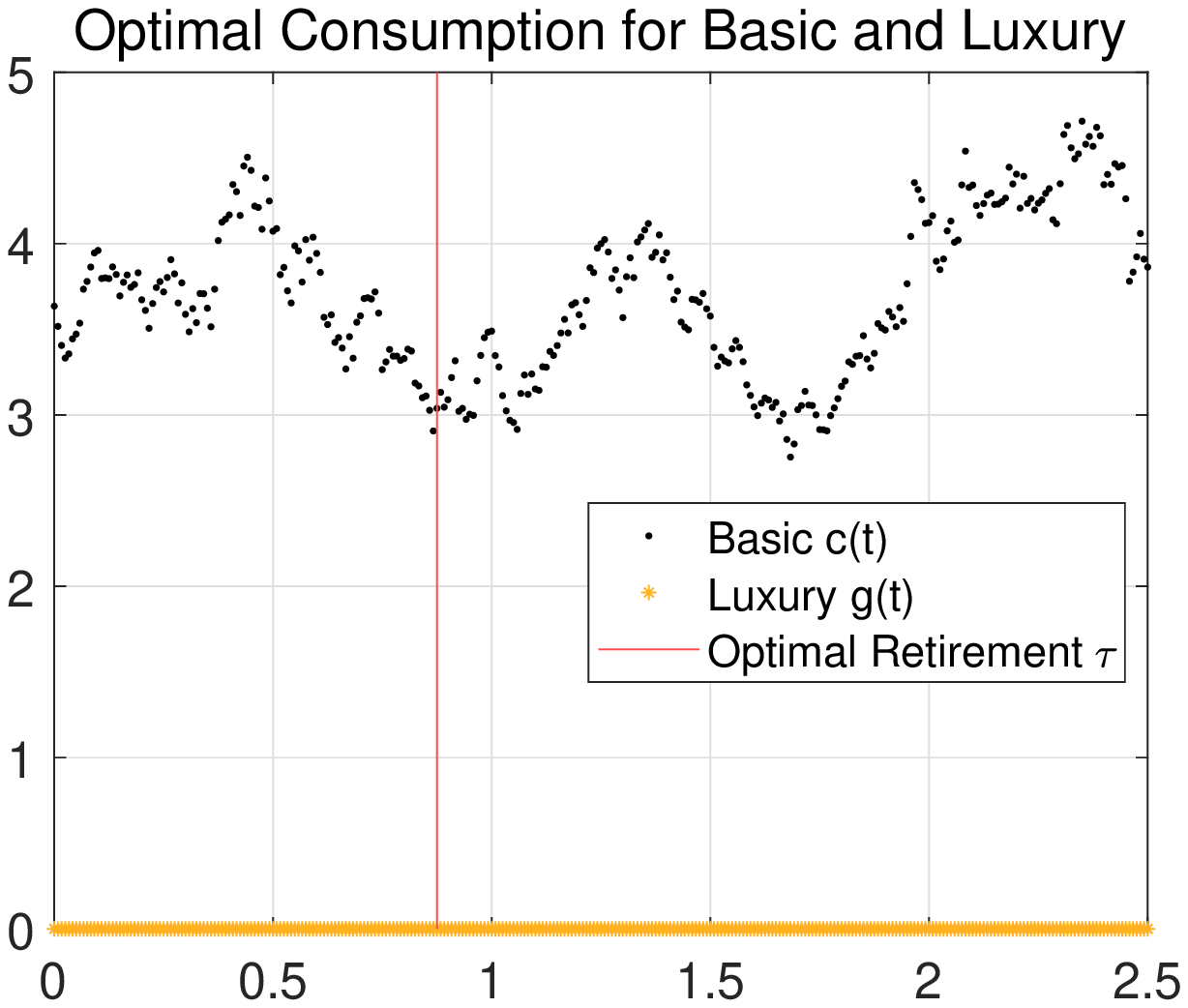}
\end{center}
\caption{A sample path of the optimal wealth $X^*(t)$, the optimal investment strategy for stock $\pi^*(t)$, the optimal total consumption $k^*(t)$ and the optimal basic $c(t)$ and luxury $g(t)$ goods corresponding to the path of $Y(t)$ generated in Figure~\ref{Fig:bt}.
}
\label{Fig:Sample_Path}
\end{figure}

\begin{figure}[H]
\begin{center}
\includegraphics[width=0.48\textwidth]{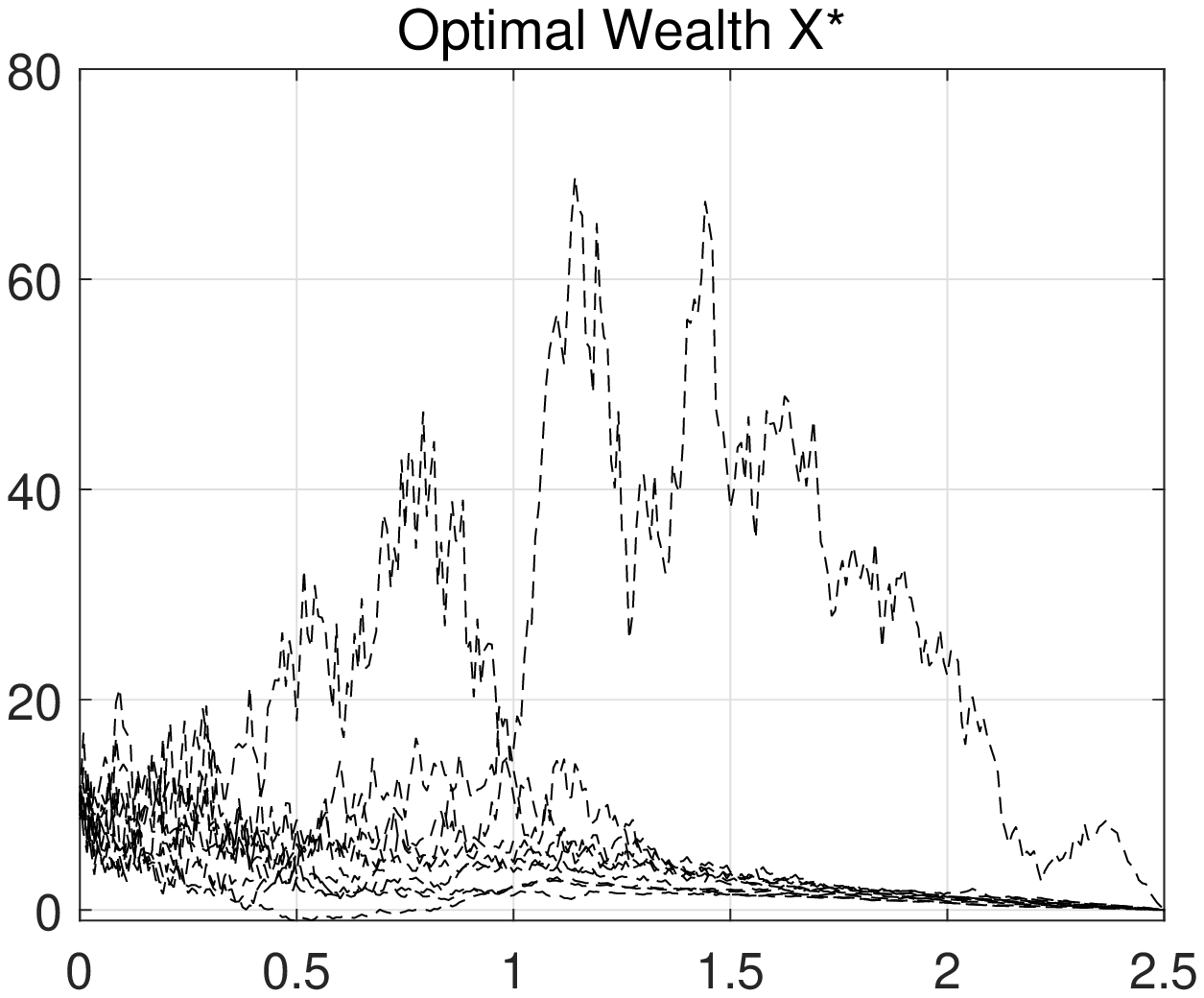}
\includegraphics[width=0.48\textwidth]{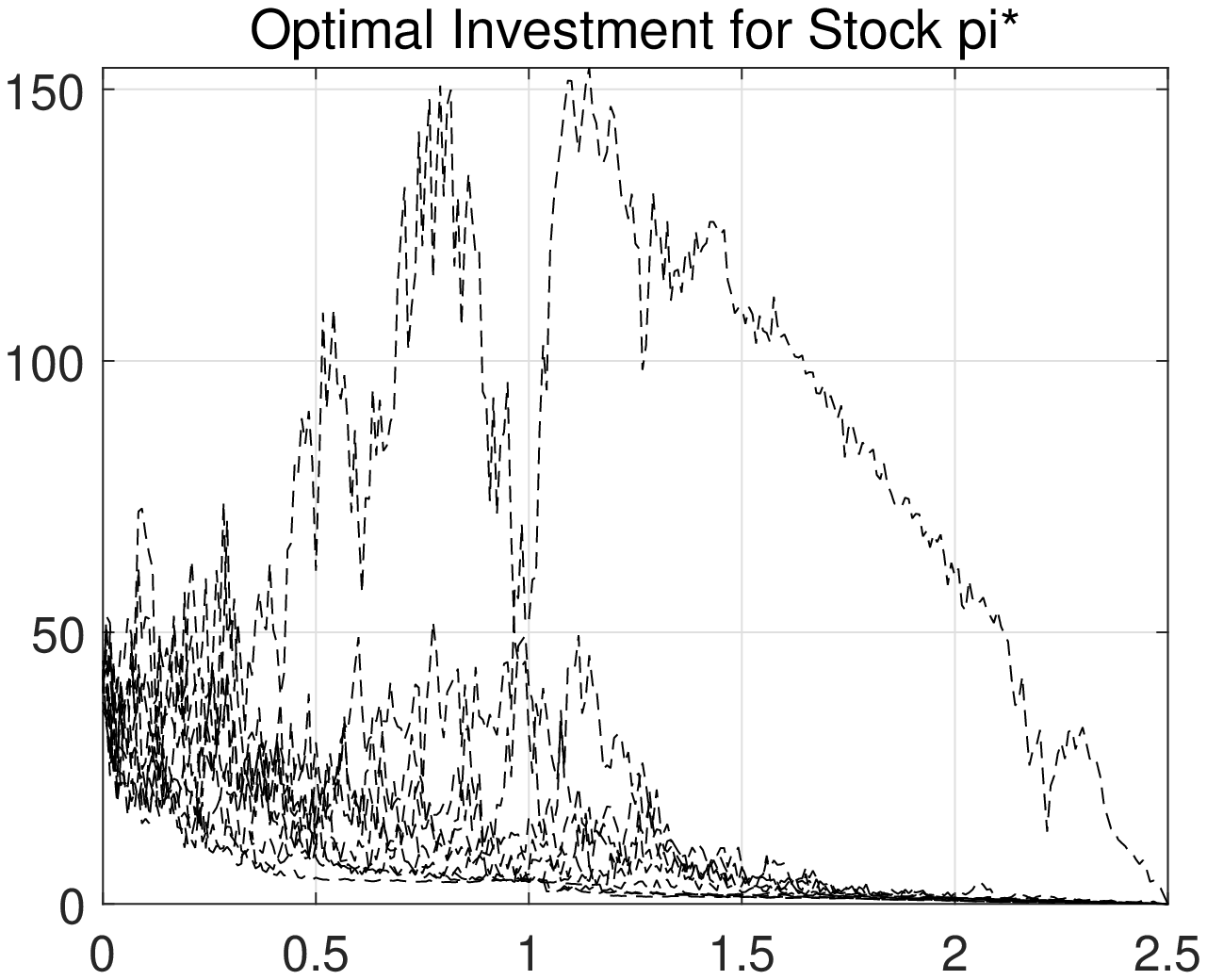}
\includegraphics[width=0.48\textwidth]{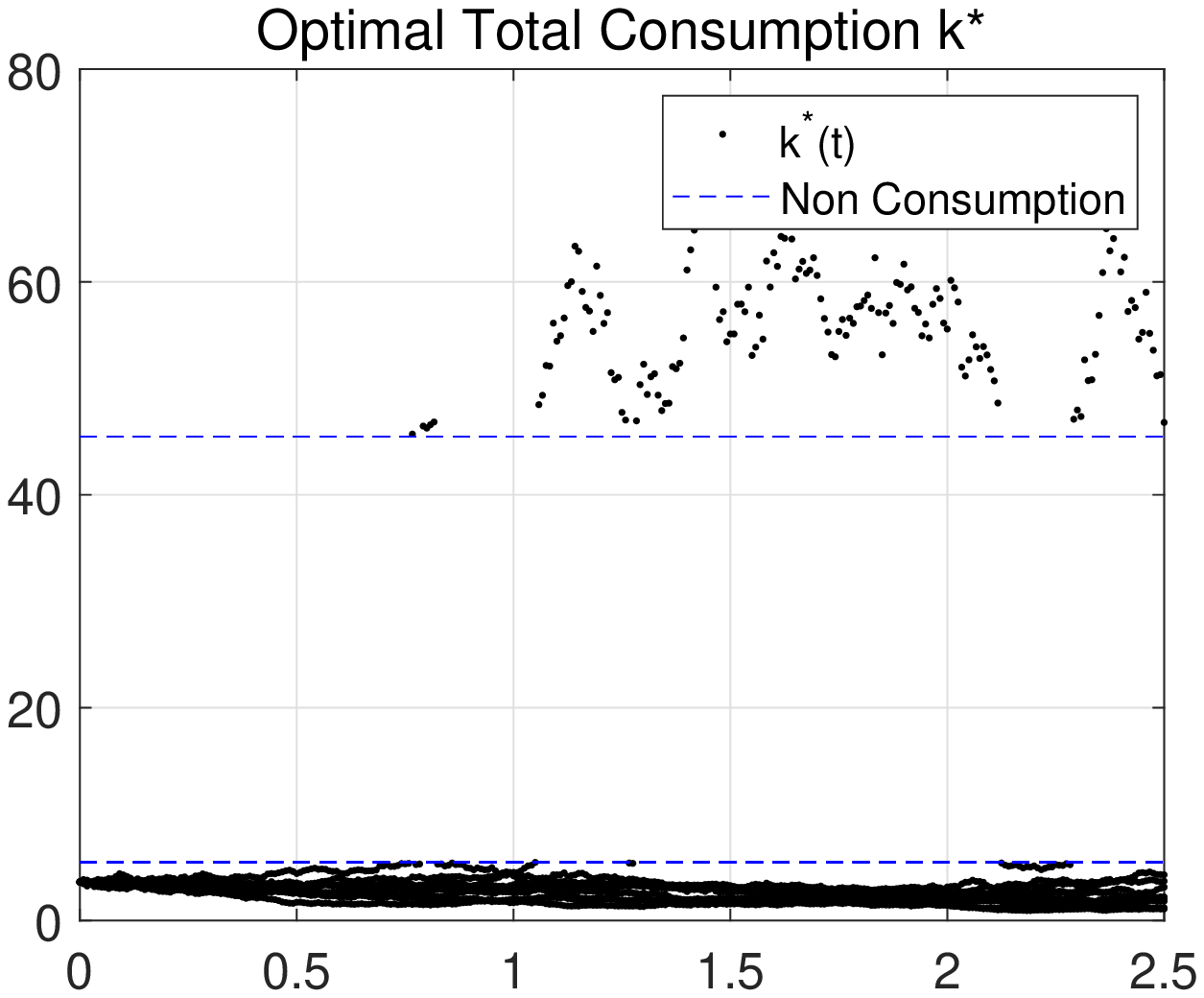}
\includegraphics[width=0.48\textwidth]{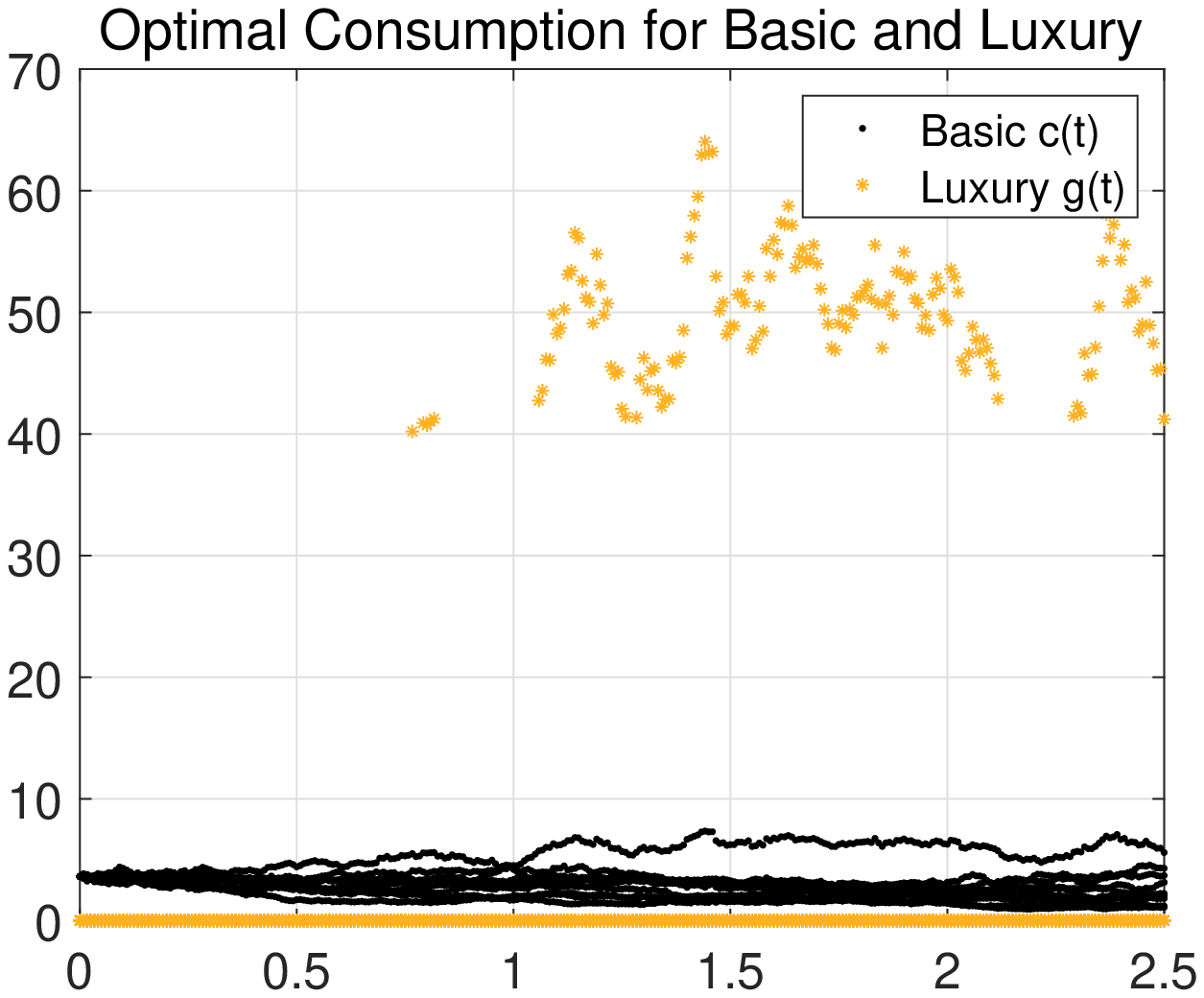}
\end{center}
\caption{Sample paths including the case of the luxury being consumed, that is, $g(t)>0$.
}
\label{Fig:Sample_Path2}
\end{figure}

Figure~\ref{Fig:Sample_Path2} illustrates ten sample paths including the case of the luxury consumed. In this scenario, the wealth reaches the level exceeding a threshold of the non-consumption band, which drives to trigger consuming luxury goods adding to consuming basic ones. 
This case contains one sample path of $g(t)>0$, which 
apparently occurs at the sample $X^*(t)$ that exceeds $k_+\approx 45.5$.
Most of the samples of $X^*(t)$ do not reach the level of $k_{+}$ since the luxury trigger $a$ is set to be relatively high compared to the initial wealth $x$.

\begin{figure}[H]
\begin{center}
\includegraphics[width=0.48\textwidth]{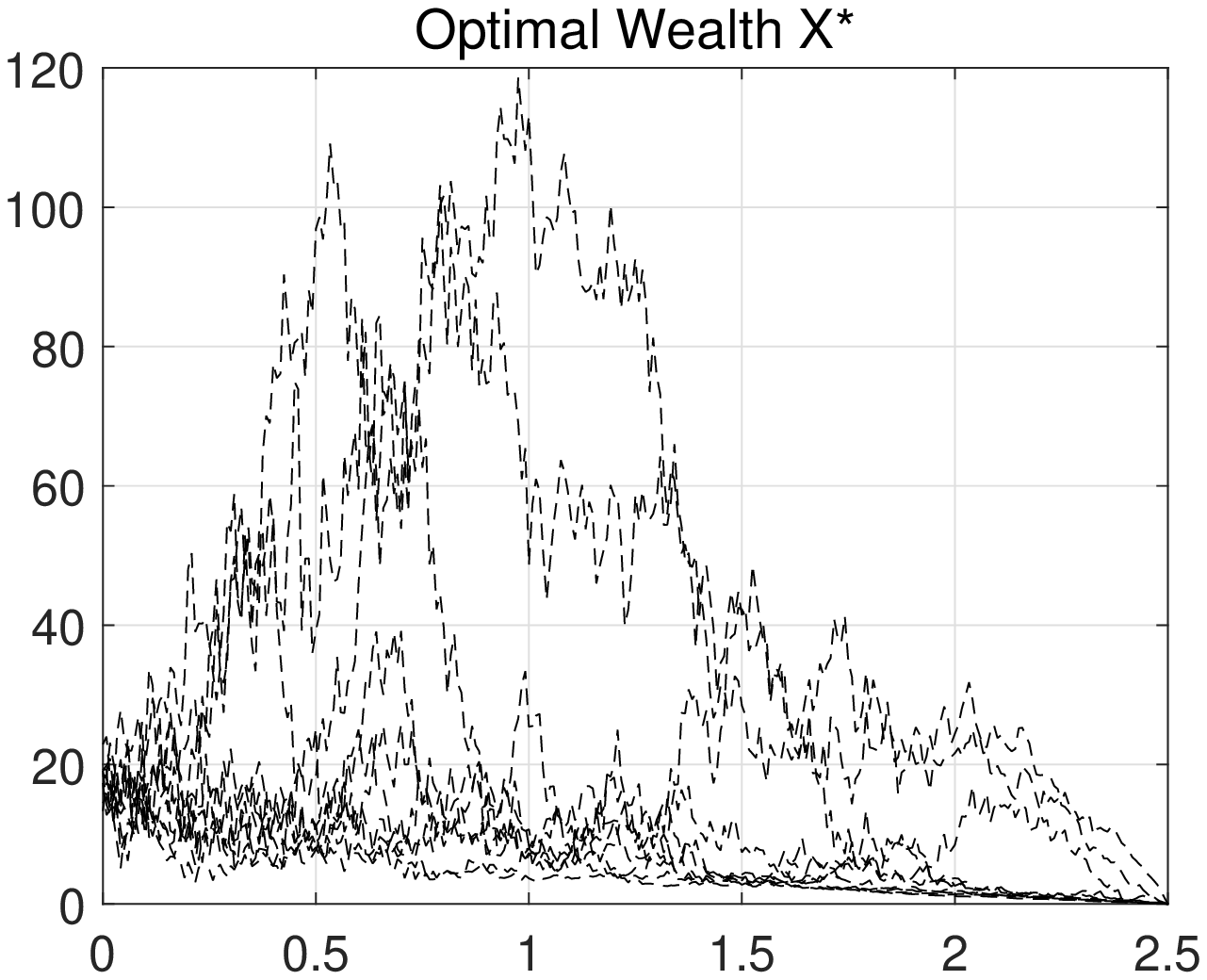}
\includegraphics[width=0.48\textwidth]{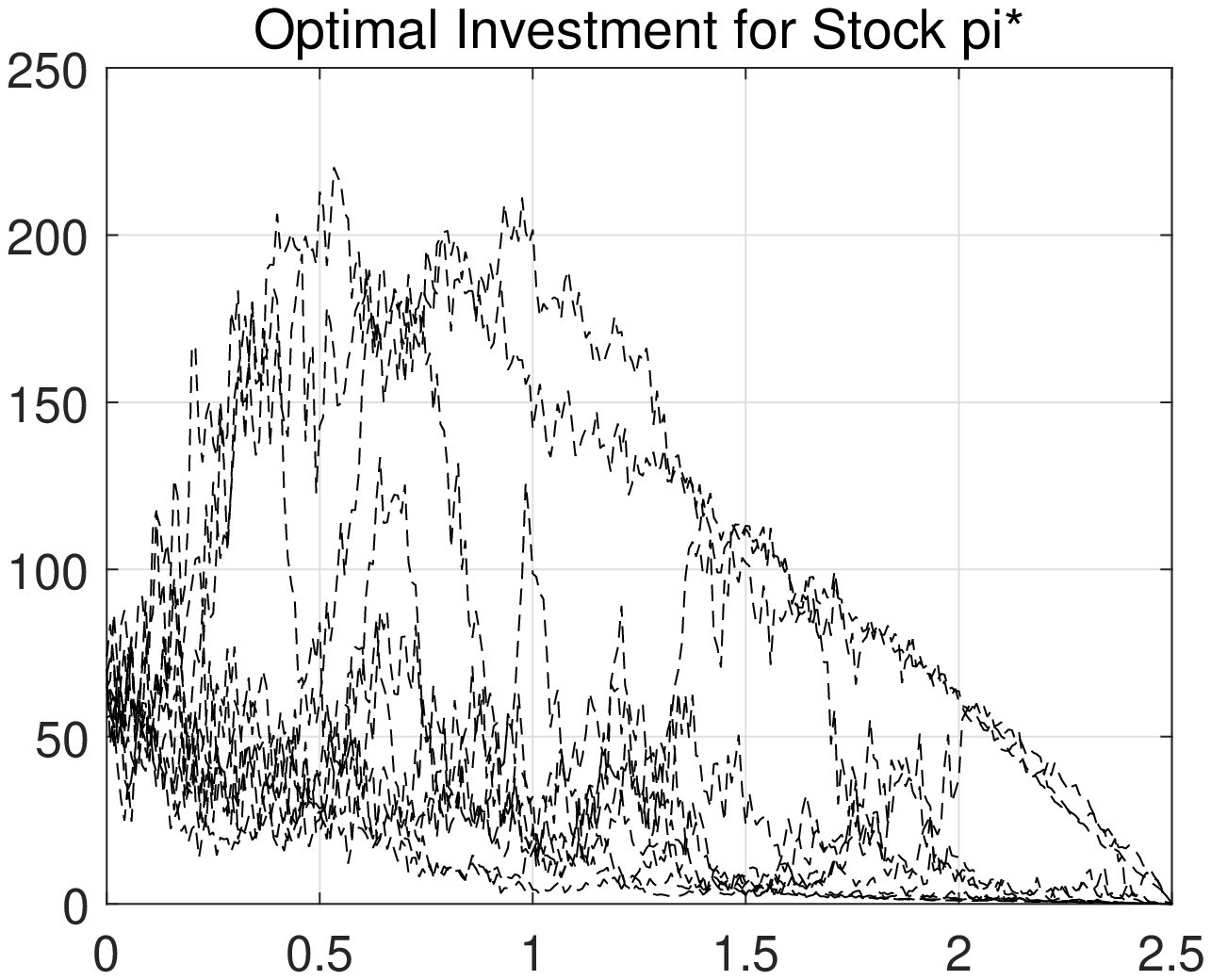}
\includegraphics[width=0.48\textwidth]{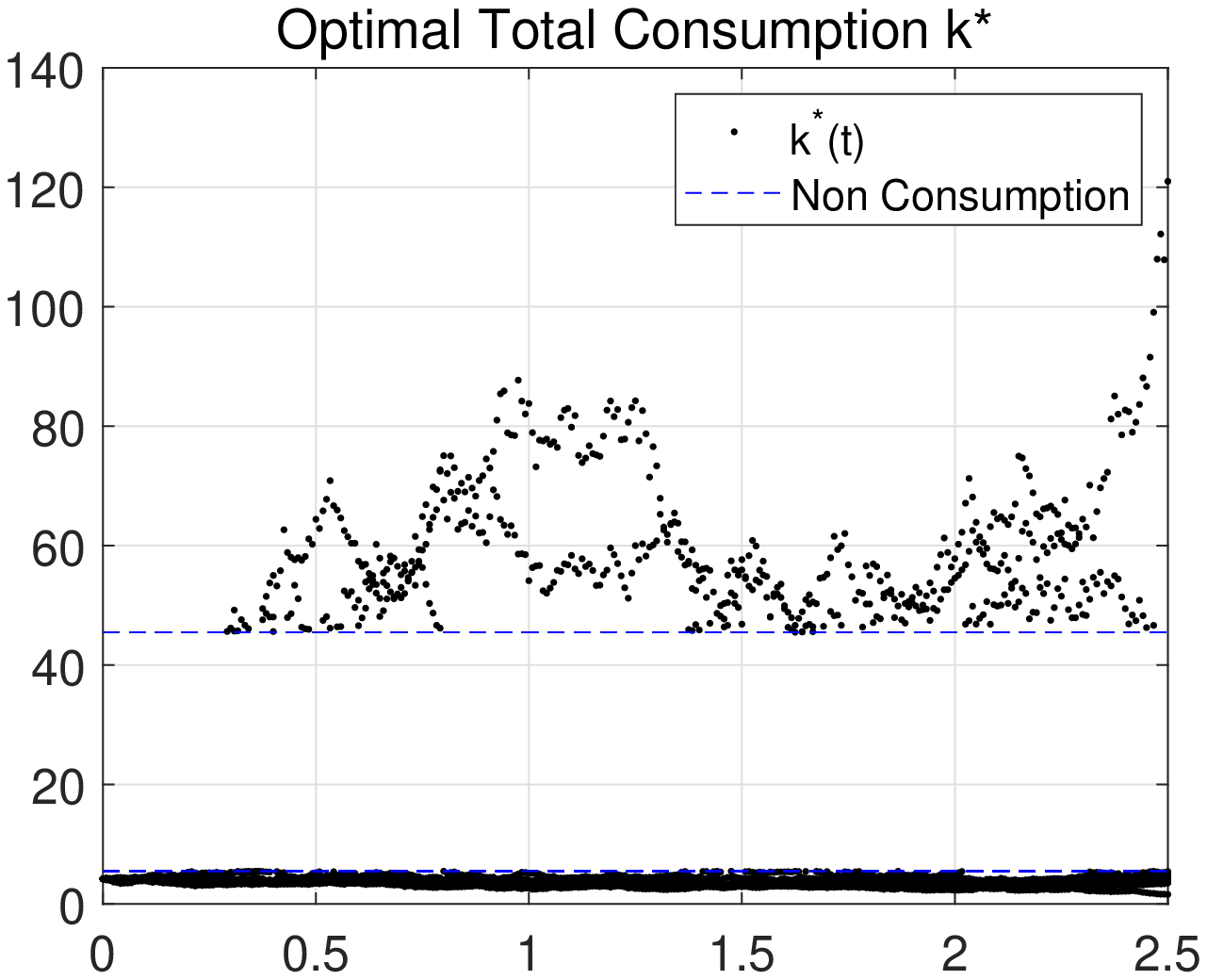}
\includegraphics[width=0.48\textwidth]{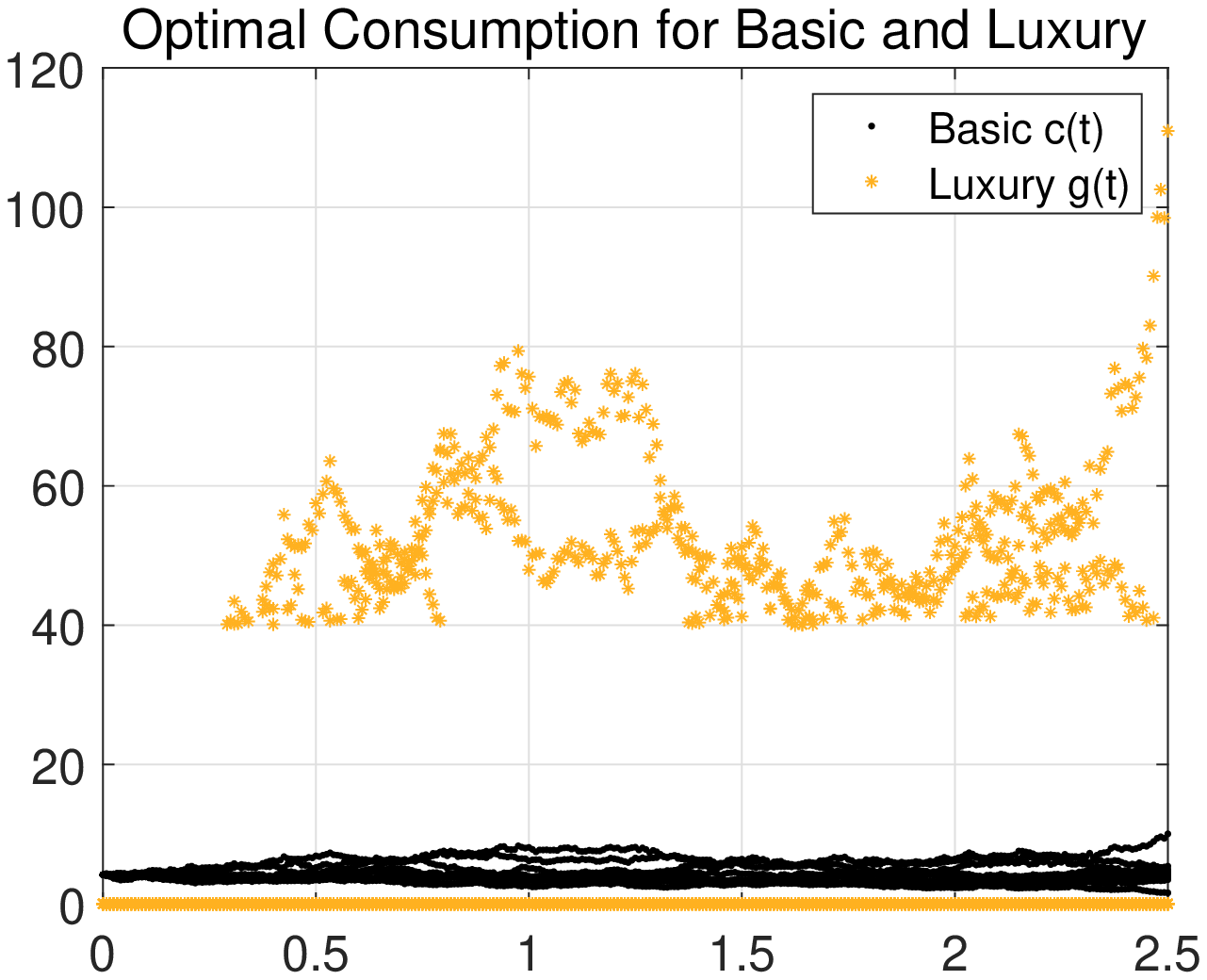}
\end{center}
\caption{Sample paths of the optimal strategies of $X^*(t)$, $\pi^*(t)$, $k^*(t)$ and $c(t)$ and $g(t)$ in the case of the initial wealth $x=20$.}
\label{Fig:Sample_Paths_x20} 
\end{figure}

\begin{figure}[H]
\begin{center}
\includegraphics[width=0.48\textwidth]{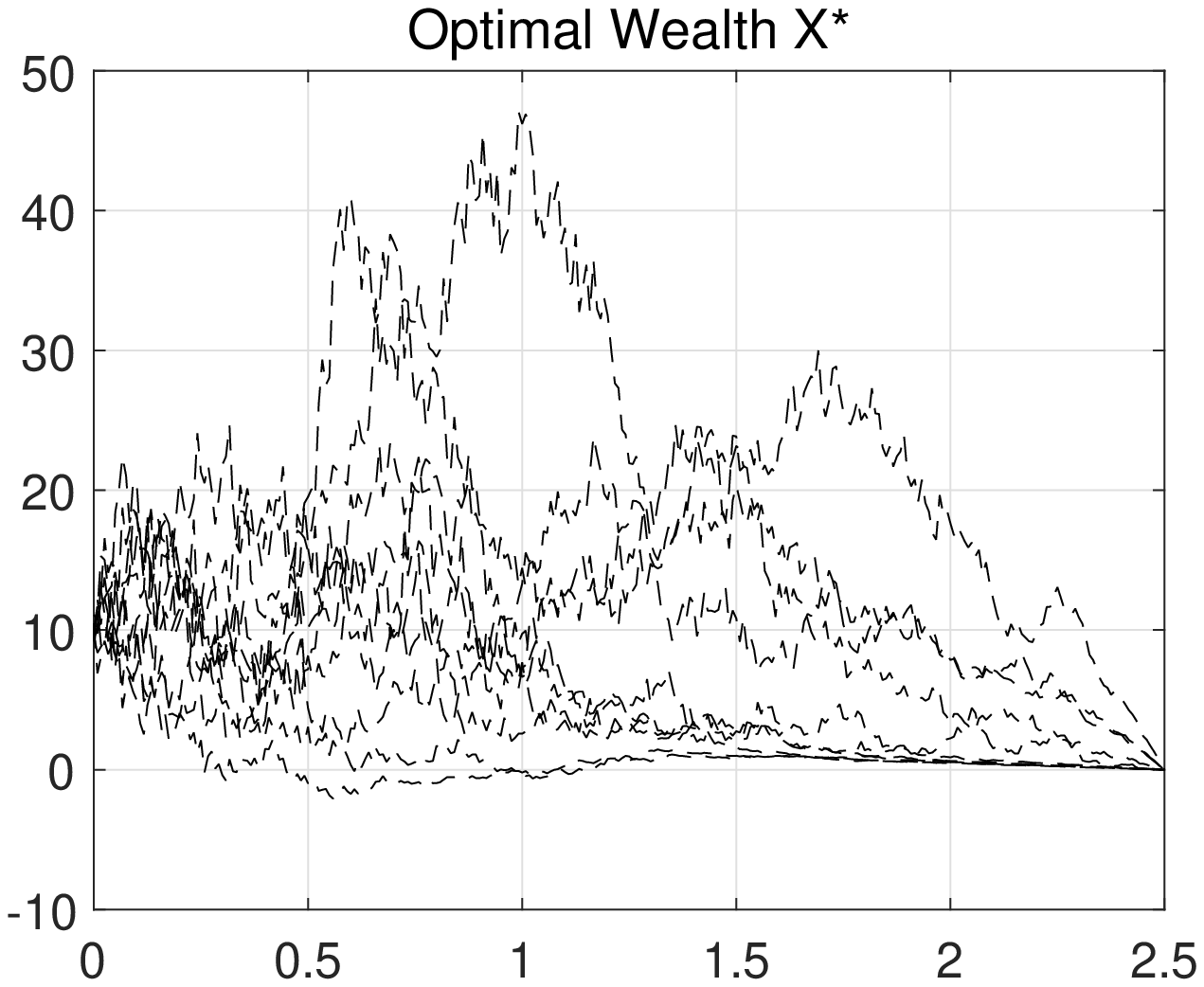}
\includegraphics[width=0.48\textwidth]{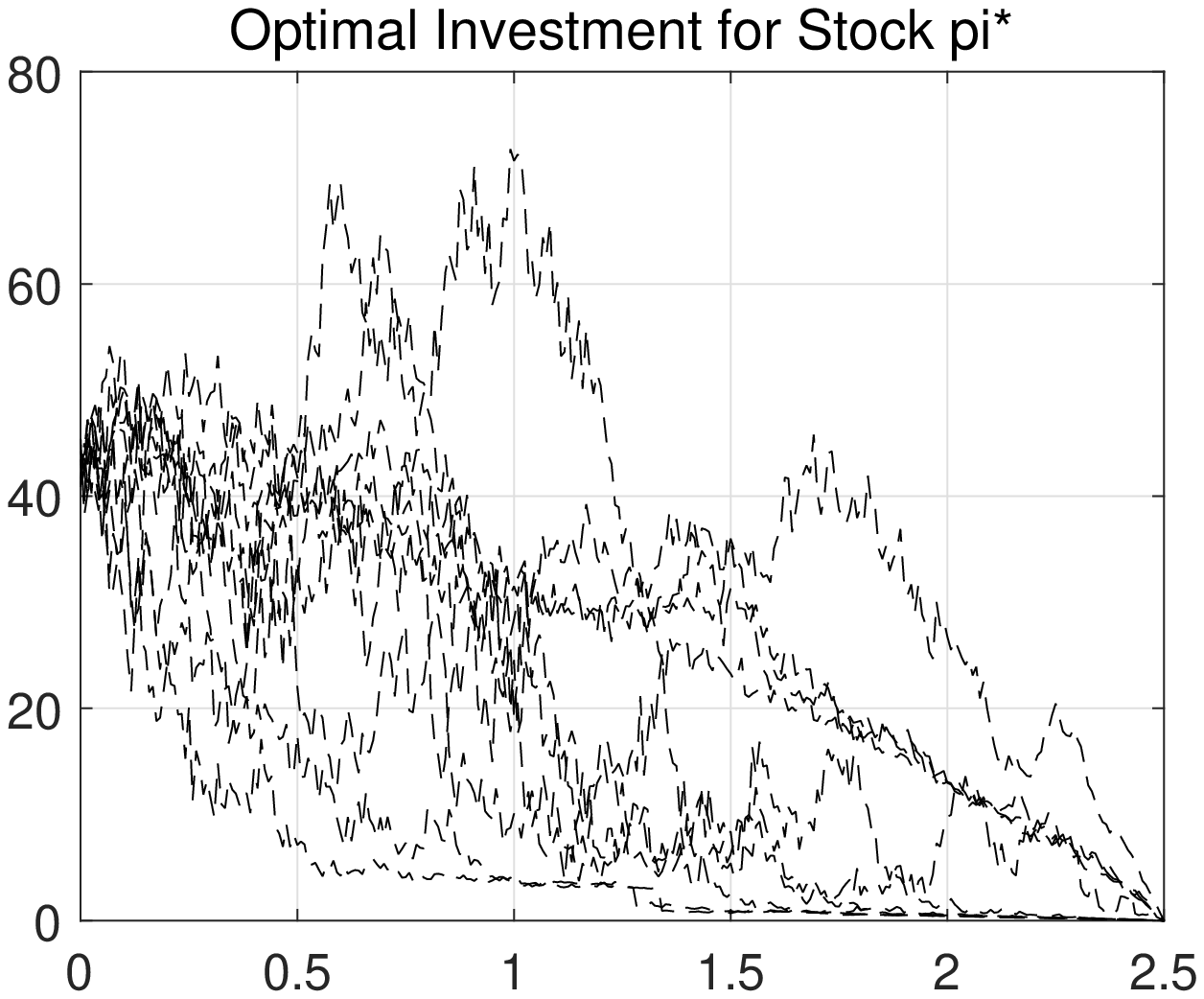}
\includegraphics[width=0.48\textwidth]{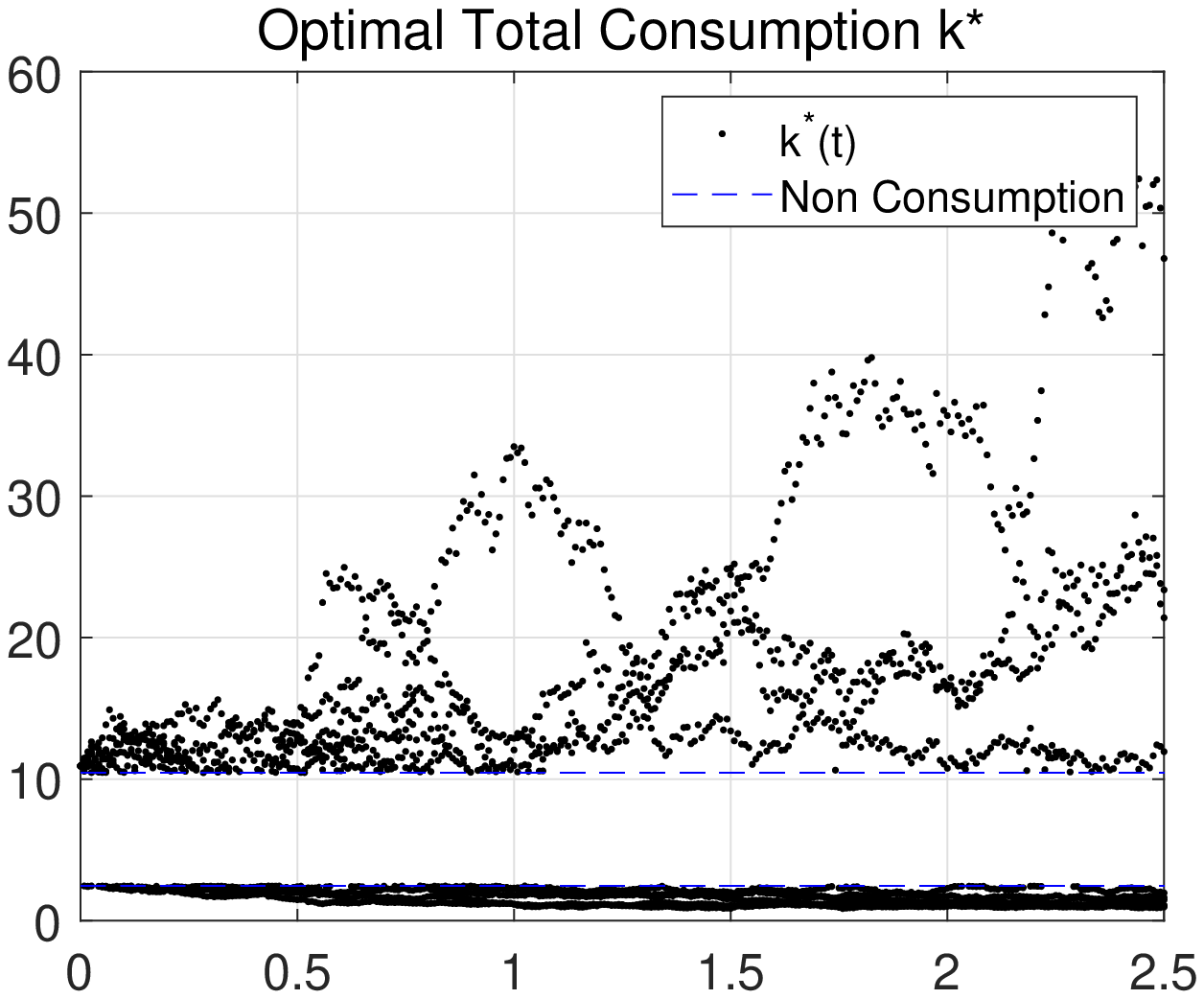}
\includegraphics[width=0.48\textwidth]{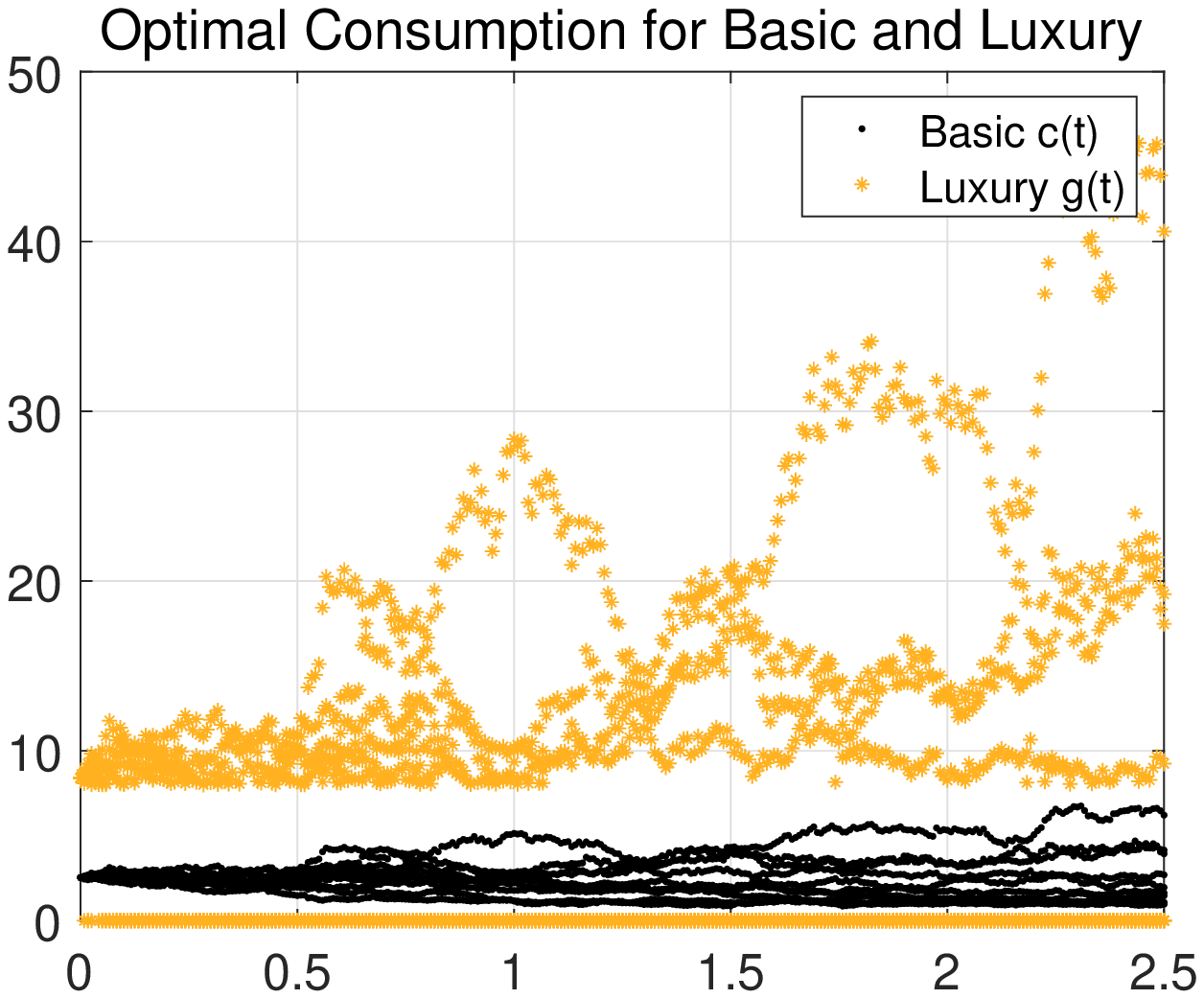}
\end{center}
\caption{Sample paths of the optimal strategies of $X^*(t)$, $\pi^*(t)$, $k^*(t)$ and $c(t)$ and $g(t)$ with the trigger level for luxury consumption $a=2$.
}
\label{Fig:Sample_Paths_lowlux}
\end{figure}

To see emergence of more frequent samples the luxury consumed, we next simulate ten sample paths by setting the high initial wealth $x$ of \$20 and low luxury trigger level $a$ of \$2, while keeping the other parameters the same with the baseline.
Both cases are illustrated in Figures~\ref{Fig:Sample_Paths_x20} and \ref{Fig:Sample_Paths_lowlux}, respectively, 
exhibiting apparently more samples the luxury consumed than the base setting.
The reason is that high initial wealth allows to consume the luxury goods without taking time to grow the current wealth by investing in the stock market.
Moreover, the trigger level $a=2$ computes the non-consumption range bounded below by $k_{-}\approx2.4$ and above by $k_{+}\approx10.4$, where scale goes down and width reduced compared to the basic setting.
It relaxes reluctance of desire to consuming the luxury even in the moderate wealth situation. 
Correspondingly, overall amount of optimal stock investment and the optimal wealth largely decreased compared to the baseline case.

\begin{figure}[H]
\begin{center}
\includegraphics[width=0.48\textwidth]{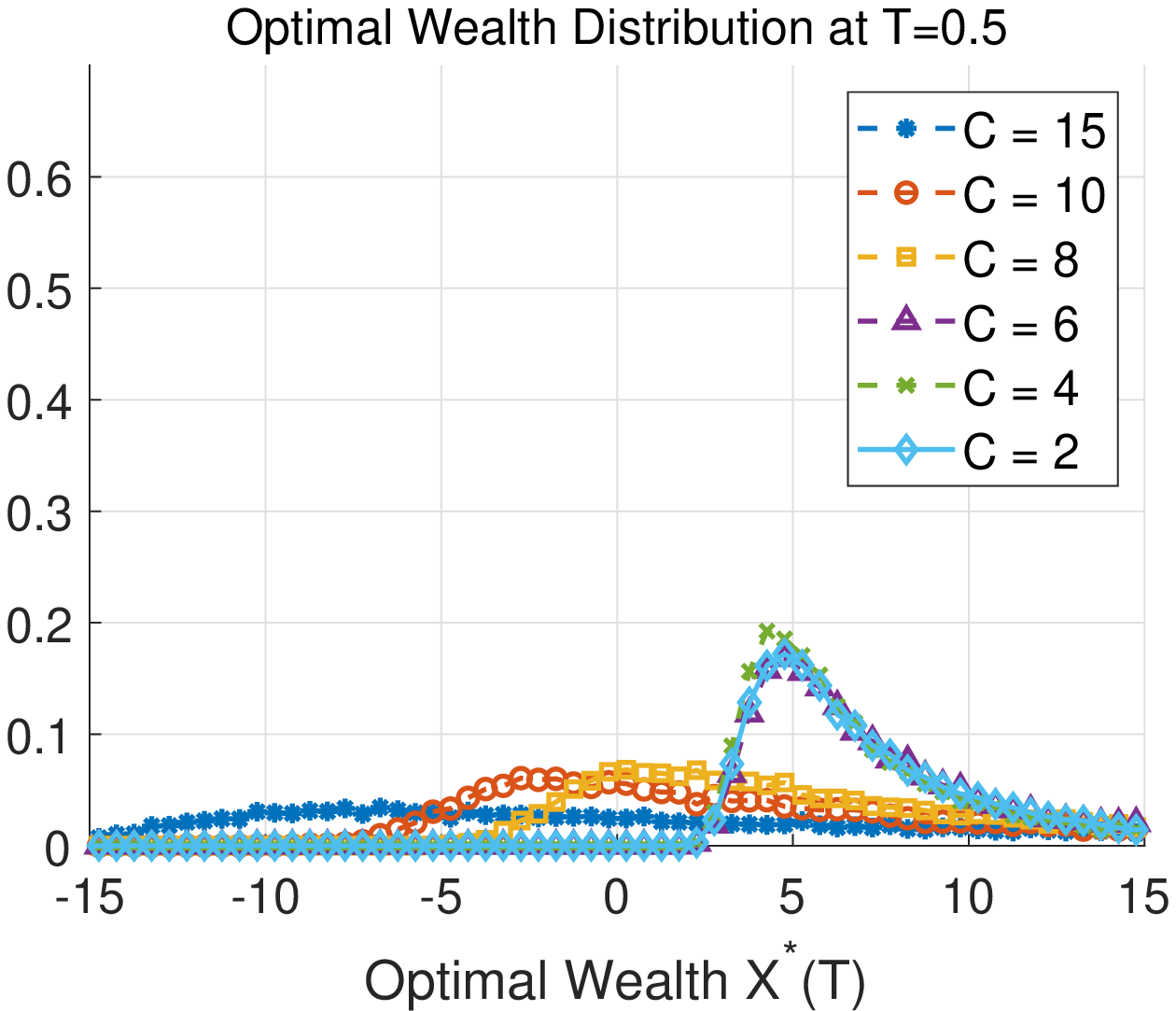}
\includegraphics[width=0.48\textwidth]{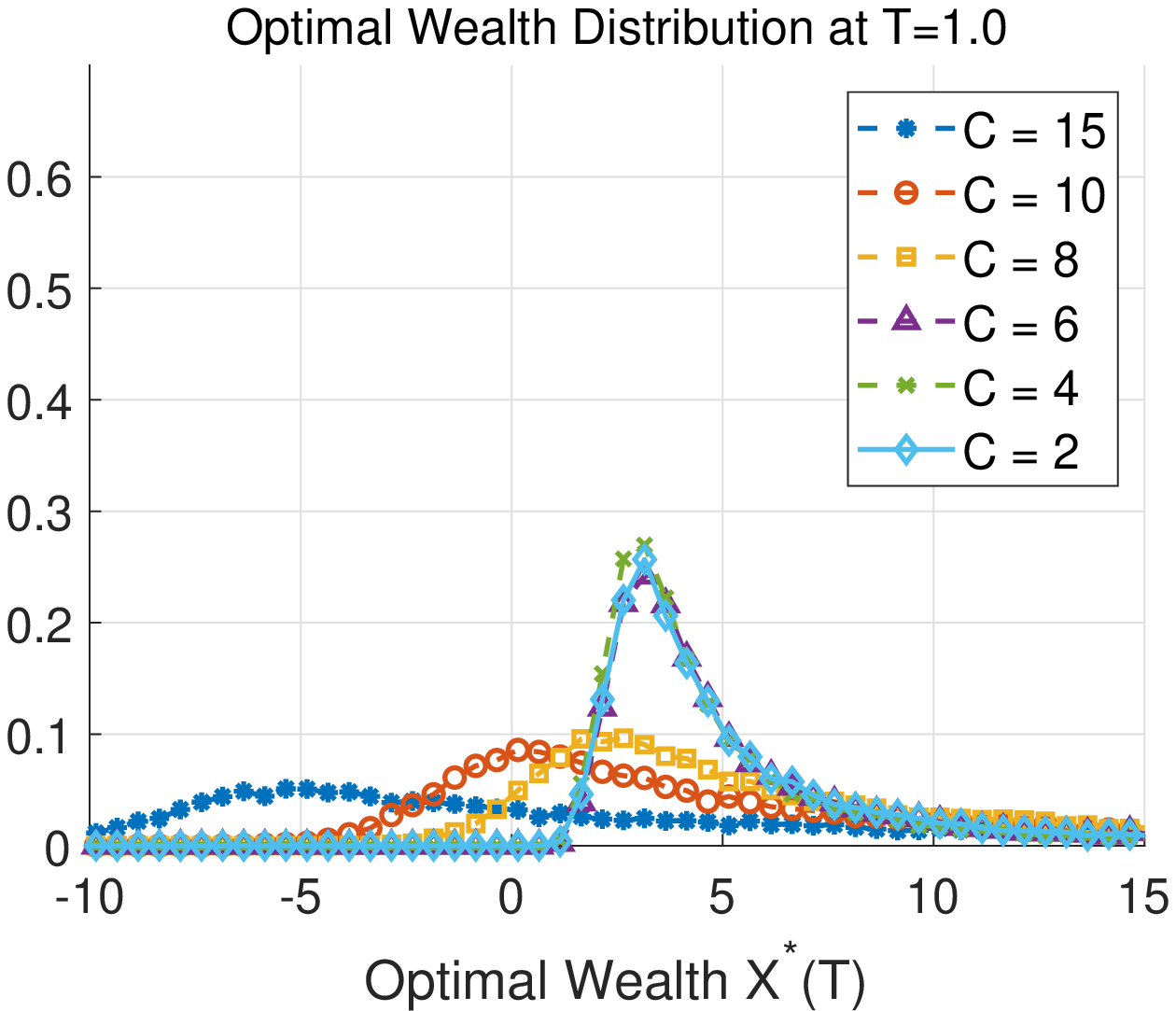}
\includegraphics[width=0.48\textwidth]{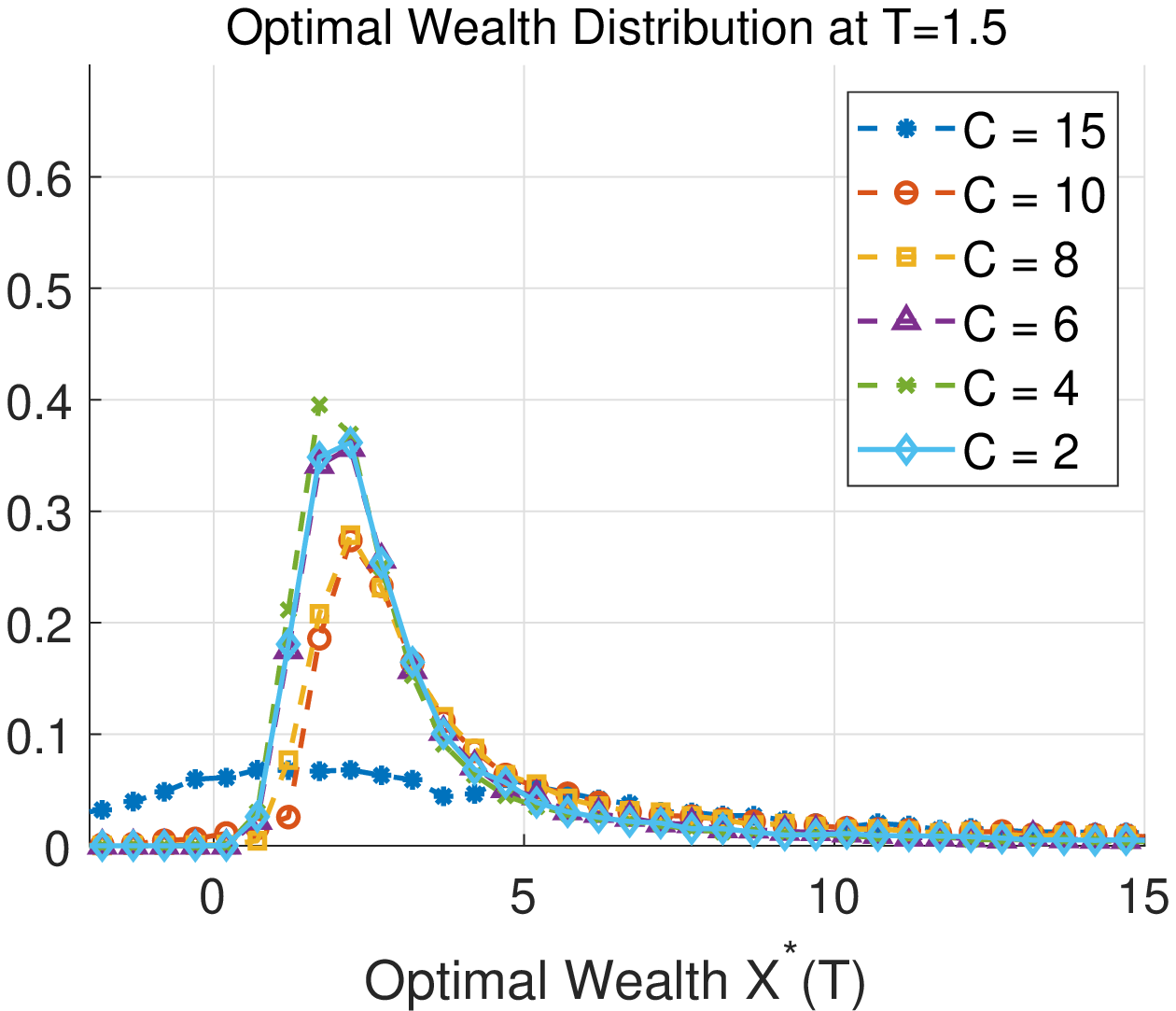}
\includegraphics[width=0.48\textwidth]{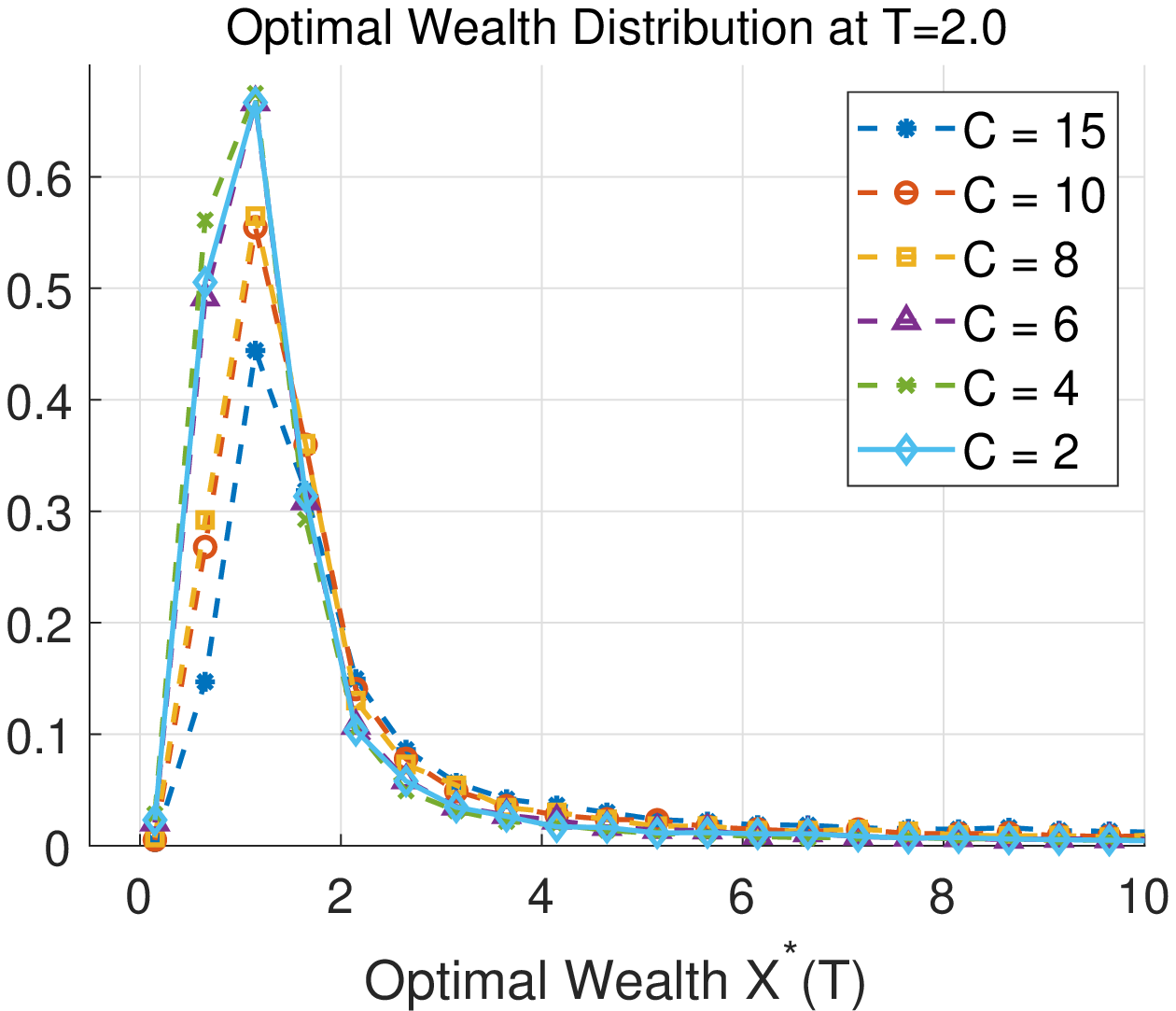}
\end{center}
\caption{Sample distributions of the optimal wealth $X^*(T)$ at $T=0.5,1.0,1.5,2.0$ (in order) for the initial income amount $C=2,4,\cdots,10$, and 15.}
\label{Fig:WealthDist} 
\end{figure}

\begin{figure}[H]
\begin{center}
\includegraphics[width=0.48\textwidth]{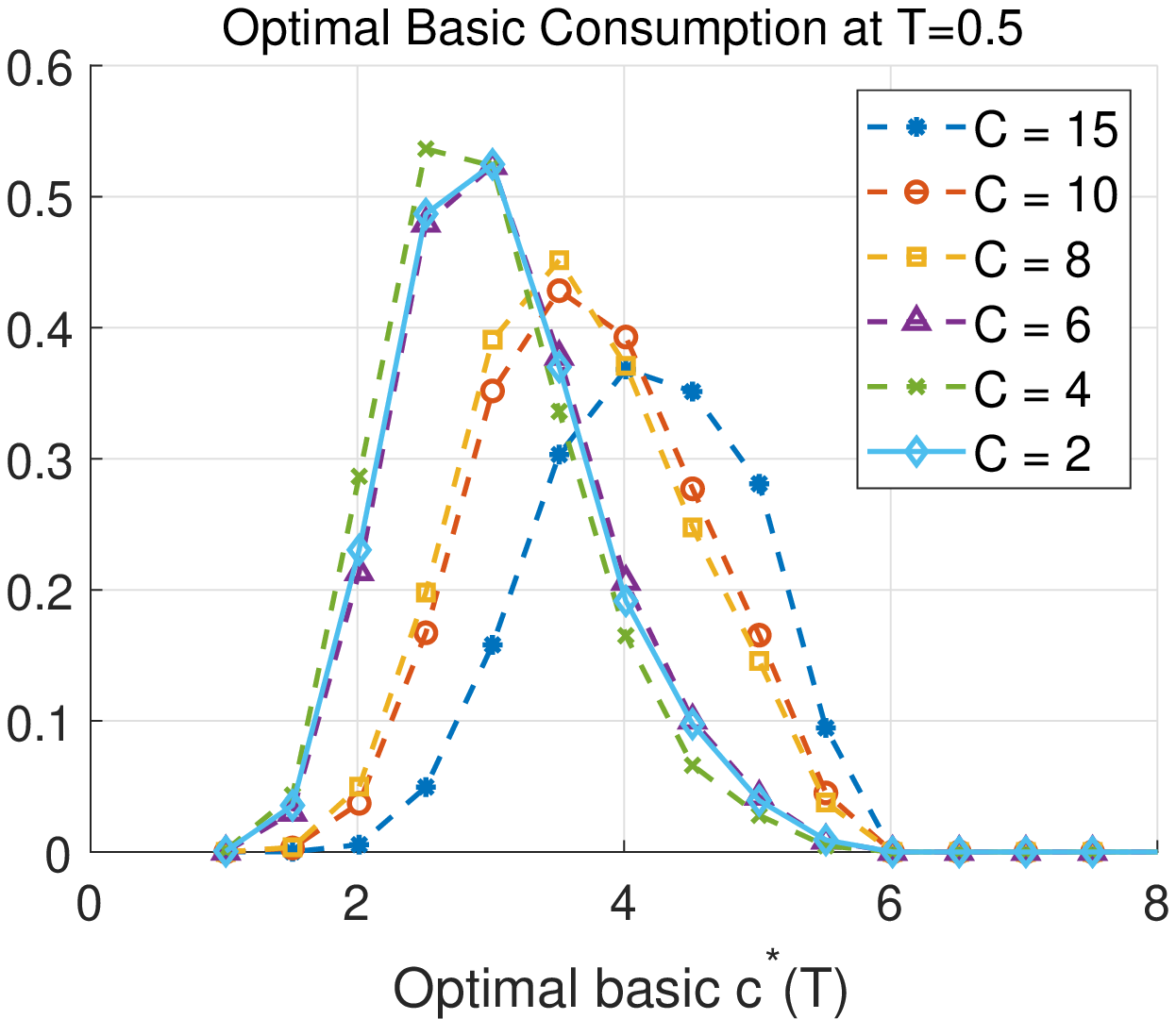}
\includegraphics[width=0.48\textwidth]{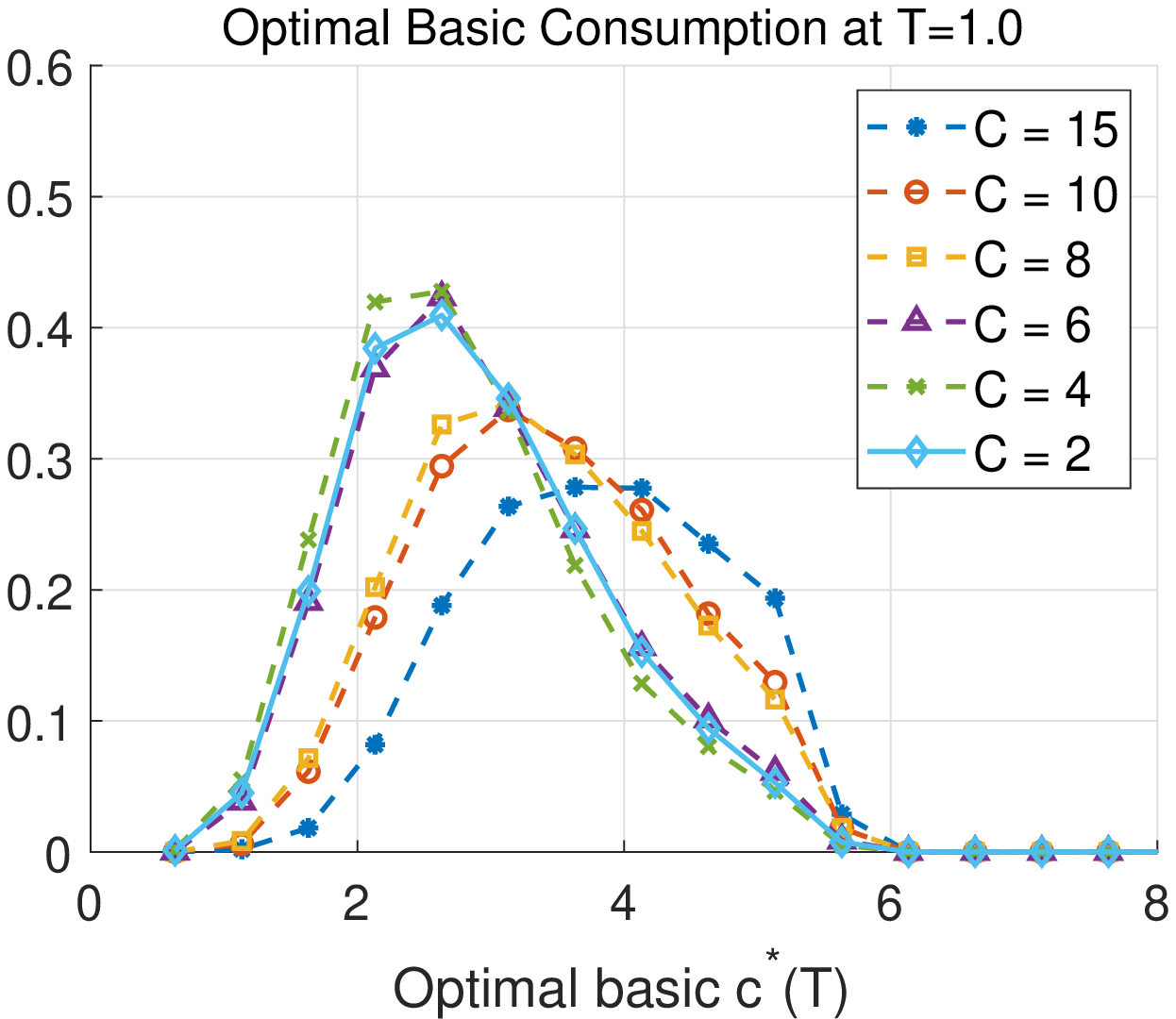}
\includegraphics[width=0.48\textwidth]{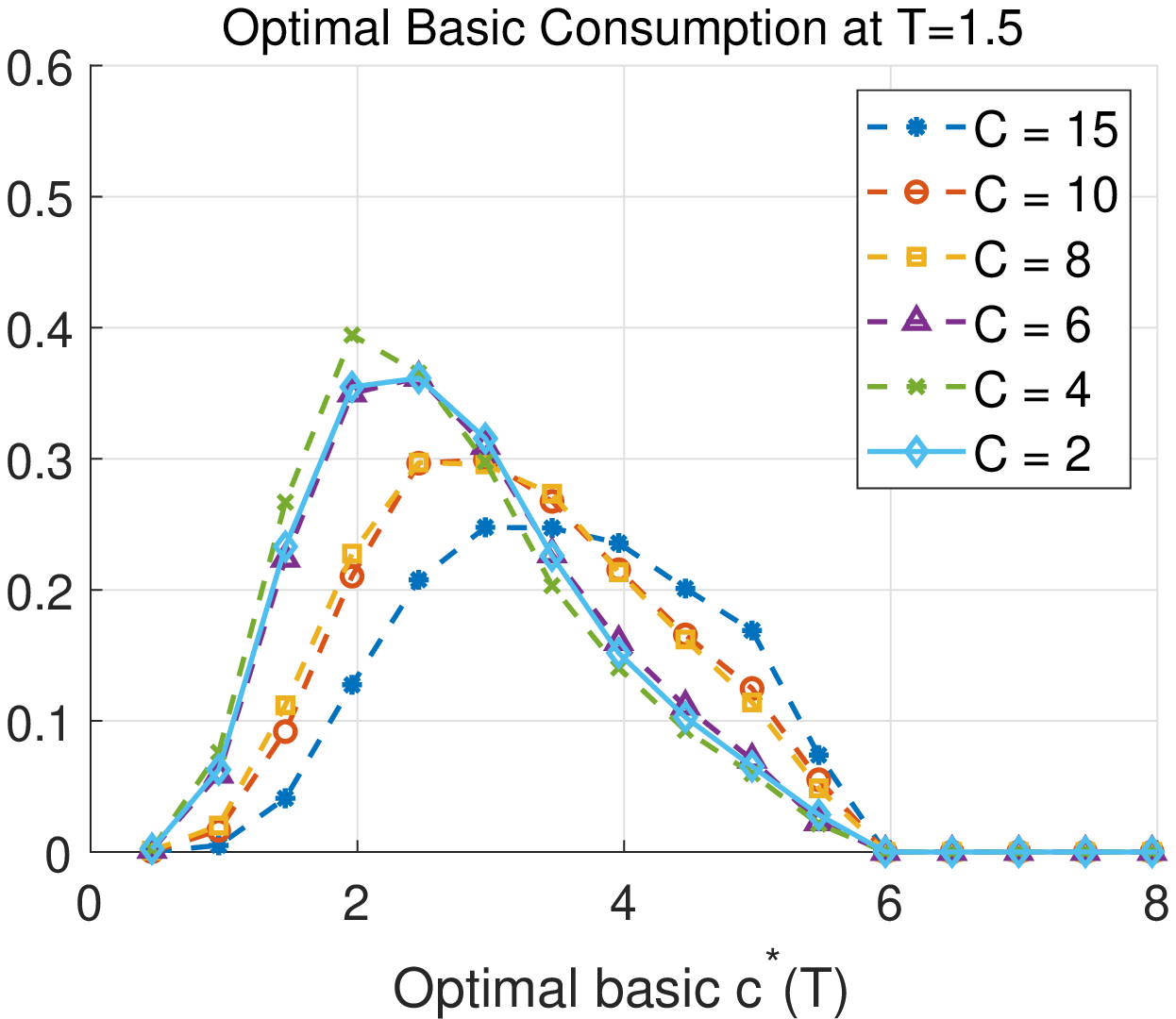}
\includegraphics[width=0.48\textwidth]{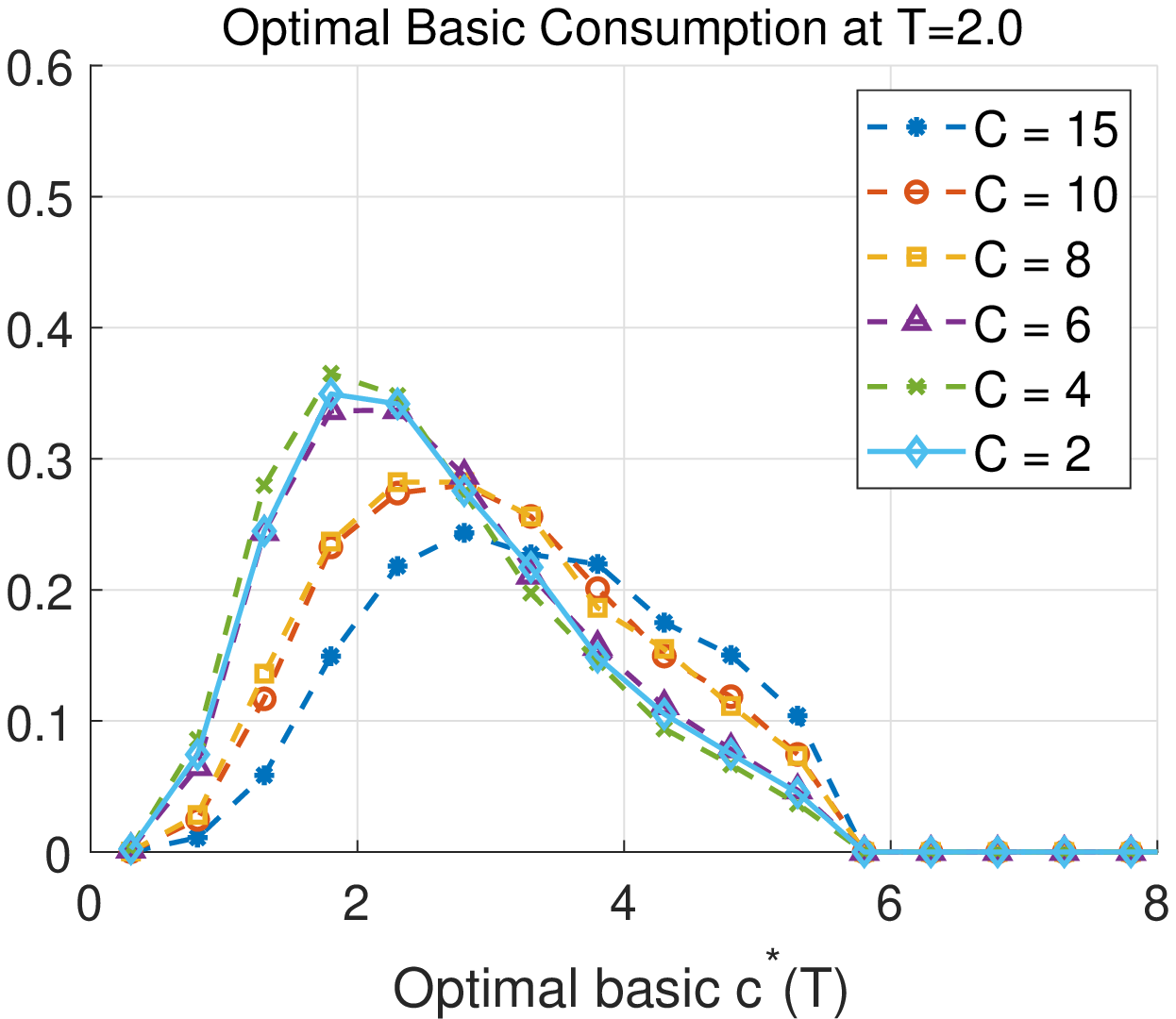}
\end{center}
\caption{Sample distributions of the optimal basic consumption distribution $c^*(T)$ at $T=0.5,1.0,1.5,2.0$ (in order) for the initial income amount $C=2,4,\cdots,10$, and 15.}
\label{Fig:BConsumeDist}
\end{figure}

\begin{figure}[H]
\begin{center}
\includegraphics[width=0.48\textwidth]{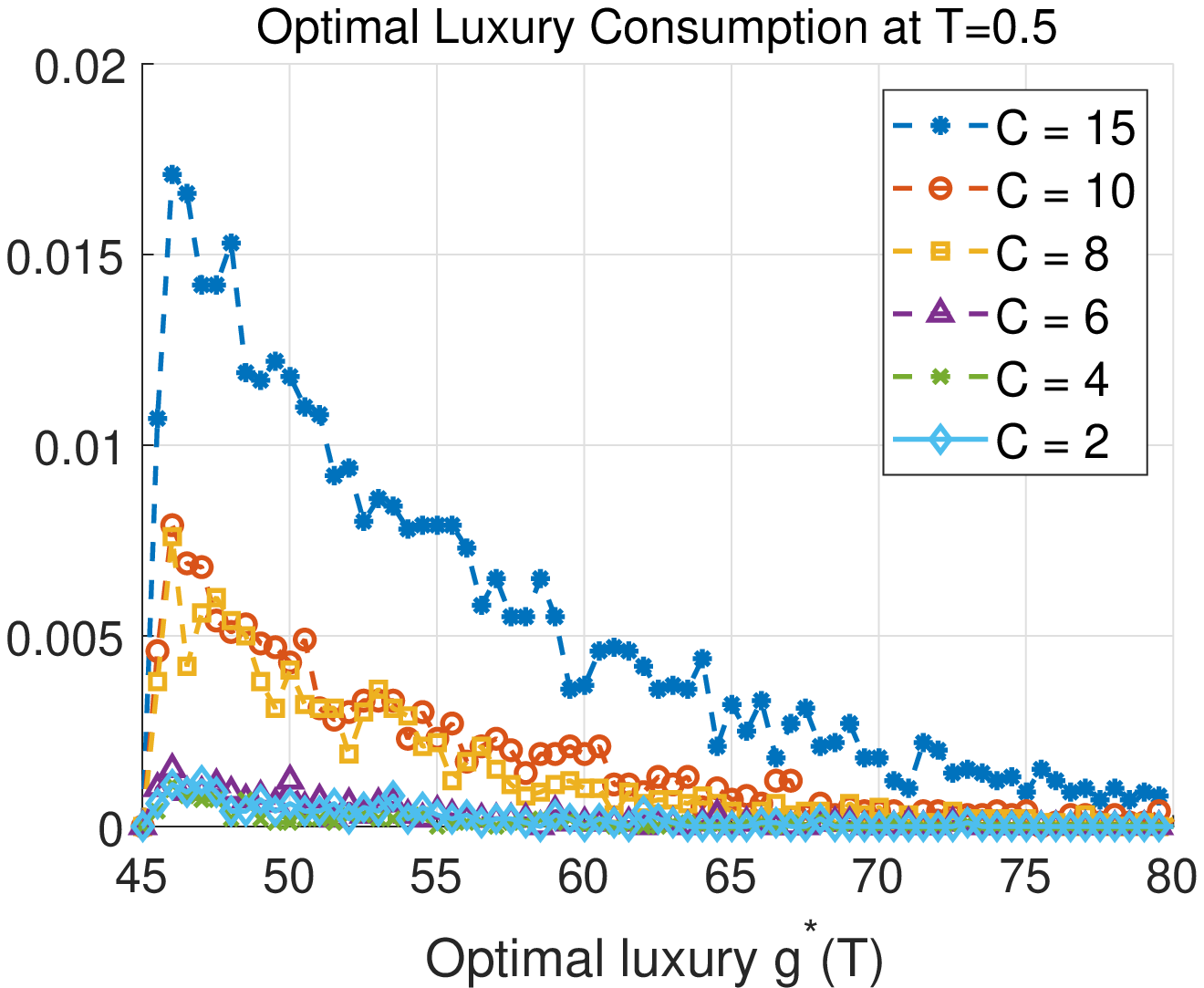}
\includegraphics[width=0.48\textwidth]{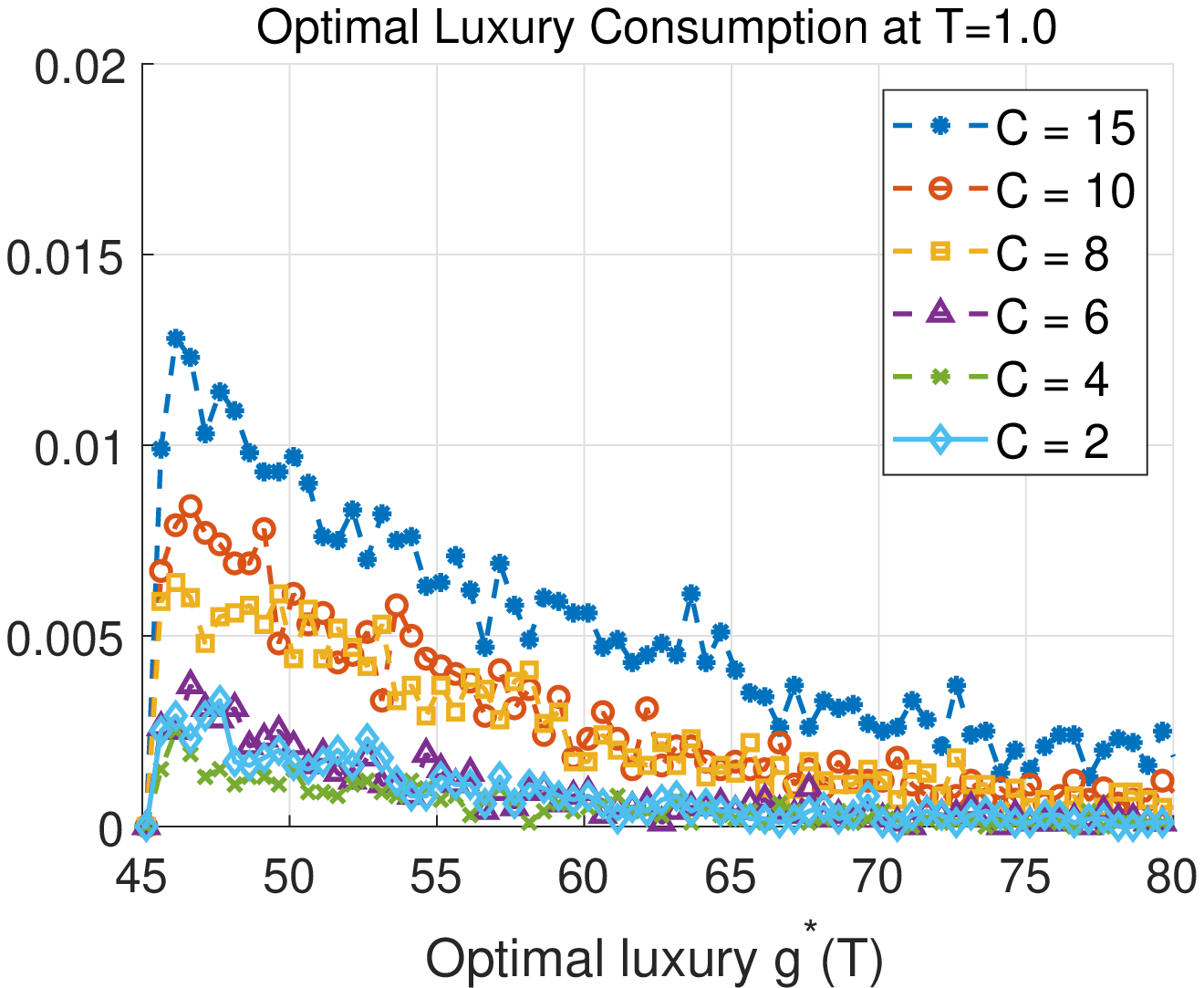}
\includegraphics[width=0.48\textwidth]{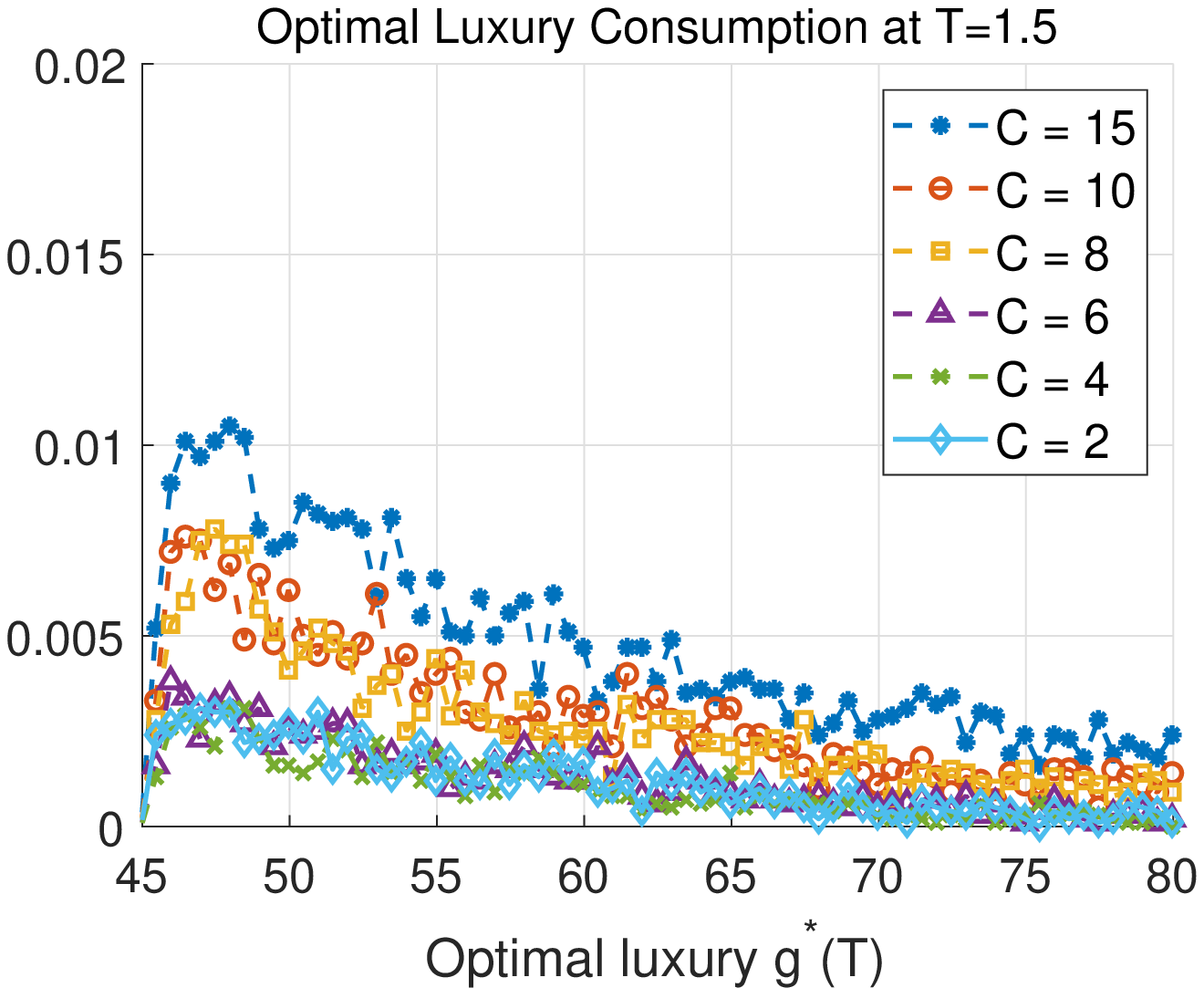}
\includegraphics[width=0.48\textwidth]{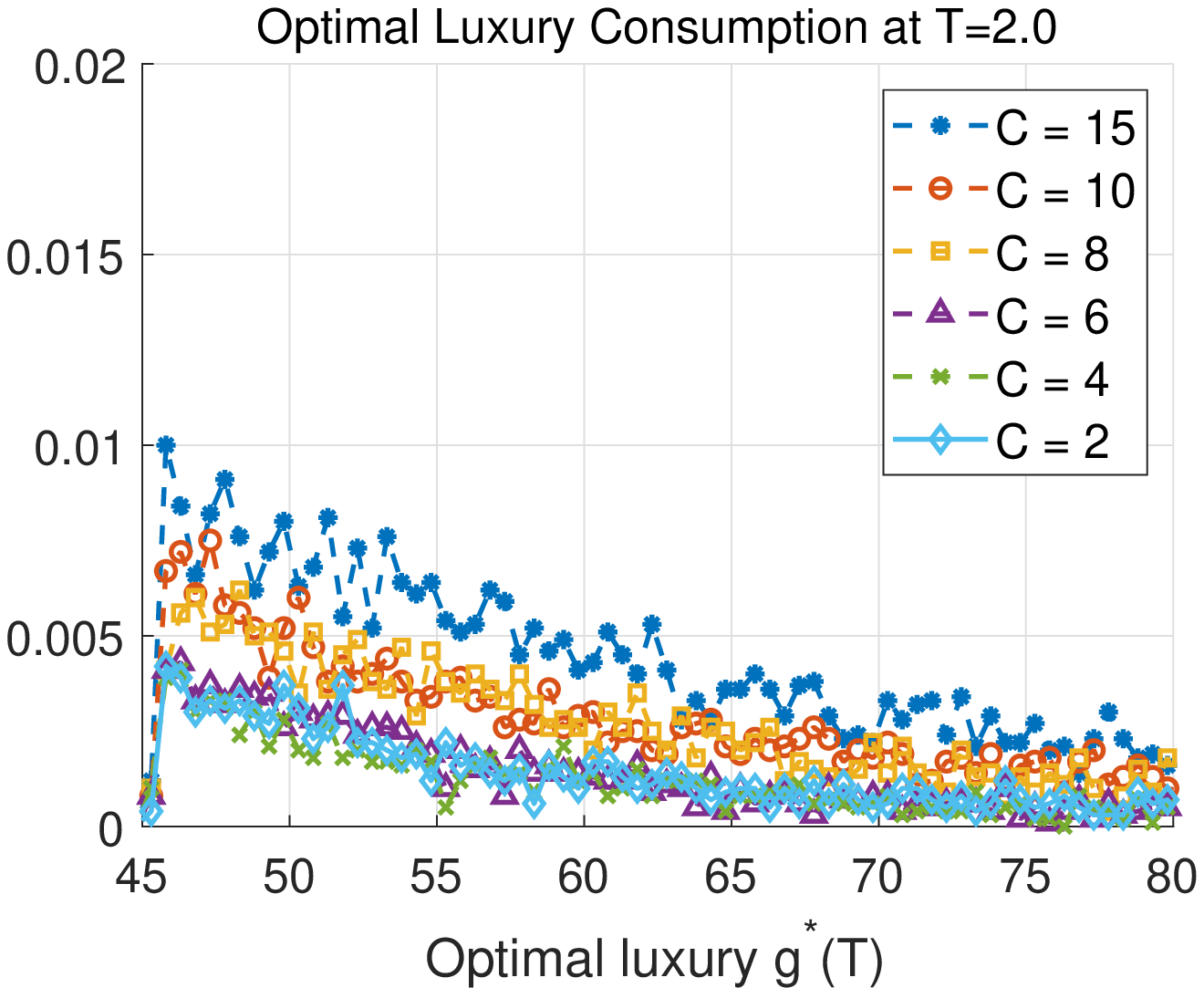}
\end{center}
\caption{Sample distributions of the optimal luxury consumption distribution $g^*(T)$ at $T=0.5,1.0,1.5,2.0$ (in order) for the initial income amount $C=2,4,\cdots,10$, and 15.}
\label{Fig:LConsumeDist}

\end{figure}

We investigate the evolution of distribution of the optimal strategies along with lifetime depending on the relative scale of income and labor cost.
Each distribution is constructed by generating 20,000 samples.
Figure~\ref{Fig:WealthDist} illustrates the sample distributions of the optimal wealth $X^*(T)$ as time $T$ reaches retirement under the chosen cases,
and Figures~\ref{Fig:BConsumeDist} and \ref{Fig:LConsumeDist} show the corresponding for the optimal basic $c^*(T)$ and luxury $g^*(T)$ consumption distributions. 
The basic consumption is apt to be normal shape of distribution on its possible range, whereas the luxury is to have exponentially decreasing shape of distribution. 
Note that this parameter set puts the non-consumption range of $(5.5, 45.5)$. 

\end{document}